\pdfoutput=1

\documentclass[11pt,a4paper]{article}

\usepackage{jcappub}
\usepackage[utf8]{inputenc}
\usepackage{amsmath}
\usepackage{amsthm}
\usepackage{amssymb}
\usepackage{enumitem}   
\usepackage[english]{babel}
\usepackage{url}
\usepackage{mathtools}
\usepackage{bbold}
\usepackage{slashed}
\usepackage{multirow}
\usepackage{lipsum}
\usepackage{xcolor}

\begin{document}


\hfill CERN-TH-2020-018


\title{New Sensitivity Curves for Gravitational-Wave Signals from Cosmological Phase Transitions}
\author{Kai~Schmitz}
\affiliation{Theoretical Physics Department, CERN, 1211 Geneva 23, Switzerland}
\emailAdd{kai.schmitz@cern.ch}


\abstract{\textit{Gravitational waves} (GWs) from \textit{strong first-order phase transitions} (SFOPTs) in the early Universe are a prime target for upcoming GW experiments.
In this paper, I construct novel \textit{peak-integrated sensitivity curves} (PISCs) for these experiments, which faithfully represent their projected sensitivities to the GW signal from a cosmological SFOPT by explicitly taking into account the expected shape of the signal.
Designed to be a handy tool for phenomenologists and model builders, PISCs allow for a quick and systematic comparison of theoretical predictions with experimental sensitivities, as I illustrate by a large range of examples.
PISCs also offer several advantages over the conventional \textit{power-law-integrated sensitivity curves} (PLISCs); in particular, they directly encode information on the expected signal-to-noise ratio for the GW signal from a SFOPT.
I provide semianalytical fit functions for the exact numerical PISCs of LISA, DECIGO, and BBO.
In an appendix, I moreover present a detailed review of the strain noise power spectra of a large number of GW experiments.
The numerical results for all PISCs, PLISCs, and strain noise power spectra presented in this paper can be downloaded from the Zenodo online repository~\cite{Schmitz:2020aaa}.
In a companion paper~\cite{Alanne:2019bsm}, the concept of PISCs is used to perform an in-depth study of the GW signal from the cosmological phase transition in the real-scalar-singlet extension of the standard model.
The PISCs presented in this paper will need to be updated whenever new theoretical results on the expected shape of the signal become available.
The PISC approach is therefore suited to be used as a bookkeeping tool to keep track of the theoretical progress in the field.}


\maketitle


\section{Introduction}
\label{sec:introduction}


Since the celebrated first direct detection of \textit{gravitational waves} (GWs) in September 2015~\cite{Abbott:2016blz}, the \textit{Advanced Laser Interferometer Gravitational-Wave Observatory} (aLIGO)~\cite{Harry:2010zz,TheLIGOScientific:2014jea} and the \textit{Advanced Virgo} (aVirgo) experiment~\cite{TheVirgo:2014hva} have observed a multitude of GW events~\cite{LIGOScientific:2018mvr,Abbott:2020uma}.
One of the signals recorded during the first two aLIGO\,/\,aVirgo observing runs~\cite{Abbott:2019ebz} originated from the coalescence of a binary neutron star~\cite{TheLIGOScientific:2017qsa}, while all other signals were due to mergers of binary black holes~\cite{Abbott:2016blz,LIGOScientific:2018mvr,Abbott:2016nmj,Abbott:2017vtc,Abbott:2017oio,Abbott:2017gyy}.
First results from the third observing run are reported on in Ref.~\cite{Abbott:2020uma}.
After the breakthrough discovery of these transient astrophysical sources, one of the next key objectives in GW astronomy is going to be the detection of a \textit{stochastic GW background} (SGWB) from cosmological sources~\cite{Maggiore:1999vm,LIGOScientific:2019vic}.
A wide range of violent phenomena in the early Universe can give rise to a primordial SGWB~\cite{Caprini:2018mtu,Christensen:2018iqi}, among which cosmological phase transitions~\cite{Mazumdar:2018dfl,Hindmarsh:2020hop} are a preeminent example.
\textit{Strong first-order phase transitions} (SFOPTs) occur in numerous extensions of the \textit{standard model} (SM) of particle physics and can potentially lead to a strong GW signal~\cite{Weir:2017wfa}.
Many scenarios especially predict a GW signal that is peaked at frequencies in the milli-Hertz range.
This makes GWs from a SFOPT a prime target for future satellite-borne interferometers in space, such as the \textit{Laser Interferometer Space Antenna} (LISA)~\cite{Audley:2017drz,Baker:2019nia}, which is scheduled for launch in the 2030's.
Consequently, many authors have recently studied the prospects of probing the dynamics of a SFOPT in scenarios \textit{beyond the standard model} (BSM) with LISA~\cite{Caprini:2015zlo,Caprini:2019egz} and other upcoming GW experiments (see, \textit{e.g.}, the incomplete list of BSM models in Refs.~\cite{Dev:2016feu,Huang:2018aja,Croon:2018erz,Okada:2018xdh,Baldes:2018nel,Chiang:2018gsn,Alves:2018oct,Baldes:2018emh,Ellis:2018mja,Madge:2018gfl,Ahriche:2018rao,Prokopec:2018tnq,Fujikura:2018duw,Beniwal:2018hyi,Brdar:2018num,Miura:2018dsy,Addazi:2018nzm,Shajiee:2018jdq,Marzo:2018nov,Breitbach:2018ddu,Angelescu:2018dkk,Alves:2018jsw,Kannike:2019wsn,Fairbairn:2019xog,Hasegawa:2019amx,Helmboldt:2019pan,Dev:2019njv,Huang:2019riv,Zhou:2020idp,Mohamadnejad:2019vzg,Kannike:2019mzk,Bian:2019szo,Paul:2019pgt,Dunsky:2019upk,Athron:2019teq,Bian:2019kmg,Brdar:2019fur,Wang:2019pet,Alves:2019igs,DeCurtis:2019rxl,Addazi:2019dqt,Greljo:2019xan,Archer-Smith:2019gzq,Aoki:2019mlt,Hall:2019ank,Brdar:2019qut,Haba:2019qol,Carena:2019une,Hall:2019rld,Heurtier:2019beu,Baker:2019ndr,Chway:2019kft,DelleRose:2019pgi,vonHarling:2019gme,Chiang:2019oms,Zhou:2020ojf,DiBari:2020bvn}).


The GW signal from a SFOPT is conventionally expressed in terms of a GW energy density spectrum $\Omega_{\rm signal}\left(f\right)$ as a function of GW frequency $f$, while the instantaneous sensitivity of a GW experiment is quantified in terms of a noise spectrum $\Omega_{\rm noise}\left(f\right)$ [see Eq.~\eqref{eq:Osignalnoise} for the precise definition of these two quantities].
With these two spectra at one's disposal, one is able to assess the chances that the predicted signal is going to be experimentally detected.
In practice, this is typically done by adopting one or both of the following two strategies:


\medskip\noindent\textbf{Strategy \#1:} Compute the associated \textit{signal-to-noise ratio} (SNR) $\varrho$ by integrating over the experiment's total observing time $t_{\rm obs}$ and accessible frequency range $\left[f_{\rm min},f_{\rm max}\right]$~\cite{Allen:1996vm,Allen:1997ad,Maggiore:1999vm},
\begin{align}
\label{eq:rho}
\varrho = \left[n_{\rm det}\, t_{\rm obs} \int_{f_{\rm min}}^{f_{\rm max}} df \left(\frac{\Omega_{\rm signal}\left(f\right)}{\Omega_{\rm noise}\left(f\right)}\right)^2\right]^{1/2} \,,
\end{align}
where $n_{\rm det}$ distinguishes between experiments that aim at detecting the SGWB by means of an auto-correlation ($n_{\rm det} = 1$) or a cross-correlation ($n_{\rm det} = 2$) measurement.
Then, if $\varrho$ turns out to be larger than some threshold value, $\varrho > \varrho_{\rm thr}$, one concludes that the GW experiment under consideration will be able to detect the predicted GW signal.

\medskip\noindent\textbf{Strategy \#2:} Construct the \textit{power-law-integrated sensitivity curve} (PLISC) $\Omega_{\rm PLIS}\left(f\right)$~\cite{Thrane:2013oya} based on the noise spectrum $\Omega_{\rm noise}\left(f\right)$ (and some $\varrho_{\rm thr}$) and compare it to the signal spectrum $\Omega_{\rm signal}\left(f\right)$.
Then, if the signal and the PLISC intersect, such that $\Omega_{\rm signal}\left(f\right) > \Omega_{\rm PLIS}\left(f\right)$ for some $f$, one typically also concludes that the experiment will be able to detect the signal.


\medskip Both strategies have several advantages and disadvantages.
The SNR approach, \textit{e.g.}, has a clearly defined statistical interpretation and is applicable for an arbitrarily shaped signal $\Omega_{\rm signal}$.
On the other hand, the SNR $\varrho$ no longer contains any spectral information because of the integral over the frequency range $\left[f_{\rm min},f_{\rm max}\right]$ in Eq.~\eqref{eq:rho}.
This is somewhat unfortunate.
Ideally, one would like to indicate whether a certain signal is going to be experimentally detected or not directly in a plot of the signal spectrum $\Omega_{\rm signal}$.
In fact, this has been one of the main motivations behind the idea of constructing PLISCs.
However, as $\varrho$ is no longer a function of $f$, there is no canonical way of including information on $\varrho$ in a plot of $\Omega_{\rm signal}$.
As a consequence, many authors resort to graphical representations of $\varrho$ as a function of the underlying model parameters $\left\{p_i\right\}$ that determine the shape of the signal, $\varrho = \varrho\left(\left\{p_i\right\}\right)$.
This typically results in plots of (a subspace of) the model parameter space that contain information on $\varrho$ in the form of contour lines or a color code.
While such information may be very useful for a number of reasons, it can easily happen that one ends up with plots that show $\varrho$ only as a function of quantities that are not directly accessible by experiments. 
This may be regarded as a disadvantage of the SNR approach, as one may prefer to indicate an experiment's sensitivity reach in terms of physical observables rather than in terms of auxiliary model parameters.
Often times, one also looses information because one is forced to restrict oneself to lower-dimensional hypersurfaces (typically two-dimensional slices) in the higher-dimensional model parameter space.
Finally, one may argue that another disadvantage consists in the fact that the SNR approach requires a larger computational effort.
While the PLISC approach is essentially a graphical one, computing the SNR $\varrho$ always involves the extra step of carrying out the frequency integration in Eq.~\eqref{eq:rho}.
This is part of the reason why many authors in the literature actually content themselves with a graphical analysis in terms of PLISCs and refrain from performing a proper SNR analysis.


The PLISC approach, by contrast, manages to convey a useful and graphical impression of an experiment's sensitivity directly in terms of plots of the GW spectrum.
However, it is important to note that, by construction, PLISCs do not encode information on the expected SNR as soon as the spectrum deviates from a pure power law.
Therefore, in realistic situations where the signal is expected to have a richer structure than just a simple power law, PLISCs should rather be regarded as a qualitative visualization than a quantitative statistical tool.


The aim of the present paper is to remedy the shortcomings of strategies \#1 and \#2 for the particular case of GWs from a SFOPT in the early Universe.
The main observation in this case is that the shape of the signal is model-independent to first approximation, such that it is actually not necessary to perform the frequency integral in Eq.~\eqref{eq:rho} over and over again.
Instead, it is possible to compute it once and for all, whereupon the numerical result may be used for the signal in any BSM model that one is interested in.
In realistic scenarios, the GW signal from a SFOPT receives contributions from several different physical sources (see Sec.~\ref{sec:signal}).
However, for illustration, let us suppose for a moment that there is just one physical source of GWs.
In this case, the signal spectrum can be schematically written as
\begin{align}
\Omega_{\rm signal}\left(f\right) = \Omega_{\rm signal}^{\rm peak}\left(\left\{p_i\right\}\right)\, \mathcal{S}\left(f,f_{\rm peak}\right) \,,
\end{align}
where $\Omega_{\rm signal}^{\rm peak}\left(\left\{p_i\right\}\right)$ denotes the peak strength of the signal at the peak frequency $f_{\rm peak}$, and where the shape function $\mathcal{S}$ describes the frequency dependence of the signal in the vicinity of the peak frequency.
Here, the shape function $\mathcal{S}$ is assumed to be model-independent, whereas the peak amplitude $\Omega_{\rm signal}^{\rm peak}$ captures the detailed dependence on the underlying model parameters $\left\{p_i\right\}$.
For a signal of this form, it is then possible to rewrite Eq.~\eqref{eq:rho} as follows,
\begin{align}
\label{eq:rhoPIS}
\varrho = \frac{\Omega_{\rm signal}^{\rm peak}\left(\left\{p_i\right\}\right)}{\Omega_{\rm PIS}\left(f_{\rm peak}\right)} \,,
\end{align}
where $\Omega_{\rm PIS}\left(f_{\rm peak}\right)$ denotes what we shall refer to as the \textit{peak-integrated sensitivity} (PIS),
\begin{align}
\label{eq:PIS}
\Omega_{\rm PIS}\left(f_{\rm peak}\right) = \left[n_{\rm det}\, t_{\rm obs} \int_{f_{\rm min}}^{f_{\rm max}} df \left(\frac{\mathcal{S}\left(f,f_{\rm peak}\right)}{\Omega_{\rm noise}\left(f\right)}\right)^2\right]^{-1/2} \,.
\end{align}
For any given experiment, the integral in Eq.~\eqref{eq:PIS} only needs to be computed once.
As soon as this has been done, one can construct a \textit{peak-integrated sensitivity curve} (PISC) by plotting $\Omega_{\rm PIS}$ as a function of the frequency $f_{\rm peak}$.
In this plot, the original (one-dimensional) spectrum $\Omega_{\rm signal}$ then reduces to a single (zero-dimensional) point $\big(f_{\rm peak},\Omega_{\rm signal}^{\rm peak}\big)$.


Thus far, the standard procedure for phenomenologists and model builders interested in studying the GW signal from a SFOPT typically involved three steps.
First, one had to consult the latest theoretical literature on the functional forms of the peak amplitude $\Omega_{\rm signal}^{\rm peak}$ and the shape function $\mathcal{S}$.
Then, in a second step, one had to consult the experimental literature on the shape of the noise spectrum $\Omega_{\rm noise}$ for a given experiment.
And finally, one had to tie both ends together and perform the last step, namely, the frequency integration in Eq.~\eqref{eq:rho}, by hand.
The PISC approach now closes the gap between the input from the theory side and the input from the experimental side by rendering the final frequency integration obsolete.
In fact, we argue that it unifies the advantages of the two standard strategies \#1 and \#2 that we outlined above, while at the same time avoiding their disadvantages:
\begin{enumerate}
\item As soon as the PISC has been constructed for a particular experiment, it is no longer necessary to perform a frequency integration on a parameter-point-by-parameter-point basis.
Instead, one can simply work out a fit function for the exact numerical PISC (see Sec.~\ref{subsec:fits}), which enables one to write down a quasianalytical expression for the SNR in Eq.~\eqref{eq:rhoPIS}.
This property turns PISCs into a handy and ready-to-use tool for the phenomenological exploration of BSM models that predict GWs from a SFOPT.
\item The PISC approach works for an arbitrary signal shape $\mathcal{S}$.
Unlike the PLISC approach, it is not restricted to signals that are (reasonably well) described by a pure power law.
\item At the same time, it results in a useful visual representation of an experiment's sensitivity in terms of physical observables (as opposed to auxiliary model parameters), namely, the peak frequency $f_{\rm peak}$ and the peak amplitude $\Omega_{\rm signal}^{\rm peak}$ of the GW spectrum.
\item According to the PISC approach, signal spectra are projected onto individual points in the $f_{\rm peak}$\,--\,$\Omega_{\rm signal}^{\rm peak}$ plane.
This can be used to generate scatter plots that allow one to identify what one may call a model's signal region.
Plotting this signal region in combination with a PISC then indicates to what extent the model is going to be probed experimentally.
This aspect is elaborated on in more detail in the companion paper~\cite{Alanne:2019bsm}.
\item If new theoretical results should require a revision of the spectral shape $\mathcal{S}$, it suffices to update the experimental PISCs, while the model-specific signal regions remain unaffected.
Similarly, new insights into the dependence of $f_{\rm peak}$ and $\Omega_{\rm signal}^{\rm peak}$ on the underlying SFOPT parameters only cause the signal regions to shift, but leave the experimental PISCs unchanged.
This facilitates the update of sensitivity plots comparing the predictions of many models at the same time (see also Sec.~\ref{subsec:update} and Ref.~\cite{Schmitz:2020rag}).
\item The fact that PISC plots correspond to \textit{projections} onto the $f_{\rm peak}$\,--\,$\Omega_{\rm signal}^{\rm peak}$ plane distinguishes them from the usual plots of the SNR $\varrho$ on two-dimensional \textit{slices} through the higher-dimensional parameter space that one often encounters.
In the PISC approach, it is not necessary to keep a subset of model parameters fixed at specific values.
\item Most importantly, the PISC approach retains the full information on the SNR.
For a data point $\big(f_{\rm peak},\Omega_{\rm signal}^{\rm peak}\big)$, the expected SNR simply corresponds to the vertical separation between this point and the PISC of interest.
This facilitates the implementation and comparison of different SNR thresholds $\varrho_{\rm thr}$.
All the relevant information is encoded on the $y$-axis of our plots; no additional color code or contour lines are needed.
In addition, our plots allow for an easy comparison of the PISCs and hence expected SNRs for different experiments.
Such a comparison would not be feasible in plots using a color code for the SNR and significantly more complicated in SNR contour plots.
\end{enumerate}


Before we turn to a more detailed presentation of the PISC approach, it is worth pointing out some similarities to alternative treatments in the literature.
Refs.~\cite{Kannike:2019mzk,Addazi:2019dqt}, \textit{e.g.}, also present scatter plots of possible peak frequencies and peak amplitudes in specific BSM models.
These predictions are, however, combined with the usual sensitivity curves, such that the final plots do not contain any information on the expected SNR.
Moreover, Ref.~\cite{Hashino:2018wee} studies the sensitivity of future GW satellite experiments based on a Fisher matrix analysis that quantifies the precision with which one will be able to reconstruct observables such as $f_{\rm peak}$ and $\Omega_{\rm signal}^{\rm peak}$ from real data.
Studies of this kind will be an important part of the data analysis once a SGWB signal has been detected.
We, however, note that, compared to Ref.~\cite{Hashino:2018wee}, the aim of the present paper is a slightly different one.
Instead of performing a $\delta\chi^2$ analysis, we actually go one step back and focus on the maximal SNR $\varrho$ at which GWs from cosmological phase transitions can be detected in upcoming experiments.
In this sense, we hope that the novel concept of PISCs will first and foremost prove to be a helpful tool for the systematic exploration and comparison of the GW phenomenology in different BSM models.


The rest of the paper is organized as follows.
In Sec.~\ref{sec:signal}, we will first review the generation of GWs during a SFOPT in the early Universe and collect all expressions that are necessary to compute the signal spectrum $\Omega_{\rm signal}$.
In Sec.~\ref{sec:pisc}, we will then introduce the concept of PISCs (see Sec.~\ref{subsec:definition}) and highlight a few possible applications.
This will include the detailed discussion of a \textit{benchmark point} (BP) in a particular SM extension (see Sec.~\ref{subsec:example}), the comparison of the GW signals predicted by different BSM models (see Sec.~\ref{subsec:comparison}), a few remarks on how to relax the assumptions underlying the construction of PISCs (see Sec.~\ref{subsec:assumptions}), a comment on runaway phase transitions in vacuum (see Sec.~\ref{subsec:runaway}), a discussion of how new theoretical results can be used to update our PISC plots (see Sec.~\ref{subsec:update}), and finally the presentation of semianalytical fit functions for the exact numerical PISCs of three specific experiments: LISA, the \textit{Deci-Hertz Interferometer Gravitational-Wave Observatory} (DECIGO)~\cite{Seto:2001qf,Kawamura:2006up,Yagi:2011wg,Isoyama:2018rjb}, and the \textit{Big-Bang Observer} (BBO)~\cite{Crowder:2005nr,Corbin:2005ny,Harry:2006fi} (see Sec.~\ref{subsec:fits}).
Sec.~\ref{sec:conclusions} contains our conclusions as well as an outlook on how to extend and generalize our approach in the future.


In Appendix~\ref{app:review}, we provide a comprehensive review of the strain noise power spectra of several current and future GW experiments.
We specifically consider the following ground-based and space-based interferometer experiments:
aLIGO, aVirgo, the \textit{Kamioka Gravitational-Wave Detector} (KAGRA)~\cite{Somiya:2011np,Aso:2013eba,Akutsu:2018axf,Akutsu:2019rba,Michimura:2019cvl}, \textit{Cosmic Explorer} (CE)~\cite{Evans:2016mbw,Reitze:2019iox}, \textit{Einstein Telescope} (ET)~\cite{Punturo:2010zz,Hild:2010id,Sathyaprakash:2012jk,Maggiore:2019uih}, DECIGO, BBO, and LISA.
Furthermore, we also consider the following \textit{pulsar timing array} (PTA) experiments~\cite{Burke-Spolaor:2018bvk}: the \textit{North American Nano-Hertz Observatory for Gravitational Waves} (NANOGrav)~\cite{McLaughlin:2013ira,Arzoumanian:2018saf,Aggarwal:2018mgp,Brazier:2019mmu}, the \textit{Parkes Pulsar Timing Array} (PPTA)~\cite{Manchester:2012za,Shannon:2015ect}, the \textit{European Pulsar Timing Array} (EPTA)~\cite{Kramer:2013kea,Lentati:2015qwp,Babak:2015lua}, the \textit{International Pulsar Timing Array} (IPTA)~\cite{Hobbs:2009yy,IPTA):2013lea,Verbiest:2016vem,Hazboun:2018wpv}, and the \textit{Square Kilometre Array} (SKA)~\cite{Carilli:2004nx,Janssen:2014dka,Bull:2018lat}.
The purpose of Appendix~\ref{app:review} is to make our presentation self-contained and to facilitate the generalization of our results in Sec.~\ref{sec:pisc} to other experiments and potentially also other types of signals.
In addition, it may serve as a useful resource beyond the actual scope of this paper.


\section{Gravitational-wave signal from a cosmological phase transition}
\label{sec:signal}


In quantum field theory, a first-order phase transition is characterized by a scalar field $\phi$ (or a set of scalar fields $\phi_i$) experiencing a discontinuous change in its vacuum expectation value, $\left<\phi\right>_{\rm false} \rightarrow \left<\phi\right>_{\rm true}$.
In the context of early-Universe cosmology, such a transition manifests itself in the nucleation of bubbles filled by the true vacuum configuration $\left<\phi\right>_{\rm true}$ in the ambient plasma, where the scalar field still resides in its false vacuum configuration $\left<\phi\right>_{\rm false}$.
The expanding and colliding scalar-field bubbles as well as their interaction with the thermal plasma then result in the production of a primordial SGWB via three different mechanisms:

\smallskip\noindent\textbf{(b)}~collisions of expanding \textbf{bubble walls}~\cite{Kosowsky:1991ua,Kosowsky:1992rz,Kosowsky:1992vn,Kamionkowski:1993fg,Caprini:2007xq,Huber:2008hg},

\smallskip\noindent\textbf{(s)}\,~compressional modes (\textit{i.e.}, \textbf{sound waves}) in the bulk plasma~\cite{Hindmarsh:2013xza,Giblin:2013kea,Giblin:2014qia,Hindmarsh:2015qta}, and

\smallskip\noindent\textbf{(t)}\,~vortical motion (\textit{i.e.}, magnetohydrodynamic \textbf{turbulence}) in the bulk plasma~\cite{Caprini:2006jb,Kahniashvili:2008pf,Kahniashvili:2008pe,Kahniashvili:2009mf,Caprini:2009yp,Kisslinger:2015hua}.

\smallskip\noindent In general, the total GW signal from a cosmological phase transition approximately follows from the linear superposition of the signals stemming from these three individual sources,
\begin{align}
\label{eq:OSFOPT}
\Omega_{\rm SFOPT}\left(f\right) \simeq \Omega_{\rm b}\left(f\right) + \Omega_{\rm s}\left(f\right) + \Omega_{\rm t}\left(f\right) \,.
\end{align}
In this section, we shall give a brief overview of these three signal contributions, following the review reports by the LISA Cosmology Working Group in Refs.~\cite{Caprini:2015zlo,Caprini:2019egz}, which represent standard references on this subject.
For more recent work on the dynamics of cosmological phase transitions and the shape of the resulting GW signal, we refer to Refs.~\cite{Hindmarsh:2016lnk,Megevand:2016lpr,Bodeker:2017cim,Hindmarsh:2017gnf,Jinno:2017fby,Konstandin:2017sat,Cutting:2018tjt,Niksa:2018ofa,Dorsch:2018pat,Ellis:2019oqb,Cutting:2019zws,Gould:2019qek,Kainulainen:2019kyp,Jinno:2019jhi,Jinno:2019bxw,Pol:2019yex,Hindmarsh:2019phv}.


We begin by pointing out that, among Refs.~\cite{Caprini:2015zlo,Caprini:2019egz}, only Ref.~\cite{Caprini:2015zlo} presents semianalytical expressions for all three GW sources, $\Omega_{\rm b}$, $\Omega_{\rm s}$, and $\Omega_{\rm t}$.
The discussion in Ref.~\cite{Caprini:2019egz} is more conservative in the sense that it solely accounts for the signal from sound waves, which is in many cases much stronger than the signal from bubble collisions and in general much better understood from a theoretical perspective than the signal from turbulence.
Ref.~\cite{Caprini:2019egz} also distinguishes between two different expressions for $\Omega_{\rm s}$, depending on whether the time of shock formation in the bulk plasma after the phase transition is longer or shorter than the Hubble time.
Meanwhile, it assumes the same spectral shape function for the signal from sound waves as the analysis in Ref.~\cite{Caprini:2015zlo}.
In view of this situation, we decide to consistently base our analysis on the expressions for $\Omega_{\rm b}$, $\Omega_{\rm s}$, and $\Omega_{\rm t}$ in Ref.~\cite{Caprini:2015zlo}.
There are two main reasons for this decision:
First of all, we intend to demonstrate how to apply our new PISC method in case there is more than just one contribution to the total signal. 
The point is that we expect to see significant progress on the theory side in the coming years, which will eventually prompt one to go again beyond the conservative approach of Ref.~\cite{Caprini:2019egz}.
Our analysis thus sets the stage for this moment when the understanding of all three sources has improved and more reliable expressions for $\Omega_{\rm b}$, $\Omega_{\rm s}$, and $\Omega_{\rm t}$ have been attained.
A second reason is that the focus of our analysis is primarily on the construction of new experimental sensitivity curves; we do not have anything new to say on the theoretical aspects of the expected GW signal.
For this reason, we refrain from participating in the ongoing debate on the correct treatment of shock formation and the corresponding energy transfer from sound waves to turbulence.
An attractive feature of our new sensitivity curves is that their construction is for the most part anyway independent of these open questions on the theory side. 
As anticipated in Eq.~\eqref{eq:PIS}, our PISCs will only require knowledge of the experimental noise spectra and spectral shape functions, but will be independent of the exact theoretical predictions for the peak amplitudes entering the GW spectrum.
Therefore, as Ref.~\cite{Caprini:2015zlo} and Ref.~\cite{Caprini:2019egz} use the same spectral shape function for the signal from sound waves, our sensitivity curves are actually not affected by the different treatment of sound waves in Ref.~\cite{Caprini:2019egz}.
For our purposes, the only noticeable consequence consists in the fact that the analysis in Ref.~\cite{Caprini:2019egz} causes some of the benchmark points to slightly shift in our PISC plots compared to the analysis in Ref.~\cite{Caprini:2015zlo}.
We will comment on these shifts in more detail in Sec.~\ref{subsec:update}, where we illustrate how an improved theoretical understanding of the peak amplitudes can be used to update our PISC plots without the need to revise any of the experimental sensitivity curves.
At the same time, we stress that, otherwise, any future update of the spectral shape functions \textit{will} require an update of the experimental PISCs.
In fact, we expect that regular updates of our PISC plots would provide a useful means to track the theoretical progress in the field.


The three contributions $\Omega_{\rm b}$, $\Omega_{\rm s}$, and $\Omega_{\rm t}$ can be parametrized in a model-independent way in terms of a set of characteristic SFOPT parameters, $\alpha$, $\beta/H_*$, $T_*$, $v_w$, $\kappa_{\rm b}$, $\kappa_{\rm s}$, and $\kappa_{\rm t}$,
\begin{align}
\label{eq:h2Omegabst}
h^2\Omega_{\rm b}\left(f\right) & =
h^2\Omega_{\rm b}^{\rm peak}\left(\alpha,\beta/H_*,T_*,v_w,\kappa_{\rm b}\right)
\mathcal{S}_{\rm b}\left(f,f_{\rm b}\right) \,,\\ \nonumber
h^2\Omega_{\rm s}\left(f\right) & =
h^2\Omega_{\rm s}^{\rm peak}\left(\alpha,\beta/H_*,T_*,v_w,\kappa_{\rm s}\right)\,
\mathcal{S}_{\rm s}\left(f,f_{\rm s}\right) \,,\\ \nonumber
h^2\Omega_{\rm t}\left(f\right) & =
h^2\Omega_{\rm t}^{\rm peak}\left(\alpha,\beta/H_*,T_*,v_w,\kappa_{\rm t}\right)\,
\mathcal{S}_{\rm t}\left(f,f_{\rm t},h_*\right) \,.
\end{align}
Let us now go through the different quantities in this equation one by one.
The dimensionless energy density parameters $\Omega_{\rm i} = \rho_{\rm i}/\rho_{\rm c}$ on the \textit{left-hand side} (LHS) of Eq.~\eqref{eq:h2Omegabst} measure the fractions of the total (critical) energy density $\rho_{\rm c} = 3\,H_0^2M_{\rm Pl}^2$ that are contained in GWs of a particular physical origin, $\textrm{i} \in \left\{\textrm{b},\textrm{s},\textrm{t}\right\}$.
Here, $H_0$ is the Hubble parameter in the present Universe, and $M_{\rm Pl} \simeq 2.44 \times 10^{18}\,\textrm{GeV}$ denotes the reduced Planck mass.
In the following, we will typically multiply all energy density parameters $\Omega$ by the square of the dimensionless Hubble parameter $h$, which is defined via the relation $H_0 = 100\,h\,\textrm{km}/\textrm{s}/\textrm{Mpc}$.
In this way, we make sure that quantities of the form $h^2\Omega$ are not affected by the experimental uncertainty in the Hubble parameter $H_0$.
The SFOPT parameter $\alpha$ in Eq.~\eqref{eq:h2Omegabst} is proportional to the change in the trace of the energy-momentum tensor, $\Delta T_\mu^\mu$, across the phase transition~\cite{Caprini:2019egz},%
\footnote{This definition relates $\alpha$ to the so-called bag parameter $\varepsilon = (\Delta u-3\Delta p)/4$ in the bag equation of state~\cite{Espinosa:2010hh}, where $\Delta u$ and $\Delta p$ denote the changes in the internal-energy density and pressure across the phase transition, respectively.
Assuming zero chemical potential for all relevant particle species (\textit{i.e.}, zero Gibbs energy\,/\,free enthalpy), constant temperature $T$, and constant volume $V$, one can show that $\alpha = \left.\varepsilon/\rho_{\rm r}\right|_{T = T_*}$.
For a more general $\alpha$ parameter, which also accounts for possibility of a varying speed of sound, see Refs.~\cite{Giese:2020rtr,Giese:2020znk}.}
\begin{align}
\label{eq:alpha}
\alpha = \frac{1}{\rho_{\rm r}^*} \left[\left(\left.\rho_{\rm v}\right|_{\rm false} - \left.\rho_{\rm v}\right|_{\rm true}\right) - \frac{T}{4}\left(\left.\frac{\partial\rho_{\rm v}}{\partial T}\right|_{\rm false} - \left.\frac{\partial\rho_{\rm v}}{\partial T}\right|_{\rm true}\right)\right]_{T = T_*}
\end{align}
Here, $\rho_{\rm v}$ and $\partial\rho_{\rm v}/\partial T$ are the temperature-dependent effective potential (\textit{i.e.}, free-energy density) in the scalar sector and its derivative with respect to temperature $T$, respectively, while $\rho_{\rm r}$ denotes the energy density of relativistic radiation.
The subscripts ``false'' and ``true'' indicate that these quantities are evaluated in the false and the true vacuum configuration in field space, respectively.
The temperature $T_*$ is the characteristic temperature at the time of GW production, which roughly equals the temperature at the time of bubble nucleation, $T_* \simeq T_{\rm n}$, unless the phase transition occurs in a strongly supercooled state.
In this case, the nucleation temperature can be significantly suppressed compared to the reheating temperature after the completion of the phase transition, $T_{\rm n} \ll T_{\rm rh} \simeq T_*$.
Thus, for strongly supercooled phase transitions, all quantities in Eq.~\eqref{eq:alpha} need to be evaluated at $T_{\rm n}$ rather than $T_*$.
The SFOPT parameter $\beta/H_*$ is the inverse of the duration of the phase transition in units of the Hubble time $H_*^{-1}$ at the time of GW production.
Formally, it is defined in terms of the derivative of the bounce action $S$ that controls the rate of bubble nucleation,
\begin{align}
\frac{\beta}{H_*} = T_* \left.\frac{dS}{dT}\right|_{T = T_*} \,.
\end{align}
Again, this definition only holds as long as there is no large hierarchy between the temperatures $T_*$ and $T_{\rm n}$.
In the case of strong supercooling (\textit{i.e.}, for $T_{\rm n} \ll T_*$), one has to work instead with $\beta/H_* = H_{\rm n} / H_*\, T_n \left.dS/dT\right|_{T=T_n}$.
The parameter $v_w$ in Eq.~\eqref{eq:h2Omegabst} represents the velocity of the bubble wall in the plasma rest frame.
$\kappa_{\rm b}$, $\kappa_{\rm s}$, and $\kappa_{\rm t}$ are three efficiency factors that characterize the fractions of the released vacuum energy that are converted into the energy of scalar-field gradients, sound waves, and turbulence, respectively.
It is customary to express $\kappa_{\rm s}$ and $\kappa_{\rm t}$ in terms of an efficiency factor $\kappa_{\rm kin}$ that characterizes the energy fraction that is converted into bulk kinetic energy and an additional parameter $\epsilon$, \textit{i.e.}, $\kappa_{\rm s} = \kappa_{\rm kin}$ and $\kappa_{\rm t} = \epsilon\,\kappa_{\rm kin}$.
The precise numerical value of $\epsilon$ is the subject of an ongoing debate in the literature.
While some authors estimate $\epsilon$ to be quite small, $\epsilon \simeq 0.05 \cdots 0.10$ (see, \textit{e.g.}, Refs.~\cite{Caprini:2015zlo,Alves:2018jsw,Alves:2019igs}), based on the results presented in Ref.~\cite{Hindmarsh:2015qta}, other authors use values as large as $\epsilon = 1$ (see, \textit{e.g.} Ref.~\cite{Axen:2018zvb}).
In our analysis, we will set $\epsilon = 0.10$.
For a more detailed discussion on the efficiency factors $\kappa_{\rm s}$ and $\kappa_{\rm t}$, see also the recent analysis in Ref.~\cite{Ellis:2019oqb}.


In order to estimate the efficiency factors $\kappa_{\rm b}$ and $\kappa_{\rm kin}$, one has to distinguish between three different types of phase transitions: (i) \textit{nonrunaway phase transitions in a plasma} (NP), (ii) \textit{runaway phase transitions in a plasma} (RP), and (iii) \textit{runaway phase transitions in vacuum} (RV).
In the first case, the bubble wall velocity saturates at a subluminal value, and the contribution to the total GW signal from bubble collisions is negligibly small, $\kappa_{\rm b} \simeq 0$.
Meanwhile, the efficiency factor $\kappa_{\rm kin}$ is well approximated by the following fit function~\cite{Espinosa:2010hh},
\begin{align}
\textrm{NP:} \qquad \kappa_{\rm kin} \simeq
\begin{cases}
\frac{\alpha}{0.73 + 0.083\,\sqrt{\alpha} + \alpha}      & ; \quad v_w \sim 1 \\
\frac{6.9\,\alpha\,v_w^{6/5}}{1.36 - 0.037\,\sqrt{\alpha} + \alpha} & ; \quad v_w \lesssim 0.1 \\
\end{cases}
\end{align}
Here, we will use in practice the large-$v_w$ expression for $\kappa_{\rm kin}$ for all velocities $v_w \geq v_w^\alpha$, where
\begin{align}
v_w^\alpha = \left[\frac{1.36 - 0.037\,\sqrt{\alpha} + \alpha}{6.9\left(0.73 + 0.083\,\sqrt{\alpha} + \alpha\right)}\right]^{5/6} \,,
\end{align}
and similarly, the small-$v_w$ expression for $\kappa_{\rm kin}$ for all velocities $v_w \leq v_w^\alpha$.
In our numerical analysis in Sec.~\ref{sec:pisc}, we will moreover fix the bubble wall velocity at $v_w = 0.95$ for all NP phase transitions.
This is in accord with the analysis in Ref.~\cite{Caprini:2015zlo}, where the choice $v_w = 0.95$ simply serves the purpose to increase the strength of the GW signal.


For RP phase transitions, the picture is a slightly different one, as in this case, the energy deposited into the scalar field is no longer negligible.
RP phase transitions occur for $\alpha$ values $\alpha > \alpha_\infty$, where $\alpha_\infty$ marks the threshold value at which the walls of the scalar-field bubbles begin to ``run away'' at the speed of light, $v_w = 1$.
The threshold value $\alpha_\infty$ is model-dependent and follows from the shifts in the bosonic and fermionic masses squared that are induced by the changing scalar-field background across the phase transition~\cite{Espinosa:2010hh},
\begin{align}
\alpha_\infty \simeq \frac{30}{24\pi^2 g_*T_*^2}\left[\sum_{\rm bosons} g_b\left(\left.m_b^2\right|_{\rm true} -\left.m_b^2\right|_{\rm false}\right) - \frac{1}{2}\sum_{\rm fermions} g_f\left(\left.m_f^2\right|_{\rm true} -\left.m_f^2\right|_{\rm false}\right)\right] \,.
\end{align}
Here, $g_*$, $g_b$, and $g_f$ represent the effective number of relativistic \textit{degrees of freedom} (DOFs) in the unbroken phase at the time of bubble nucleation, the internal DOFs of boson species $b$, and the internal DOFs of fermion species $f$, respectively.
For a \textit{strong first-order electroweak phase transition} (SFOEWPT) in a model with SM-like particle content, one finds
\begin{align}
\label{eq:ainfSM}
\alpha_\infty \simeq 4.9 \times 10^{-3} \left(\frac{\phi_*}{T_*}\right)^2 \,,
\end{align}
where $\phi_*$ denotes the scalar field value in the broken phase at the time of bubble nucleation.
For simplicity, we shall estimate $\alpha_\infty$ based on Eq.~\eqref{eq:ainfSM} in the case of all RP phase transitions that we are going to be interested in.
In SFOEWPT scenarios where the SM Higgs field is only weakly coupled to the new-physics sector, this is typically a reasonable approximation.
The ratio of $\alpha_\infty$ and the actual $\alpha$ value of a RP phase transition determines the efficiency of energy transfer from the vacuum to the scalar field and to the bulk plasma, respectively,
\begin{align}
\label{eq:kappaRP}
\textrm{RP:} \qquad
\kappa_{\rm b} = 1 - \frac{\alpha_\infty}{\alpha} \,, \quad \kappa_{\rm kin} = \frac{\alpha_\infty}{\alpha}\,\kappa_\infty \,, \quad \kappa_{\rm therm} = \left(1 - \kappa_\infty\right)\frac{\alpha_\infty}{\alpha} \,,
\end{align}
where $\kappa_{\rm therm}$ accounts for the fraction of vacuum energy that is converted to thermal energy (which does not source the production of GWs) and where the parameter $\kappa_\infty$ is given by
\begin{align}
\kappa_\infty = \frac{\alpha_\infty}{0.73 + 0.083\,\sqrt{\alpha_\infty} + \alpha_\infty} \,.
\end{align}
The expression for $\kappa_{\rm b}$ in Eq.~\eqref{eq:kappaRP} has recently been updated in Ref.~\cite{Ellis:2020nnr} based on the new all-orders calculation in Ref.~\cite{Hoeche:2020rsg}.
We will come back to this point in Sec.~\ref{subsec:update}.


The case of a RV phase transition, finally, corresponds to the limit of very large $\alpha$, such that the energy transferred to the plasma becomes negligible.
In this case, the only remaining free parameters are $\beta/H_*$ and $T_*$.
All other parameters are fixed at characteristic values,
\begin{align}
\label{eq:rvpt}
\textrm{RV:} \qquad \alpha \rightarrow \infty \,, \quad v_w = 1 \,, \quad \kappa_{\rm b} = 1 \,,\quad \kappa_{\rm kin} = 0 \,.
\end{align}


Having introduced the parameters $\alpha$, $\beta/H_*$, $T_*$, $v_w$, $\kappa_{\rm b}$, $\kappa_{\rm s}$, and $\kappa_{\rm t}$, we are now able to spell out how the different contributions to the GW signal in Eq.~\eqref{eq:h2Omegabst} can be parametrized in terms of these quantities.
We begin by writing down the three peak amplitudes~\cite{Caprini:2015zlo},
\begin{align}
\label{eq:Opeak}
h^2\Omega_{\rm b}^{\rm peak} & \simeq
1.67 \times 10^{-5} \left(\frac{v_w}{\beta/H_*}\right)^2
\left(\frac{100}{g_*\left(T_*\right)}\right)^{1/3}
\left(\frac{\kappa_{\rm b}\,\alpha}{1+\alpha}\right)^2
\left(\frac{0.11\,v_w}{0.42 + v_w^2}\right)
\,, \\ \nonumber
h^2\Omega_{\rm s}^{\rm peak} & \simeq 
2.65 \times 10^{-6} \left(\frac{v_w}{\beta/H_*}\right)^{\phantom{2}}
\left(\frac{100}{g_*\left(T_*\right)}\right)^{1/3}
\left(\frac{\kappa_{\rm s}\,\alpha}{1+\alpha}\right)^2
\,, \\ \nonumber
h^2\Omega_{\rm t}^{\rm peak} & \simeq
3.35 \times 10^{-4} \left(\frac{v_w}{\beta/H_*}\right)^{\phantom{2}}
\left(\frac{100}{g_*\left(T_*\right)}\right)^{1/3}
\left(\frac{\kappa_{\rm t}\,\alpha}{1+\alpha}\right)^{3/2} \,.
\end{align}
Here, we emphasize that the effective number of relativistic DOFs, $g_*$, needs to be evaluated as a function of $T_*$.
In our numerical analysis in Sec.~\ref{sec:pisc}, we will approximate $g_*$ by its SM value, making use of the numerical data tabulated in Ref.~\cite{Saikawa:2018rcs} in order to correctly describe its temperature dependence.
The spectral shape functions $\mathcal{S}_{\rm b}$, $\mathcal{S}_{\rm s}$, and $\mathcal{S}_{\rm t}$ are given as~\cite{Caprini:2015zlo}
\begin{align}
\label{eq:S}
\mathcal{S}_{\rm b} & = \left(\frac{f}{f_{\rm b}}\right)^{2.8} \left[\frac{3.8}{1+2.8\left(f/f_{\rm b}\right)^{3.8}}\right]
\,, \\ \nonumber
\mathcal{S}_{\rm s} & = \left(\frac{f}{f_{\rm s}}\right)^3 \:\:\: \left[\frac{7}{4+3\left(f/f_{\rm s}\right)^2}\right]^{7/2}
\,, \\ \nonumber
\mathcal{S}_{\rm t} & = \left(\frac{f}{f_{\rm t}}\right)^3 \:\:\: \left[\frac{1}{1+\left(f/f_{\rm t}\right)}\right]^{11/3} \frac{1}{1+8\pi\,f/h_*} \,.
\end{align}
Note that $\mathcal{S}_{\rm b}$ and $\mathcal{S}_{\rm s}$ are normalized to unity at the respective peak frequencies, whereas the value of $\mathcal{S}_{\rm t}$ at $f = f_{\rm t}$ depends on what we shall refer to as the Hubble frequency $h_*$,
\begin{align}
\mathcal{S}_{\rm b} \left(f = f_{\rm b}\right) = 1 \,, \quad \mathcal{S}_{\rm s} \left(f = f_{\rm s}\right) = 1 \,, \quad \mathcal{S}_{\rm t} \left(f = f_{\rm t}\right) = \frac{1}{2^{11/3}\left(1+8\pi\,f_{\rm t}/h_*\right)} \,.
\end{align}
This normalization is merely a matter of convention.
Alternatively, one could simply rescale both $h^2\Omega_{\rm t}^{\rm peak}$ and $\mathcal{S}_{\rm t}$ by a factor $2^{11/3}\left(1+8\pi\,f_{\rm t}/h_*\right)$, such that $\mathcal{S}_{\rm t} \left(f = f_{\rm t}\right) = 1$.
The Hubble frequency $h_*$ corresponds to the particular wavenumber $k_*$ that equals the Hubble rate $H_*$ at the time of GW production.
Its redshifted, present-day value is solely controlled by $T_*$,
\begin{align}
\label{eq:hstar}
h_* = \frac{a_*}{a_0}\,H_* = 1.6 \times 10^{-2}\,\textrm{mHz} \left(\frac{g_*\left(T_*\right)}{100}\right)^{1/6}\left(\frac{T_*}{100\,\textrm{GeV}}\right) \,.
\end{align}
Finally, we state the three peak frequencies~\cite{Caprini:2015zlo}, which completes our discussion of Eq.~\eqref{eq:h2Omegabst},
\begin{align}
\label{eq:fbst}
f_{\rm b} & =
1.6 \times 10^{-2}\,\textrm{mHz}
\left(\frac{g_*\left(T_*\right)}{100}\right)^{1/6}
\left(\frac{T_*}{100\,\textrm{GeV}}\right)
\left(\frac{\beta/H_*}{v_w}\right)
\left(\frac{0.62\,v_w}{1.8-0.1\,v_w+v_w^2}\right)
\,, \\ \nonumber
f_{\rm s} & =
1.9 \times 10^{-2}\,\textrm{mHz}
\left(\frac{g_*\left(T_*\right)}{100}\right)^{1/6}
\left(\frac{T_*}{100\,\textrm{GeV}}\right)
\left(\frac{\beta/H_*}{v_w}\right)
\,, \\ \nonumber
f_{\rm t} & =
2.7 \times 10^{-2}\,\textrm{mHz}
\left(\frac{g_*\left(T_*\right)}{100}\right)^{1/6}
\left(\frac{T_*}{100\,\textrm{GeV}}\right)
\left(\frac{\beta/H_*}{v_w}\right)
\,.
\end{align}
%


\section{Peak-integrated sensitivity curves}
\label{sec:pisc}


\subsection{Definition}
\label{subsec:definition}


Our idea of \textit{peak-integrated sensitivity curves} (PISCs) is based on the following observation:
For a GW signal from a SFOPT and within the approximations outlined in Sec.~\ref{sec:signal}, the frequency integration in Eq.~\eqref{eq:rho} becomes independent of any model details.
Therefore, by combining the expressions in Eqs.~\eqref{eq:rho}, \eqref{eq:OSFOPT}, and \eqref{eq:h2Omegabst}, the SNR can always be written as
\begin{align}
\label{eq:rhoMaster}
\varrho = \sqrt{\frac{t_{\rm obs}}{1\,\textrm{yr}}\left(\varrho_{\rm b}^2 + \varrho_{\rm s}^2 + \varrho_{\rm t}^2 + \varrho_{\rm b/s}^2 + \varrho_{\rm b/t}^2 + \varrho_{\rm s/t}^2\right)} \,,
\end{align}
with the partial SNRs $\varrho_{\rm i/j}$ on the \textit{right-hand side} (RHS) of this relation being defined as
\begin{align}
\label{eq:rhoij}
\varrho_{\rm i/j} = \frac{\Omega_{\rm i/j}^{\rm peak}}{\Omega_{\rm PIS}^{\rm i/j}} \,,\quad \varrho_{\rm i/i} \equiv \varrho_{\rm i} \,, \quad \textrm{i},\textrm{j} \in  \left\{\textrm{b},\textrm{s},\textrm{t}\right\} \,.
\end{align}
Here, the numerator is defined as the geometric mean of the corresponding peak amplitudes,
\begin{align}
\Omega_{\rm i/j}^{\rm peak} = \sqrt{\Omega_{\rm i}^{\rm peak}\,\Omega_{\rm j}^{\rm peak}} \,, \quad \Omega_{\rm i/i}^{\rm peak} \equiv \Omega_{\rm i}^{\rm peak} \,,
\end{align}
while the denominator represents what we will refer to as the \textit{peak-integrated sensitivity} (PIS),
\begin{align}
\label{eq:OPIS}
\Omega_{\rm PIS}^{\rm i/j} = \left[\left(2 - \delta_{\rm ij}\right) n_{\rm det}\, 1\,\textrm{yr} \int_{f_{\rm min}}^{f_{\rm max}} df\:\frac{\mathcal{S}_{\rm i}\left(f\right)\mathcal{S}_{\rm j}\left(f\right)}{\Omega_{\rm noise}^2\left(f\right)}\right]^{-1/2} \,, \quad \Omega_{\rm PIS}^{\rm i/i} \equiv \Omega_{\rm PIS}^{\rm i} \,,
\end{align}
We normalize $\Omega_{\rm PIS}^{\rm i/j}$ to an observing time of one average year in the Gregorian calendar, 
\begin{align}
1\,\textrm{yr} = 3.1556952 \times 10^7\,\textrm{Hz}^{-1} \,,
\end{align}
such that the actual observing time $t_{\rm obs}$ appears as a simple rescaling factor in Eq.~\eqref{eq:rhoMaster}.%
\footnote{Alternatively, this rescaling factor could also be absorbed in the definition of $\Omega_{\rm PIS}^{\rm i/j}$, making this quantity explicitly $t_{\rm obs}$-dependent.
In the following, we will, however, stick to our conventions in Eqs.~\eqref{eq:rhoMaster} to \eqref{eq:OPIS}.}


A remarkable property of the six sensitivities $\Omega_{\rm PIS}^{\rm i/j}$ is that they can be explicitly computed without ever referring to a particular BSM model.
In particular, they can be fully parametrized in terms of the peak frequencies $f_{\rm b}$, $f_{\rm s}$, $f_{\rm t}$ and the Hubble frequency $h_*$ without ever specifying the values of the SFOPT parameters $\alpha$, $\beta/H_*$, $T_*$, $v_w$, $\kappa_{\rm b}$, $\kappa_{\rm s}$, and $\kappa_{\rm t}$,
\begin{align}
\label{eq:OPISbst}
\Omega_{\rm PIS}^{\rm b} & = \Omega_{\rm PIS}^{\rm b}\left(f_{\rm b}\right) \,, & 
\Omega_{\rm PIS}^{\rm s} & = \Omega_{\rm PIS}^{\rm s}\left(f_{\rm s}\right) \,, & 
\Omega_{\rm PIS}^{\rm t} & = \Omega_{\rm PIS}^{\rm t}\left(f_{\rm t},h_*\right) \,, \\ \nonumber
\Omega_{\rm PIS}^{\rm b/s} & = \Omega_{\rm PIS}^{\rm b/s}\left(f_{\rm b},f_{\rm s}\right) \,, & 
\Omega_{\rm PIS}^{\rm b/t} & = \Omega_{\rm PIS}^{\rm b/t}\left(f_{\rm b},f_{\rm t},h_*\right) \,, & 
\Omega_{\rm PIS}^{\rm s/t} & = \Omega_{\rm PIS}^{\rm s/t}\left(f_{\rm s},f_{\rm t},h_*\right) \,.
\end{align}
This renders them a handy and model-independent tool for discussing the sensitivity of future searches for GWs from a SFOPT.
The sensitivities $\Omega_{\rm PIS}^{\rm i/j}$ can especially be used to construct \textit{peak-integrated sensitivity curves} (PISCs) and \textit{peak-integrated sensitivity bands} (PISBs) by plotting $\Omega_{\rm PIS}^{\rm b}$ as a function of $f_{\rm b}$, $\Omega_{\rm PIS}^{\rm s}$ as a function of $f_{\rm s}$, etc.
Then, once these curves and bands are known, one can fit the exact numerical results by semianalytical fit functions, which allows one to write down quasianalytic expressions for the total SNR as functions of the SFOPT parameters $\alpha$, $\beta/H_*$, $T_*$, $v_w$, $\kappa_{\rm b}$, $\kappa_{\rm s}$, and $\kappa_{\rm t}$.
In this way, the concept of PISCs and PISBs closes the gap between the numerical modeling of SFOPTs on the theory side and the instrumental properties of future GW searches on the experimental side.


In the remainder of this paper, we will now discuss the idea of PISCs and PISBs in more detail and highlight a few possible applications.
In doing so, we will focus on the sensitivities of three proposed space-borne GW interferometers: LISA, DECIGO, and BBO.
Among these three experiments, LISA is the most mature one, which was approved by the European Space Agency as its \textit{third large-class} (L3) mission in 2017.
According to the L3 mission concept, LISA will consist of three identical spacecraft in an equilateral triangular formation separated by 2.5 million km and connected by six active laser links~\cite{Audley:2017drz}.
In this configuration, LISA will be able to search for a SGWB signal by performing an auto-correlation measurement~\cite{Cutler:1997ta}, which means that we have to set $n_{\rm det} = 1$ in Eq.~\eqref{eq:OPIS}.
The design concepts of DECIGO and BBO envision, by contrast, a hexagonal configuration of two triangular detectors (\textit{i.e.}, a ``Star-of-David''-like configuration).
Each of these two experiments will hence effectively represent a two-detector network, which will enable DECIGO and BBO to search for a SGWB signal by performing a cross-correlation measurement.
This implies $n_{\rm det} = 2$ in Eq.~\eqref{eq:OPIS}.


\subsection{Benchmark point in the scalar-singlet extension}
\label{subsec:example}

\begin{figure}
\begin{center}

\includegraphics[width=0.95\textwidth]{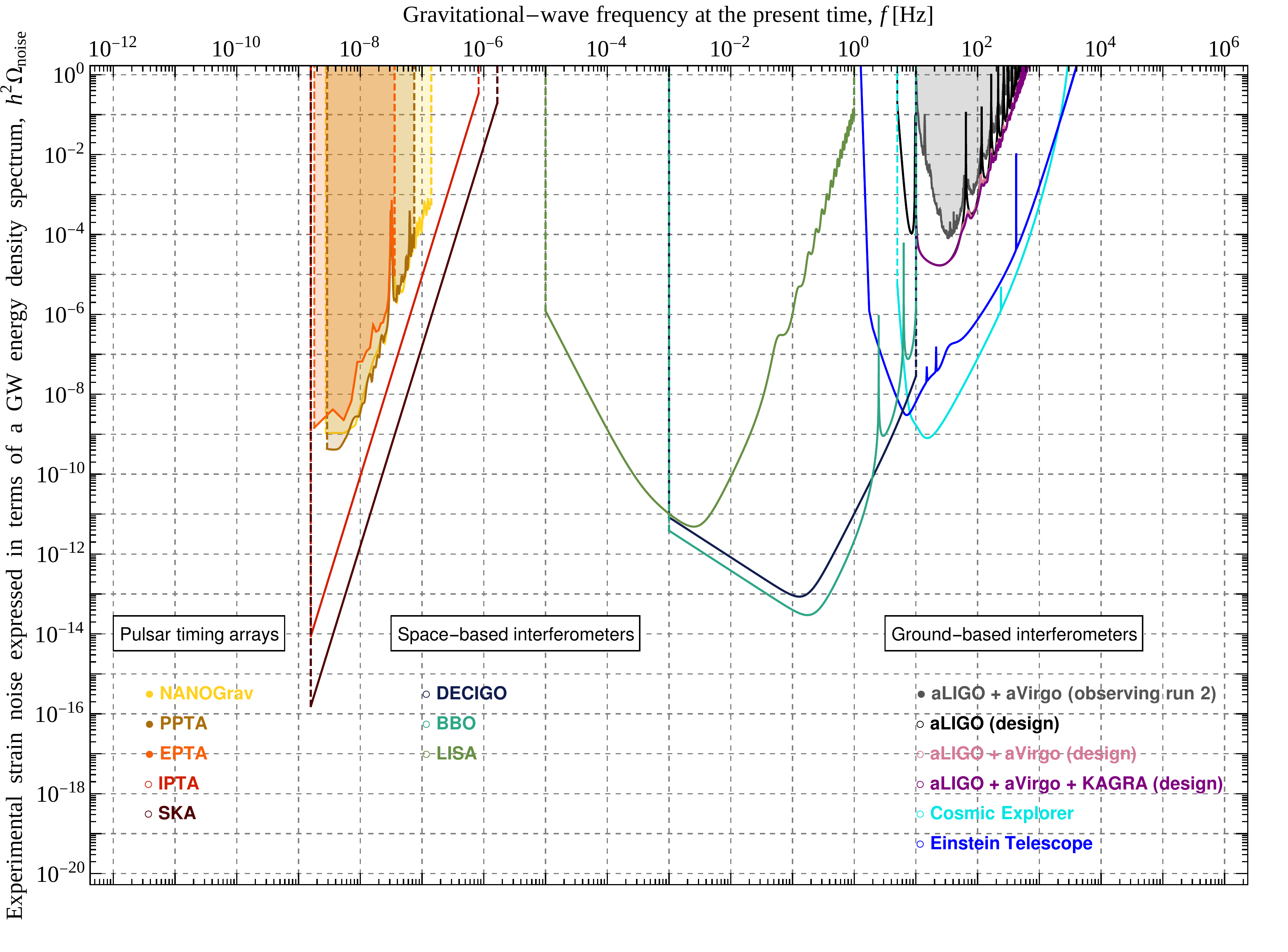}

\vspace{-0.8cm}

\includegraphics[width=0.95\textwidth]{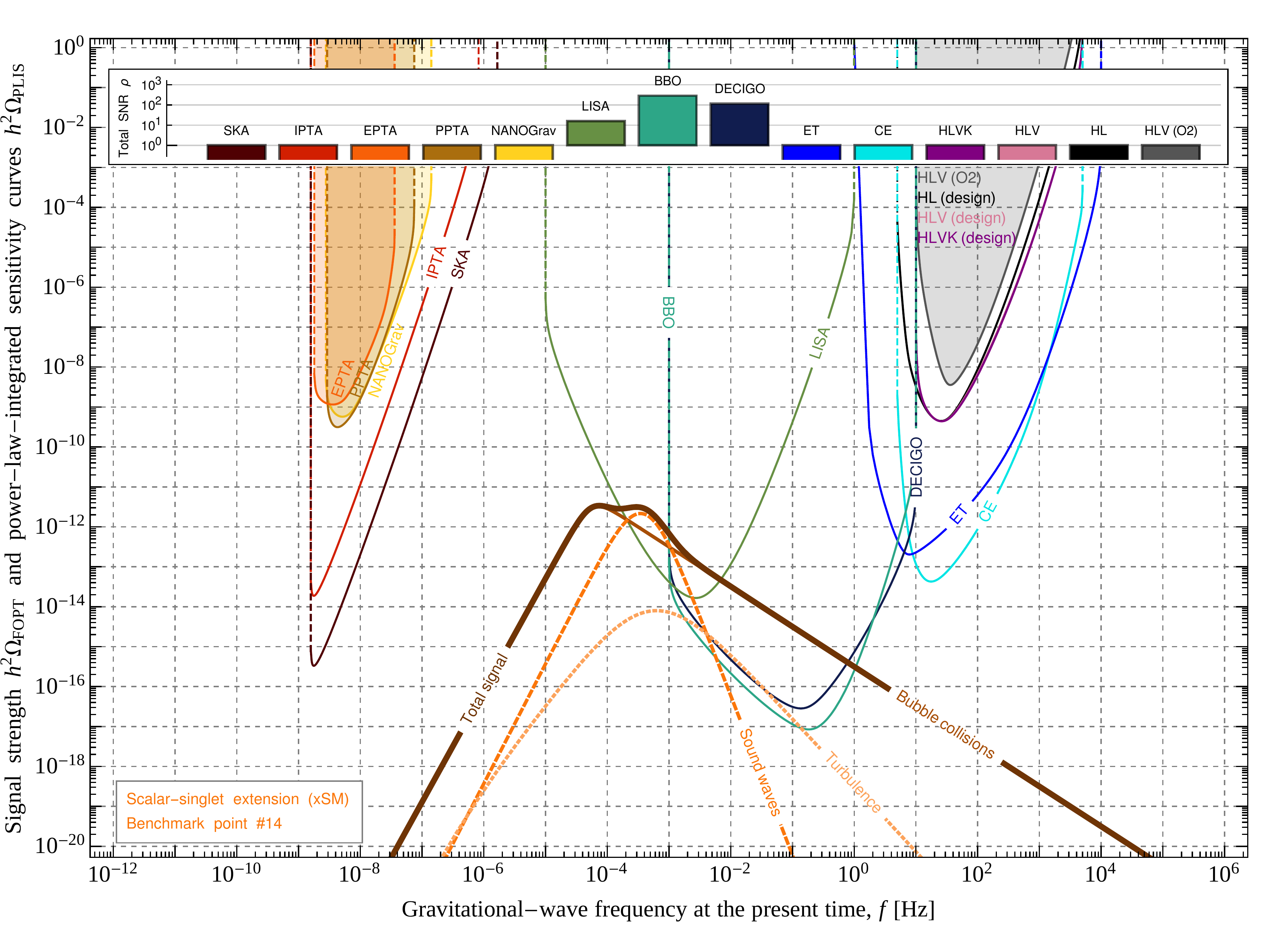}
\caption{Top: Strain noise spectra.
Bottom: PLISCs and GW signal for BP \#14. See text.}
\label{fig:plis}
\end{center}
\end{figure}

First, let us illustrate the philosophy behind our PISC method by means of a single benchmark point in a particular SFOPT scenario.
To this end, we shall consider the simplest example of a BSM model giving rise to GWs from a SFOEWPT, namely, the \textit{real-scalar-singlet extension of the standard model} (xSM).
This model, also known as the (scalar) Higgs portal scenario, features an extra gauge singlet in the scalar sector, which couples to the SM Higgs boson and which may or may not be charged under a $\mathbb{Z}_2$ symmetry.
More details on the xSM as well as a more complete list of references are contained in the companion paper~\cite{Alanne:2019bsm}, where we apply our PISC method to investigate the GW phenomenology of this model.
In Ref.~\cite{Alanne:2019bsm}, we include all renormalizable operators in the scalar potential that are allowed by gauge invariance, \textit{i.e.}, we do not require the scalar singlet to be charged under a $\mathbb{Z}_2$ symmetry.
However, for the purposes of the following discussion, it will suffice to restrict ourselves to the $\mathbb{Z}_2$-symmetric formulation of the xSM, which can also result in a SFOEWPT~\cite{Espinosa:2011ax}.
The xSM benchmark point that we are going to be interested in corresponds to benchmark point~B in Sec.~4.2.2 of Ref.~\cite{Caprini:2015zlo}.
It is characterized by the following SFOPT parameter values,
\begin{align}
\label{eq:xSM14}
T_* = 65.2\,\textrm{GeV} \,,\quad \alpha = 0.12 \,,\quad \frac{\beta}{H_*} = 29.96 \,,\quad \frac{\phi_*}{T_*} = 3.70 \,,
\end{align}
and describes a RP phase transition, such that all three GW sources during the phase transition (bubble collisions, sound waves, and turbulence) contribute to the total signal.%
\footnote{According to the new results in Refs.~\cite{Ellis:2020nnr,Hoeche:2020rsg}, the GW signal from bubble collisions is actually strongly suppressed in the xSM (see also Refs.~\cite{Bodeker:2017cim,Ellis:2019oqb}).
We nevertheless stick to the benchmark point in Eq.~\eqref{eq:xSM14} and its interpretation as a RP phase transition.
On the one hand, this facilitates the direct comparison between our results and the sensitivity plots in Ref.~\cite{Caprini:2015zlo}.
On the other hand, it is expected that similar values of the SFOPT parameters $T_*$, $\alpha$, $\beta/H_*$, and $\phi_*/T_*$ can be easily obtained for a RP phase transition in a hidden scalar sector that does not couple to the SM (see, \textit{e.g.}, Ref.~\cite{Breitbach:2018ddu}).
In this sense, ``xSM'' is understood to refer to such a hidden-sector equivalent of the actual xSM in the following (see also our discussion in Sec.~\ref{subsec:update}).}
In the next section, where we will compare different SFOPT scenarios to each other, we will refer to this point as benchmark point \#14 (see Tab.~\ref{tab:bp}).
The numerical values in Eq.~\eqref{eq:xSM14} are all we need to evaluate the expressions for $\Omega_{\rm b}$, $\Omega_{\rm s}$, and $\Omega_{\rm t}$ in Sec.~\ref{sec:signal} and plot the total GW signal as well as its individual contributions as functions of frequency $f$ (see Fig.~\ref{fig:plis}).


In view of Fig.~\ref{fig:plis}, several comments are in order.
In the upper panel of Fig.~\ref{fig:plis}, we show the strain noise spectra $\Omega_{\rm noise}$ of all current and future GW experiments that we consider in this paper.
These strain noise spectra, which we review in more detail in Appendix~\ref{app:review}, are the starting point for constructing both power-law- and peak-integrated sensitivity curves.
As for the second-generation ground-based interferometers (aLIGO, aVirgo, and KAGRA), we consider three different detector networks that can be formed by these experiments:

\smallskip\noindent\textbf{\textit{Hanford-Livingston} (HL):}~\textit{aLIGO Hanford $+$ Livingston Observatories} (aLHO $+$ aLLO) 

\smallskip\noindent\textbf{\textit{Hanford-Livingston-Virgo} (HLV):} aLHO $+$ aLLO $+$ aVirgo

\smallskip\noindent\textbf{\textit{Hanford-Livingston-Virgo-KAGRA} (HLVK):}~aLHO $+$ aLLO $+$ aVirgo $+$ KAGRA

\smallskip\noindent For the HL and HLVK networks, we indicate the respective design noise spectra~\cite{Ligo:2018aaa,Virgo:2018aaa}, while for the HLV network, we also indicate, in addition to the design noise spectrum, a noise spectrum representative for \textit{observing run 2} (O2)~\cite{LHO:2017aaa,LLO:2017aaa,Virgo:2017aaa}.
This is also reflected in our use of color in Fig.~\ref{fig:plis}.
Projected noise spectra based on sensitivity estimates are represented by simple lines, whereas noise spectra based on existing data are highlighted by a color shading.


In the lower panel of Fig.~\ref{fig:plis}, we show the power-law-integrated sensitivities $\Omega_{\rm PLIS}$ that can be constructed from the strain noise spectra $\Omega_{\rm noise}$ according to the algorithm outlined in Appendix~\ref{app:review} [see Eq.~\eqref{eq:plis} for the precise definition of $\Omega_{\rm PLIS}$].
All PLISCs in Fig.~\ref{fig:plis} are normalized to an SNR threshold $\varrho_{\rm thr} = 1$; for all future interferometers, we set the observing time to $t_{\rm obs} = 1\,\textrm{yr}$; and for all future PTA experiments, we use $t_{\rm obs} = 20\,\textrm{yr}$.
These values are not necessarily realistic (see, \textit{e.g.}, the discussion and references in Appendix~B of Ref.~\cite{Breitbach:2018ddu}).
Our main motivation for setting $\varrho_{\rm thr}$ and $t_{\rm obs}$ to these values rather is to guarantee an equal normalization of our power-law- and peak-integrated sensitivity curves.
For the purposes of the present paper, this is a reasonable strategy, which allows for a more direct comparison of our PLISC and PISC plots.
In addition to the experimental PLISCs, the lower panel of Fig.~\ref{fig:plis} also displays the GW signal $\Omega_{\rm SFOPT}$ for the xSM benchmark point as well as its three individual contributions, $\Omega_{\rm b}$, $\Omega_{\rm s}$, and $\Omega_{\rm t}$ [see Eqs.~\eqref{eq:OSFOPT} and \eqref{eq:h2Omegabst}].
The bar diagram in this plot indicates that this signal is within the sensitivity reach of LISA, DECIGO, and BBO.
Carrying out the frequency integral in Eq.~\eqref{eq:rho} for all three experiments, we obtain
\begin{align}
\label{eq:SNRxSM14}
\varrho^{\rm LISA} \simeq 16 \,,\quad \varrho^{\rm DECIGO} \simeq 110 \,,\quad \varrho^{\rm BBO} \simeq 290 \,.
\end{align}


The PLISCs in Fig.~\ref{fig:plis} convey a useful impression of the different sensitivities of ongoing and planned GW experiments.
The PLISCs for LISA, DECIGO, and BBO leave in particular no doubt that these three experiments would have (very) good chances to detect the GW signal from the SFOPT in our xSM benchmark scenario.
This is the important, qualitative message of the PLISC plot in Fig.~\ref{fig:plis}.
Its quantitative information content, on the other hand, is somewhat limited.
Just by inspecting the GW signal and experimental PLISCs in Fig.~\ref{fig:plis}, it is, \textit{e.g.}, impossible to precisely infer the expected SNRs in Eq.~\eqref{eq:SNRxSM14}.
To see this, recall that the numerical results in Eq.~\eqref{eq:SNRxSM14} follow from the frequency integral over the strain noise spectra in the \textit{upper} panel of Fig.~\ref{fig:plis} [see Eq.~\eqref{eq:rho}].
It is therefore notoriously difficult to include information on the expected SNR in the \textit{lower} panel of Fig.~\ref{fig:plis}.
Strictly speaking, the only type of signal for which a PLISC plot lends itself to a statistical interpretation is a pure power law; hence the name.
In this case, the factor by which the signal curve needs to be rescaled (\textit{i.e.}, the amount by which it needs to be vertically shifted) in order to align it with an equally sloped tangent of a PLISC can be interpreted as the corresponding SNR (see Appendix~\ref{app:review}).%
\footnote{This interpretation relies on the fact that our PLISCs are normalized to an SNR threshold $\varrho_{\rm thr} = 1$.
Also note that our normalization agrees with the one chosen by Romano and Thrane in their original work~\cite{Thrane:2013oya}.}
However, for a GW signal from a SFOPT, the assumption of a pure power law is maximally violated in the most relevant part of the spectrum, \textit{i.e.}, close to the dominating peak(s) in the spectrum.
This is also nicely illustrated by the signal in the xSM benchmark scenario, which features a double-peak structure at $f \sim f_{\rm b}, f_{\rm s}$ and thus clearly deviates from a pure power law in the frequency range where the signal strength is the largest.


In general, one should therefore take PLISC plots such as the one in Fig.~\ref{fig:plis} with a grain of salt.
They typically represent a helpful qualitative visualization; however, for signals that notably deviate from a power law, they no longer contain useful information on the SNR.
This observation is one of the main motivations behind our new PISC method.
We argue that, as soon as more information on the expected shape of the signal is available, this information should also be made use of in the construction of sensitivity curves.
This is exactly the situation in which we find ourselves in the context of GWs from a SFOPT, where the spectral shape of the signal is controlled by $\mathcal{S}_{\rm b}$, $\mathcal{S}_{\rm s}$, and $\mathcal{S}_{\rm t}$ in Eq.~\eqref{eq:S}.
Given the large and increasing interest in this type of signal, we therefore deem it justified to construct new sensitivity curves\,---\,PISCs\,---\,that incorporate knowledge of the expected signal shape.


\begin{figure}
\begin{center}

\includegraphics[width=0.45\textwidth]{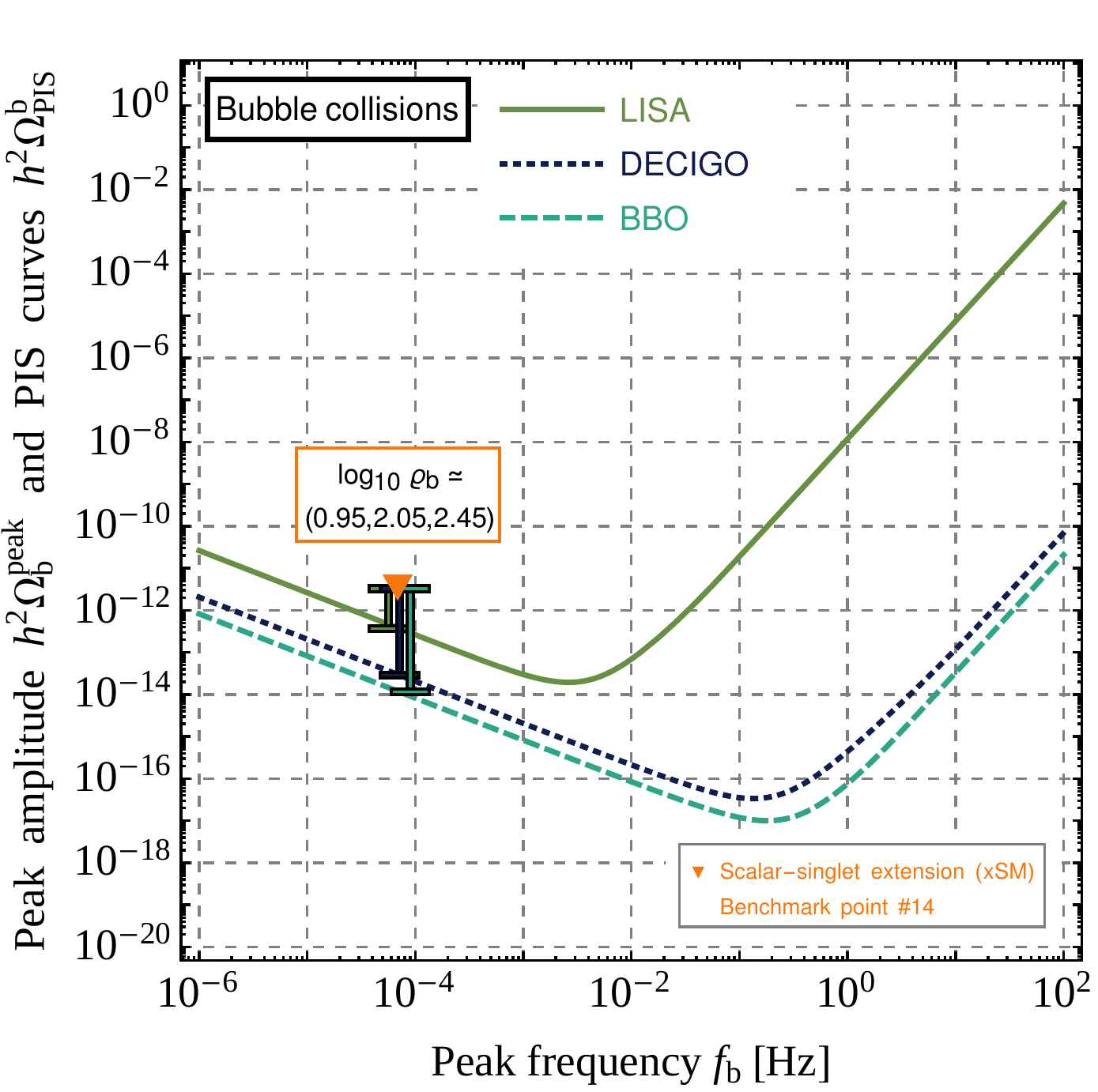}\hfill
\includegraphics[width=0.45\textwidth]{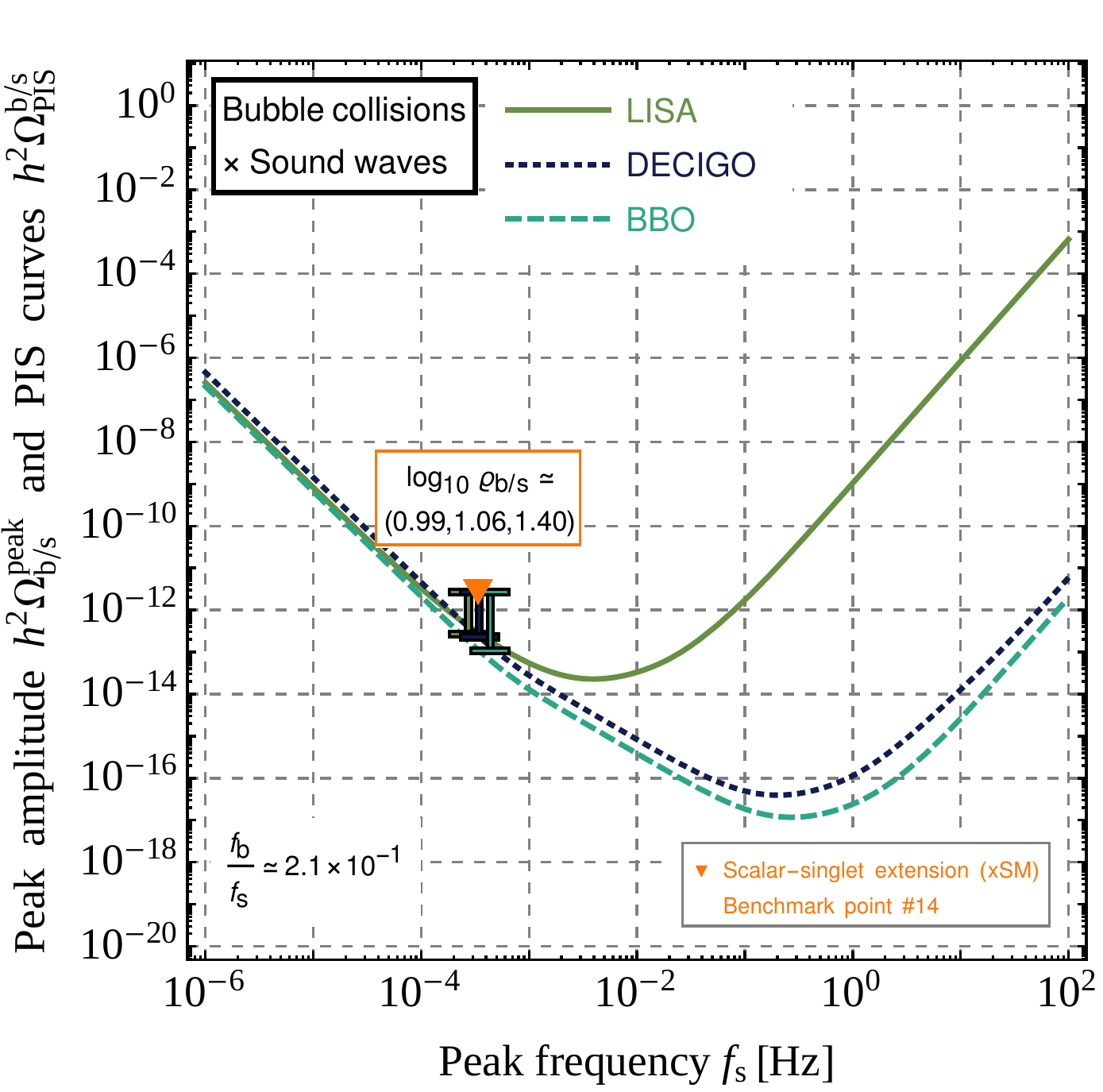}

\includegraphics[width=0.45\textwidth]{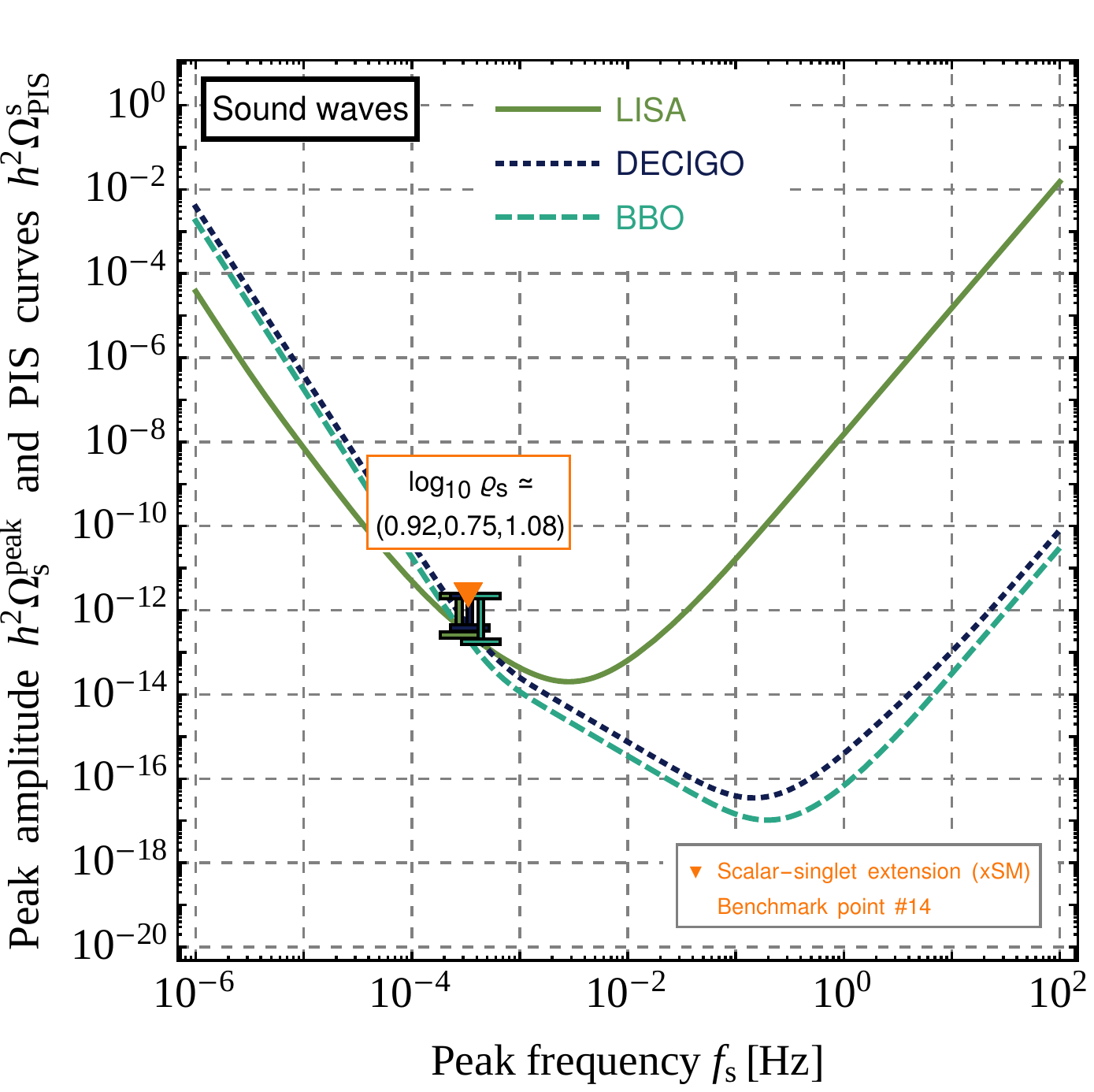}\hfill
\includegraphics[width=0.45\textwidth]{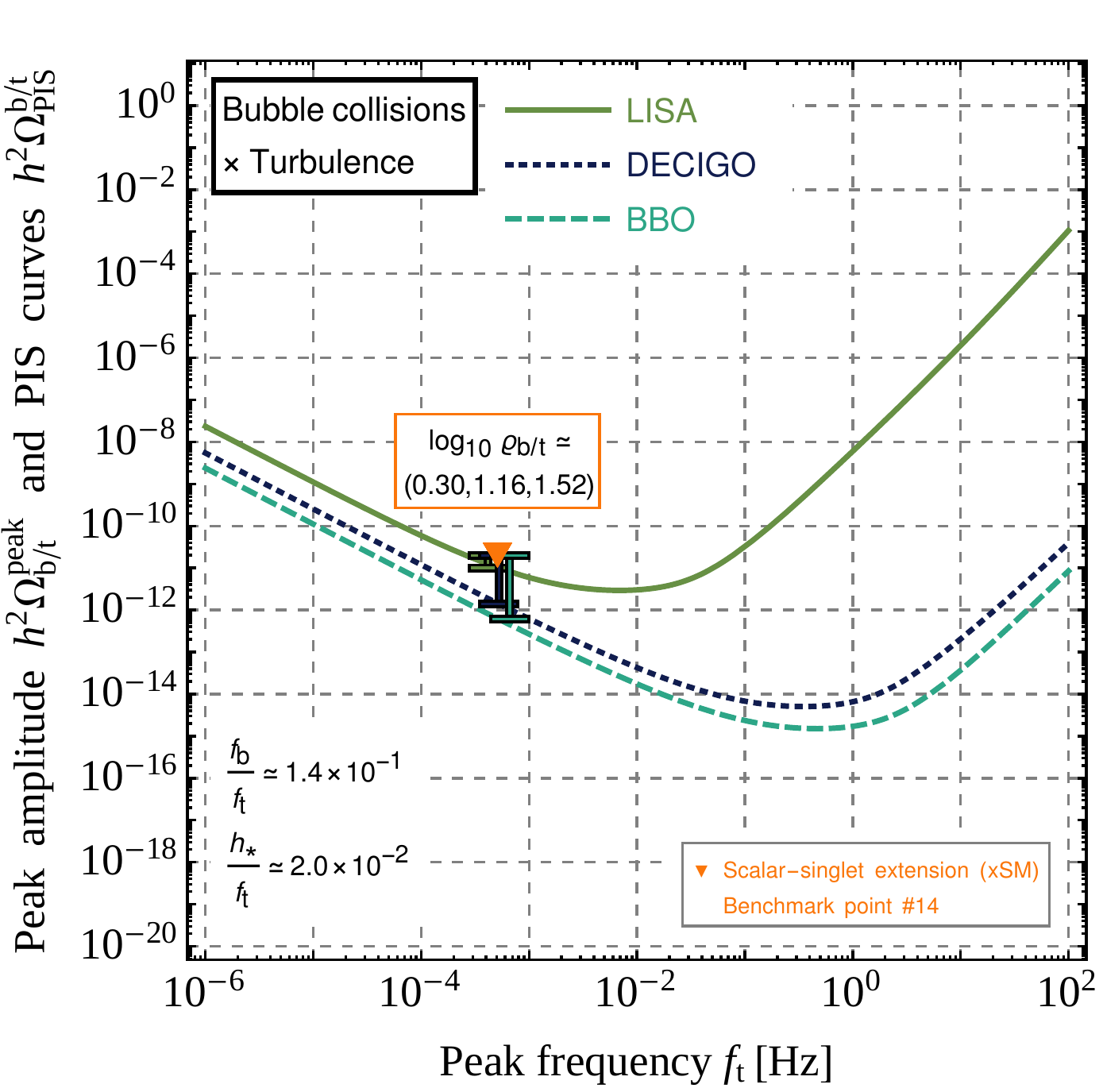}

\includegraphics[width=0.45\textwidth]{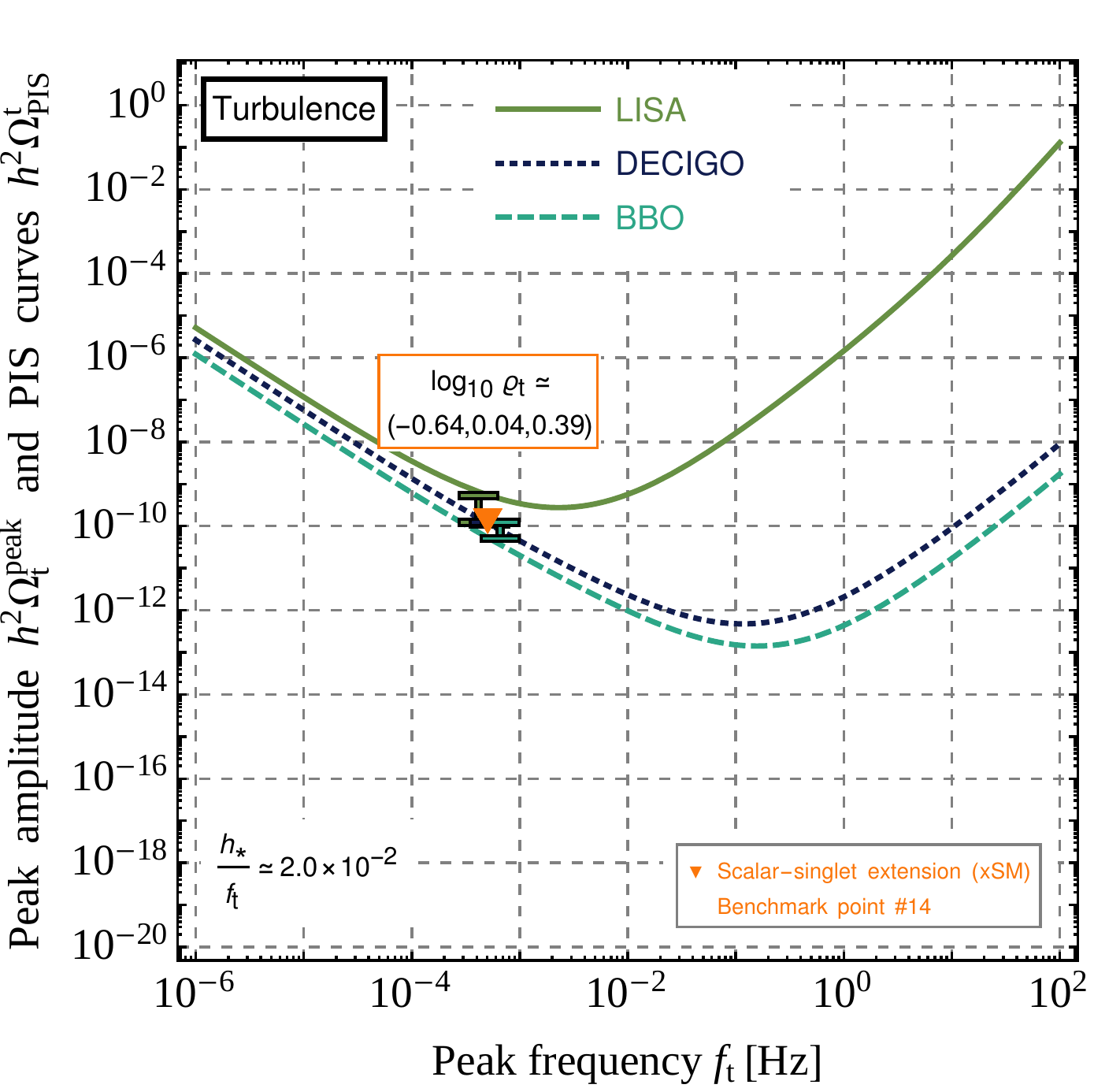}\hfill
\includegraphics[width=0.45\textwidth]{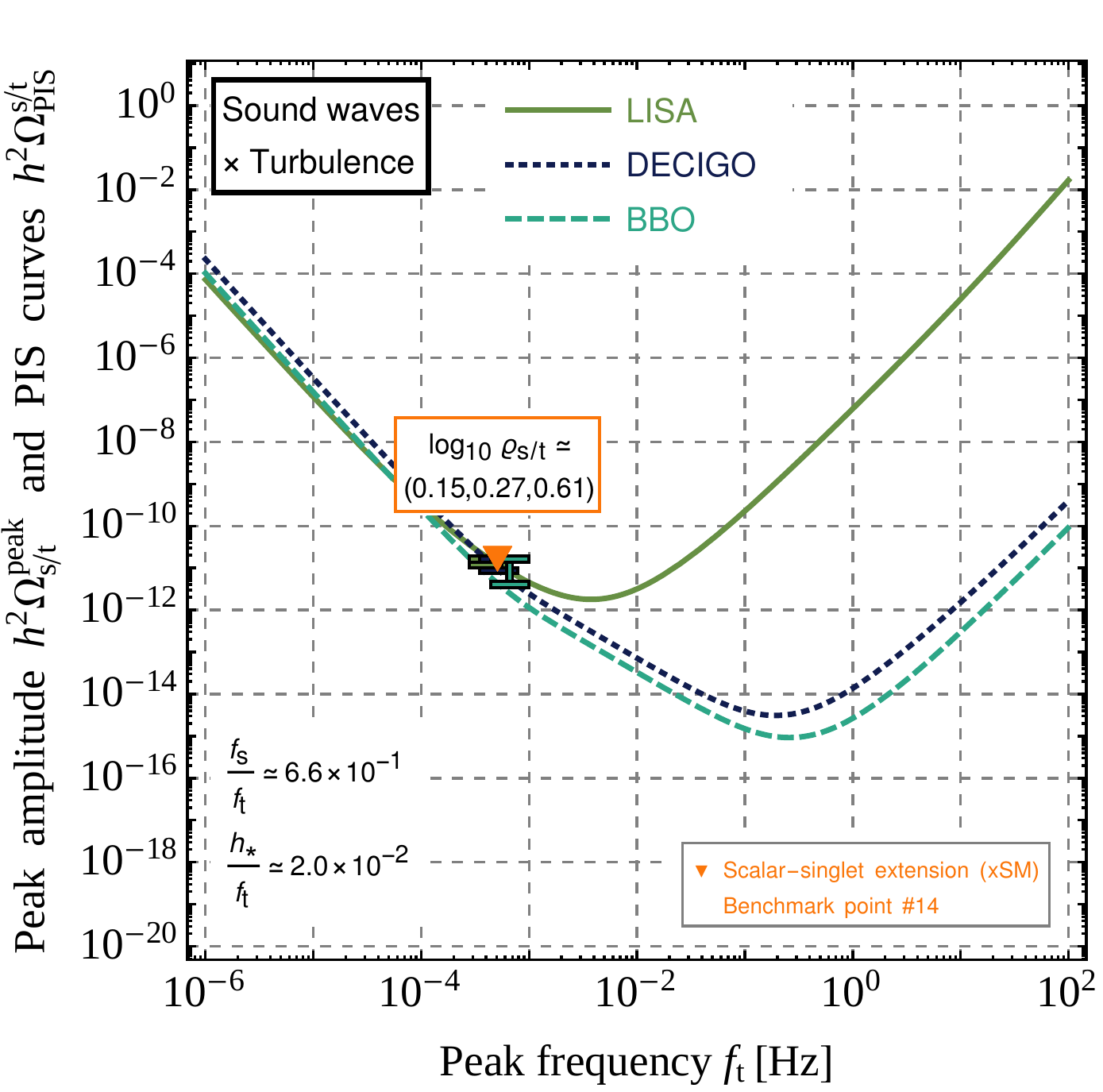}
\caption{PISCs for LISA, DECIGO, and BBO together with the predictions of BP \#14. See text.}
\label{fig:pis}
\end{center}
\end{figure}


In Fig.~\ref{fig:pis}, we present the PISCs for LISA, DECIGO, and BBO [see Eq.~\eqref{eq:OPIS}] in combination with the predictions for the peak frequencies and peak amplitudes in the xSM benchmark scenario.
For each PISC that depends on more than just one frequency, we fix the remaining frequencies at their respective benchmark values.
In each of the six plots in Fig.~\ref{fig:pis}, we also indicate the respective partial SNRs in the format $\log_{10}\big(\varrho_{\rm i/j}^{\rm LISA},\varrho_{\rm i/j}^{\rm DECIGO},\varrho_{\rm i/j}^{\rm BBO}\big)$ [see Eq.~\eqref{eq:rhoij}].
These partial SNRs illustrate a characteristic feature of our PISC plots:
They retain the full information on the SNR, encoding it on the $y$-axis.
That is, in each of the six plots, the vertical separations between the benchmark point and the PISCs directly correspond to the respective partial SNRs.
The total SNRs in Eq.~\eqref{eq:SNRxSM14} then follow from adding these partial SNRs in quadrature [see Eq.~\eqref{eq:rhoMaster}].
The fact that our method splits the total SNR into six different contributions also enables one to easily combine and compare these contributions.
From Fig.~\ref{fig:pis}, we can, \textit{e.g.}, read off that LISA will be most sensitive to the b/s-channel, \textit{i.e.}, the overlap of the two signals from bubble collisions and sound waves, while DECIGO and BBO will be most sensitive to the b-channel, \textit{i.e.}, the signal from bubble collisions only, without any inference from other sources.
At the same time, all three experiments will be least sensitive to the t-channel, \textit{i.e.}, the signal from turbulence.
Similar qualitative conclusions can also be drawn from Fig.~\ref{fig:plis}, just by looking at the relation of the signal curve and the three sensitivity curves for LISA, DECIGO, and BBO; however, the advantage of our PISC plots in Fig.~\ref{fig:pis} is that they make these conclusions quantitatively more precise.
Thanks to the partial SNRs in Fig.~\ref{fig:pis}, it now possible to assess \textit{precisely} by how much a signal is dominated by a certain contribution or to what extent an experiment will be sensitive to the six individual signal channels.
Likewise, it is straightforward to restrict oneself to a particular channel in our approach, while neglecting all others.
Suppose, \textit{e.g.}, we were only interested in the signal from sound waves because we wanted to compare our analysis with the one in Ref.~\cite{Caprini:2019egz}.
In this case, the full information on the projected sensitivities of LISA, DECIGO, and BBO to the GW signal from a phase transition would be encoded in the middle left panel of Fig.~\ref{fig:pis}, \textit{i.e.}, the plot of the three $\Omega_{\rm PIS}^{\rm s}$ curves as functions of $f_{\rm s}$.
We argue that essentially all sensitivity plots presented in Ref.~\cite{Caprini:2019egz} can be mapped onto this single PISC plot.


It is also interesting to compare our PISC plots to the usual scatter or contour plots of the total SNR as a function of a subset of model parameters, $\varrho = \varrho\left(\left\{p_i\right\}\right)$, that one frequently encounters in the literature (see also Sec.~6 in Ref.~\cite{Alanne:2019bsm}).%
\footnote{This class of plots includes the SNR plots that can be generated by the online tool \textsc{PTPlot} [\href{http://www.ptplot.org}{ptplot.org}].}
Such SNR plots typically show the dependence of the SNR on auxiliary SFOPT parameters, such as $\alpha$, $\beta/H_*$, $T_*$, etc., on two-dimensional hypersurfaces in parameter space.
In principle, one may produce arbitrarily many SNR plots because of the arbitrarily many possibilities to slice the higher-dimensional parameter space, and still, one would fail to really capture the \textit{entire} available information.
In our case, the model parameter space is, by contrast, projected onto only six frequency\,--\,amplitude planes without any loss of information.
In consequence, we do not need to keep any SFOPT parameters fixed at specific values.
In Refs.~\cite{Caprini:2015zlo,Caprini:2019egz}, most sensitivity plots assume, \textit{e.g.}, a specific phase transition temperature, such as $T_* = 50\,\textrm{GeV}$ or $T_* = 100\,\textrm{GeV}$.
In the case of our PISC plots, this is not necessary.
All benchmark points are simply projected onto the same six planes, irrespective of their associated $T_*$ value.
The sensitivity curves in these planes (\textit{i.e.}, our PISCs) are moreover completely independent of the SFOPT parameters.
By construction, they only depend on the experimental noise spectra and spectral shape functions in Eq.~\eqref{eq:S}.
In this sense, they represent truly \textit{experimental} quantities that can be studied without worrying much about questions related to theory and model building.%
\footnote{Of course, this may change in the future when more refined computations of the spectral shape functions should require one to incorporate a dependence on the SFOPT parameters in one way or another.
In this case, one will be able to indicate the range of possible spectral shapes, or the uncertainty in the spectral shapes, by sensitivity \textit{bands} just like the PISBs in Fig.~\ref{fig:bps} (see also Fig.~1 in Ref.~\cite{Schmitz:2020rag} for an illustration of this point).}
This property goes hand in hand with the fact that our PISCs are formulated in terms of physical observables that are experimentally accessible and that will likely play an important role in the experimental data analysis.
SNR scatter and contour plots, on the other hand, do \textit{not} disentangle experimental from theoretical uncertainties, as they are subject to \textit{all} uncertainties entering the computation of the SNR.
Still they provide useful information from a model builder's perspective.
In the end, SNR and PISC plots are complementary to each other, with their combination being the most powerful approach.


\subsection{Comparison of different BSM models}
\label{subsec:comparison}


\begin{table}
\begin{center}
\renewcommand{\arraystretch}{1.22}
\caption{BSM models and benchmark points used in our model comparison in Sec.~\ref{subsec:comparison}.
See text.}
\label{tab:bp}
\bigskip
\begin{tabular}{|c|c|c||cccc|}
\hline
Model & Type & BP & $T_*$ [GeV] & $\alpha$ & $\beta/H_*$ & $\phi_*/T_*$
$\vphantom{a_{\big(}^{\big(}}$\\
\hline\hline
   \parbox[t]{3mm}{\multirow{3}{*}{\rotatebox[origin=c]{90}{2HDM}}}
 & \parbox[t]{3mm}{\multirow{3}{*}{\rotatebox[origin=c]{90}{NP}}}
    & \#01 & 51.64 & 0.111 &  663 & 4.53 \\
 &  & \#02 & 61.25 & 0.070 & 1383 & 3.69 \\
 &  & \#03 & 68.71 & 0.046 & 2446 & 3.15 \\
\hline
   \parbox[t]{3mm}{\multirow{8}{*}{\rotatebox[origin=c]{90}{NMSSM}}}
 & \parbox[t]{3mm}{\multirow{4}{*}{\rotatebox[origin=c]{90}{NP}}}
    & \#04 &  76.4 & 0.143 &   6.0 & 3.12 \\
 &  & \#05 &  82.5 & 0.105 &  33.2 & 2.83 \\
 &  & \#06 &  94.7 & 0.066 & 105.9 & 2.40 \\
 &  & \#07 & 112.3 & 0.037 & 277   & 1.89 \\
\cline{2-7}
 & \parbox[t]{3mm}{\multirow{4}{*}{\rotatebox[origin=c]{90}{RP}}}
    & \#08 &  76.4 & 0.143 &   6.0 & 3.12 \\
 &  & \#09 &  82.5 & 0.105 &  33.2 & 2.83 \\
 &  & \#10 &  94.7 & 0.066 & 105.9 & 2.40 \\
 &  & \#11 & 112.3 & 0.037 & 277   & 1.89 \\
\hline
   \parbox[t]{3mm}{\multirow{4}{*}{\rotatebox[origin=c]{90}{xSM}}}
 & \parbox[t]{3mm}{\multirow{4}{*}{\rotatebox[origin=c]{90}{RP}}}
    & \#12 & 56.4 & 0.20 &  6.42 & 4.32\\
 &  & \#13 & 59.6 & 0.17 & 12.54 & 4.07\\
 &  & \#14 & 65.2 & 0.12 & 29.96 & 3.70\\
 &  & \#15 & 70.6 & 0.09 & 47.35 & 3.39\\
\hline
   \parbox[t]{3mm}{\multirow{4}{*}{\rotatebox[origin=c]{90}{SMEFT}}}
 & \parbox[t]{3mm}{\multirow{2}{*}{\rotatebox[origin=c]{90}{NP}}}
    & \#16 & 26 & 2.3  &   5 & 9.5 \\
 &  & \#17 & 63 & 0.13 & 160 & 4   \\
\cline{2-7}
 & \parbox[t]{3mm}{\multirow{2}{*}{\rotatebox[origin=c]{90}{RP}}}
    & \#18 & 26 & 2.3  &   5 & 9.5 \\
 &  & \#19 & 63 & 0.13 & 160 & 4   \\
\hline
   \parbox[t]{3mm}{\multirow{12}{*}{\rotatebox[origin=c]{90}{DM}}}
 & \parbox[t]{3mm}{\multirow{8}{*}{\rotatebox[origin=c]{90}{NP}}}
    & \#20 &   10 & 0.1 &  10 & --- \\
 &  & \#21 &   10 & 0.5 & 100 & --- \\
 &  & \#22 &   50 & 0.1 &  10 & --- \\
 &  & \#23 &   50 & 0.5 & 100 & --- \\
 &  & \#24 &  100 & 0.1 &  10 & --- \\
 &  & \#25 &  100 & 0.5 & 100 & --- \\
 &  & \#26 & 1000 & 0.1 &  10 & --- \\
 &  & \#27 & 1000 & 0.5 & 100 & --- \\
\cline{2-7}
 & \parbox[t]{3mm}{\multirow{4}{*}{\rotatebox[origin=c]{90}{RV}}}
    & \#28 &    10 & --- & 100 & --- \\
 &  & \#29 &   100 & --- & 100 & --- \\
 &  & \#30 &  1000 & --- & 100 & --- \\
 &  & \#31 & 10000 & --- & 100 & --- \\
\hline
   \parbox[t]{3mm}{\multirow{3}{*}{\rotatebox[origin=c]{90}{Dilaton}}}
 & \parbox[t]{3mm}{\multirow{3}{*}{\rotatebox[origin=c]{90}{RV}}}
    & \#32 & 100 & --- &  3 & --- \\
 &  & \#33 & 100 & --- & 15 & --- \\
 &  &      &     &     &    &     \\
\hline
\end{tabular}
\end{center}
\end{table}


Another advantage of our PISC plots is that they set the stage for a systematic comparison of the GW phenomenology in different BSM models.
To illustrate this point, we shall revisit the full set of BSM models and benchmark points investigated in Ref.~\cite{Caprini:2015zlo} in this section, recasting their respective predictions in terms of our new PISC method.
An equivalent analysis for all models and benchmark points studied in Ref.~\cite{Caprini:2019egz} can be found in Ref.~\cite{Schmitz:2020rag}.
The present work and Ref.~\cite{Schmitz:2020rag} therefore cover together the full set of results in Refs.~\cite{Caprini:2015zlo,Caprini:2019egz}, illustrating how the sensitivity plots in these two papers can be translated into our PISC language.
Comparing the present paper with Ref.~\cite{Caprini:2015zlo} and Ref.~\cite{Schmitz:2020rag} with Ref.~\cite{Caprini:2019egz} thus highlights how improvements in modeling the GW signal are reflected in the different types of sensitivity plots, \textit{i.e.}, PISC and SNR scatter plots, respectively (see also Sec.~\ref{subsec:update}). 
In addition to the xSM, Ref.~\cite{Caprini:2015zlo} also considers SFOPTs in:
(i) the \textit{two-Higgs-doublet model} (2HDM)~\cite{Kakizaki:2015wua,Dorsch:2016tab},
(ii) the \textit{next-to-minimal supersymmetric standard model} (NMSSM)~\cite{Huber:2015znp},
(iii) the \textit{standard model effective field theory} (SMEFT)~\cite{Huber:2007vva},%
\footnote{Here, ``SMEFT'' refers to an extension of the SM Higgs potential by an additional dimension-six operator.}
(iv) a strongly coupled hidden sector giving rise to composite \textit{dark matter} (DM)~\cite{Cornell:2013rza,Boddy:2014yra}, and
an approximately conformal \textit{dilaton} model in the context of warped extra dimensions~\cite{Nardini:2007me,Konstandin:2011dr}.
A detailed discussion of these models and SFOPT scenarios can be found in Refs.~\cite{Caprini:2015zlo,Caprini:2019egz}, which we shall not reproduce here.
For our purposes, the entire relevant data describing these models is contained in Tab.~\ref{tab:bp}, where we list the values of $T_*$, $\alpha$, $\beta/H_*$, and $\phi_*/T_*$ for all benchmark points that we shall include in our analysis.
For each phase transition, we also indicate whether we expect it to be of NP, RP, or RV type (see Sec.~\ref{sec:signal}).
Here, note that, for the NMSSM and SMEFT, we consider both NP and RP phase transitions, which explains why the corresponding benchmark points are duplicated in Tab.~\ref{tab:bp}.
As for the DM and dilaton models, more quantitative analyses of the phase transition dynamics are still pending.
The corresponding benchmark points in Tab.~\ref{tab:bp} therefore represent educated guesses rather than precise numerical predictions.


\begin{figure}
\begin{center}

\includegraphics[width=0.95\textwidth]{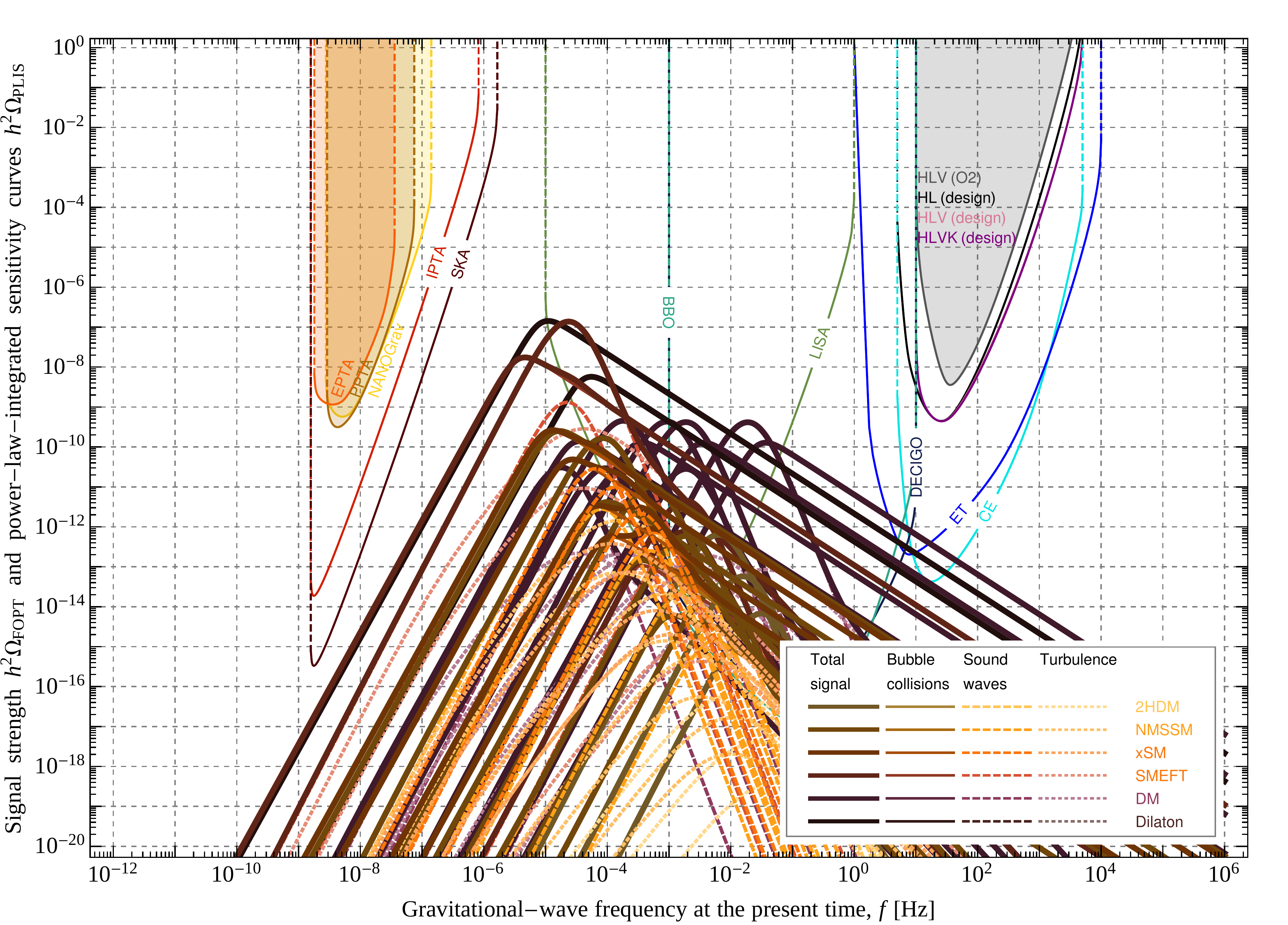}
\caption{PLISCs and GW signals for all benchmark points listed in Tab.~\ref{tab:bp}.}
\label{fig:spectra}
\end{center}
\end{figure}


Based on the data in Tab.~\ref{tab:bp}, we are able to repeat our analysis in the previous section for all 33 benchmark points that we are interested in.
First of all, we plot again the total GW signal and its three individual contributions as functions of frequency for all 33 benchmark points.
This results in the busy plot in Fig.~\ref{fig:spectra}, which highlights another limitation of the standard PLISC approach.
PLISC plots such as those in Figs.~\ref{fig:plis} and \ref{fig:spectra} are not well suited for comparing a large number of spectra to each other.
This is also part of the reason why most PLISC plots in the literature only show a handful of spectra.
As soon as one intends to study $\mathcal{O}\left(10\right)$ or more spectra at the same time, the PLISC approach becomes highly impractical.
One may argue that a possible way out of this problem might consist in restricting oneself to just plotting points of the form $\big(f_{\rm i},\Omega_{\rm i}^{\rm peak}\big)$.
In this case, our busy collection of signal curves would reduce to a simple scatter plot that could be compared more easily to the various PLISCs in Fig.~\ref{fig:spectra}.
Indeed, this is a strategy that one sometimes encounters in the literature.
We, however, argue that such an approach corresponds to comparing apples and oranges.
PLISCs do represent useful sensitivity curves; but there is no reason to believe that they are also automatically the best choice for indicating experimental sensitivities to an ensemble of peak frequencies and peak amplitudes, a purpose they are not specifically designed for.
In the case of GWs from a cosmological phase transition, PLISC plots simply no longer encode information on the expected SNR (see our discussion in Sec.~\ref{subsec:example}), irrespective of whether one decides to combine them with a busy collection of signal curves or a $\big(f_{\rm i},\Omega_{\rm i}^{\rm peak}\big)$ scatter plot.


\begin{figure}
\begin{center}

\includegraphics[width=0.45\textwidth]{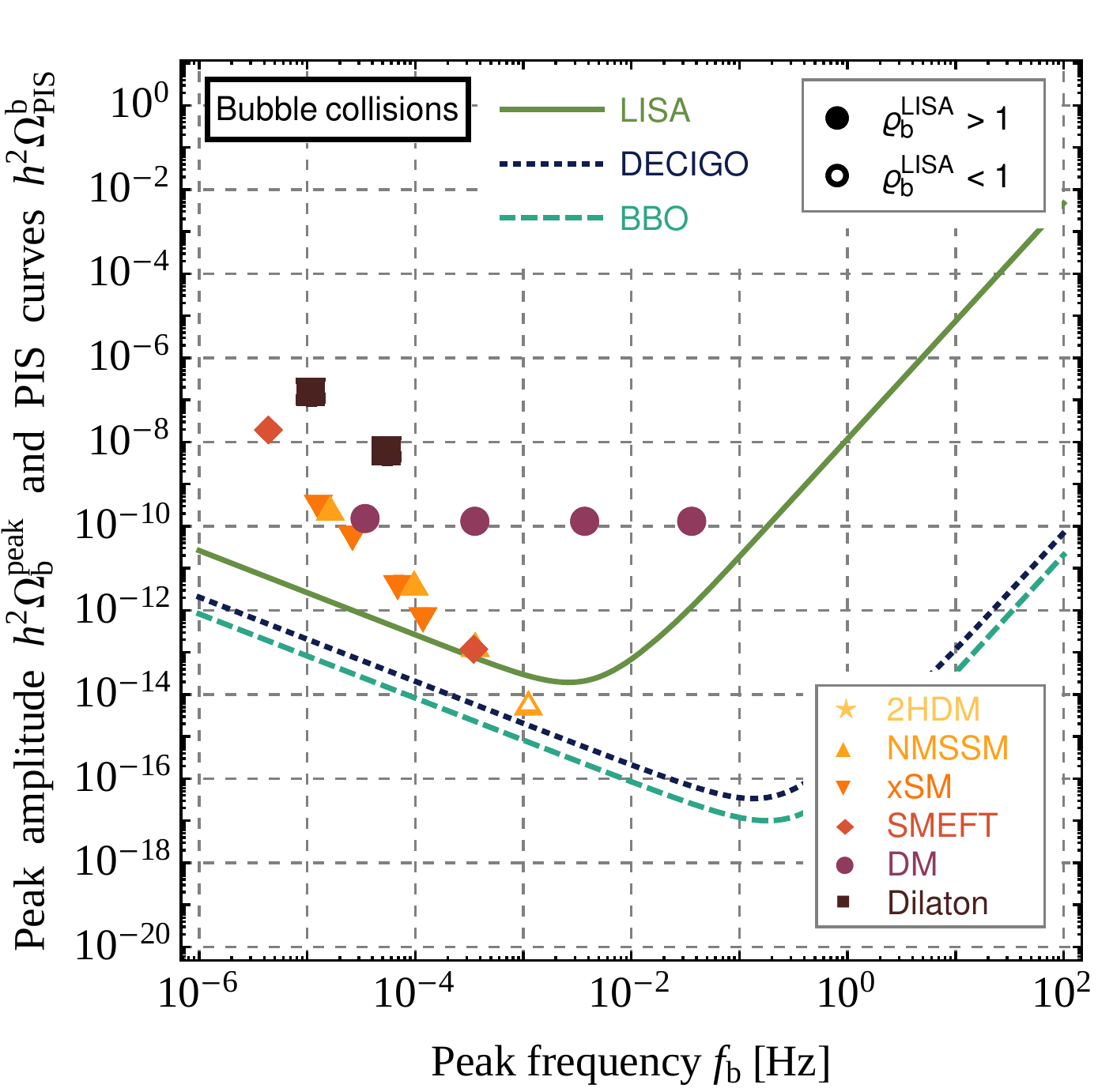}\hfill
\includegraphics[width=0.45\textwidth]{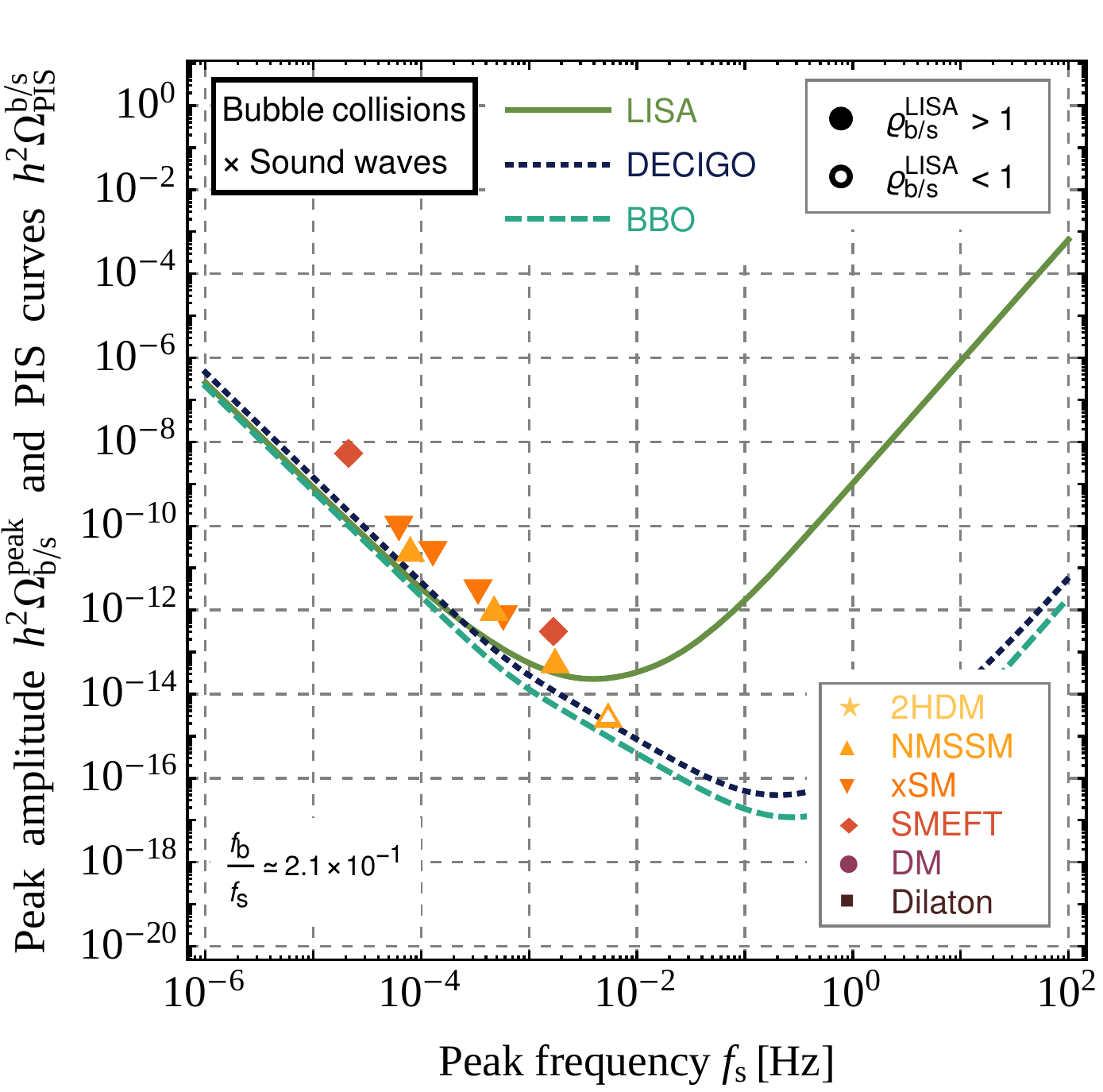}

\includegraphics[width=0.45\textwidth]{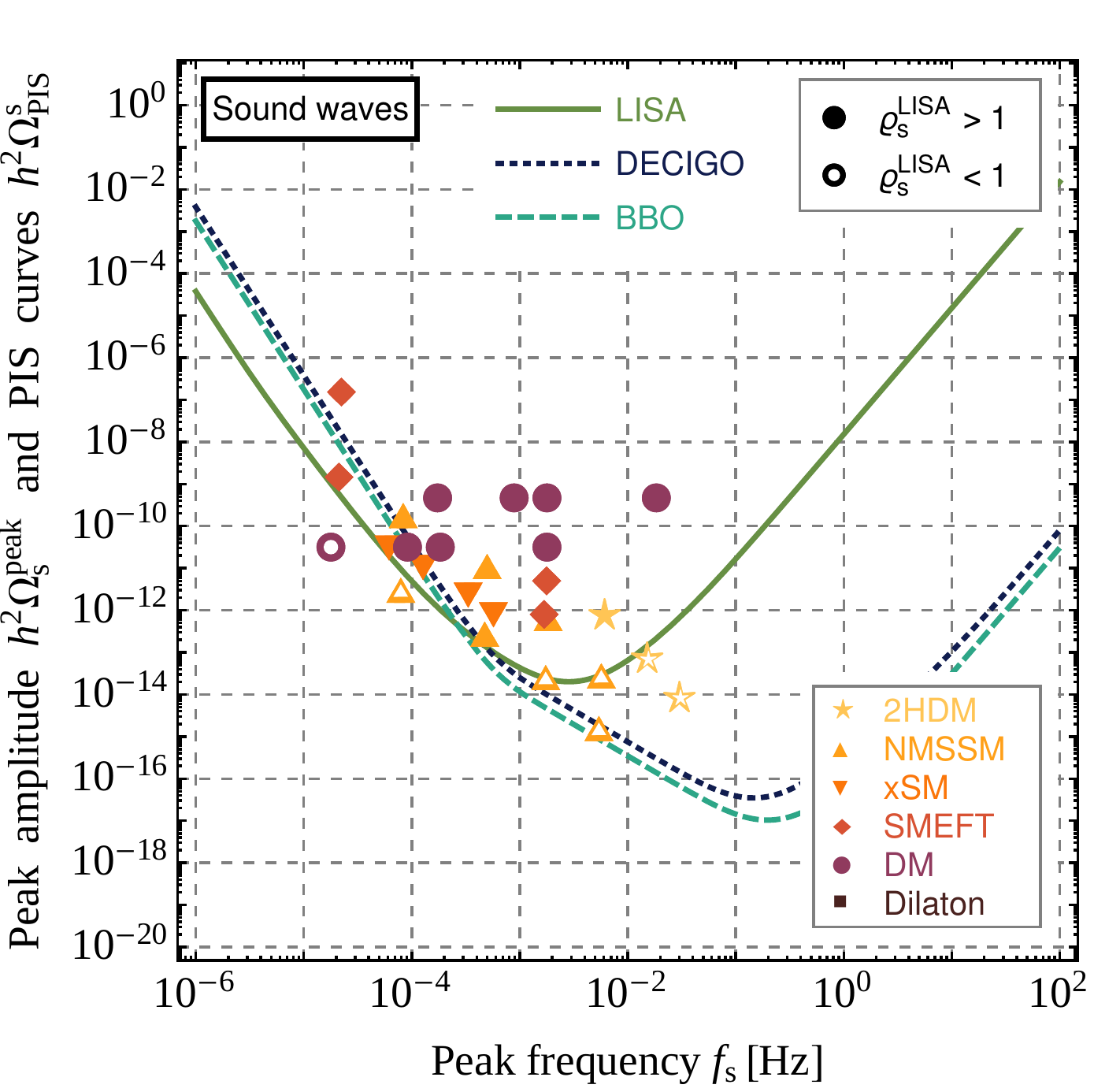}\hfill
\includegraphics[width=0.45\textwidth]{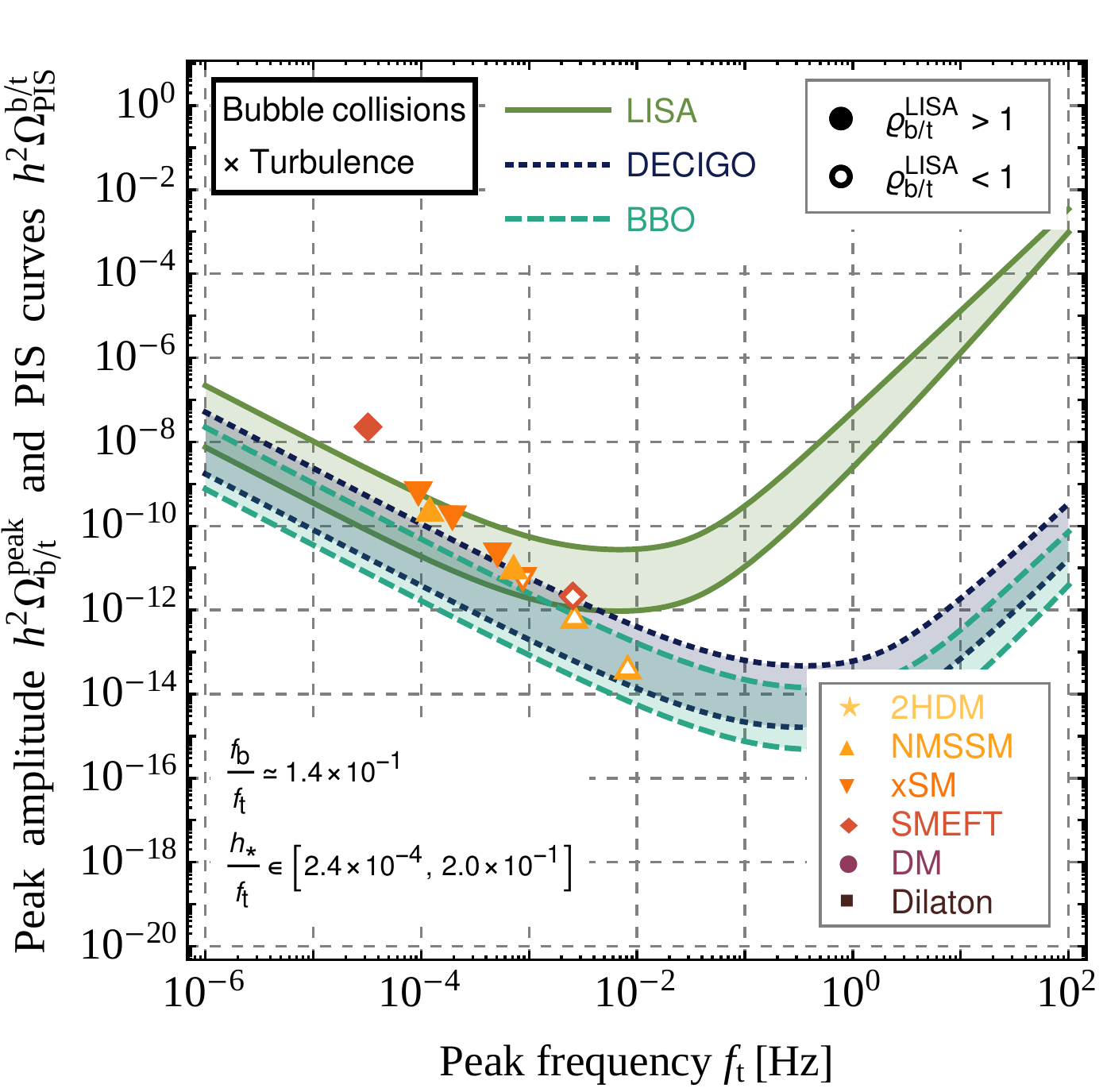}

\includegraphics[width=0.45\textwidth]{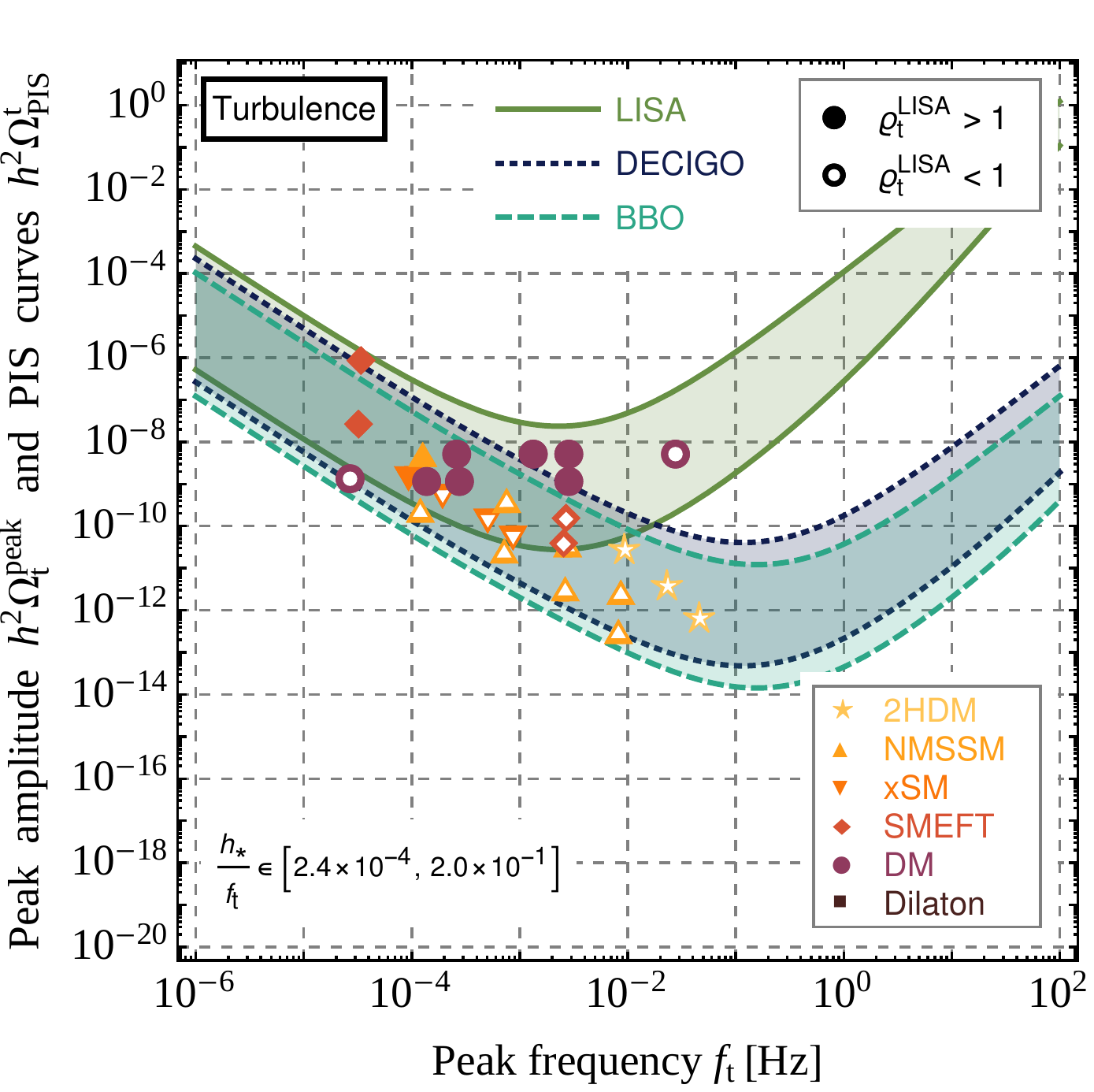}\hfill
\includegraphics[width=0.45\textwidth]{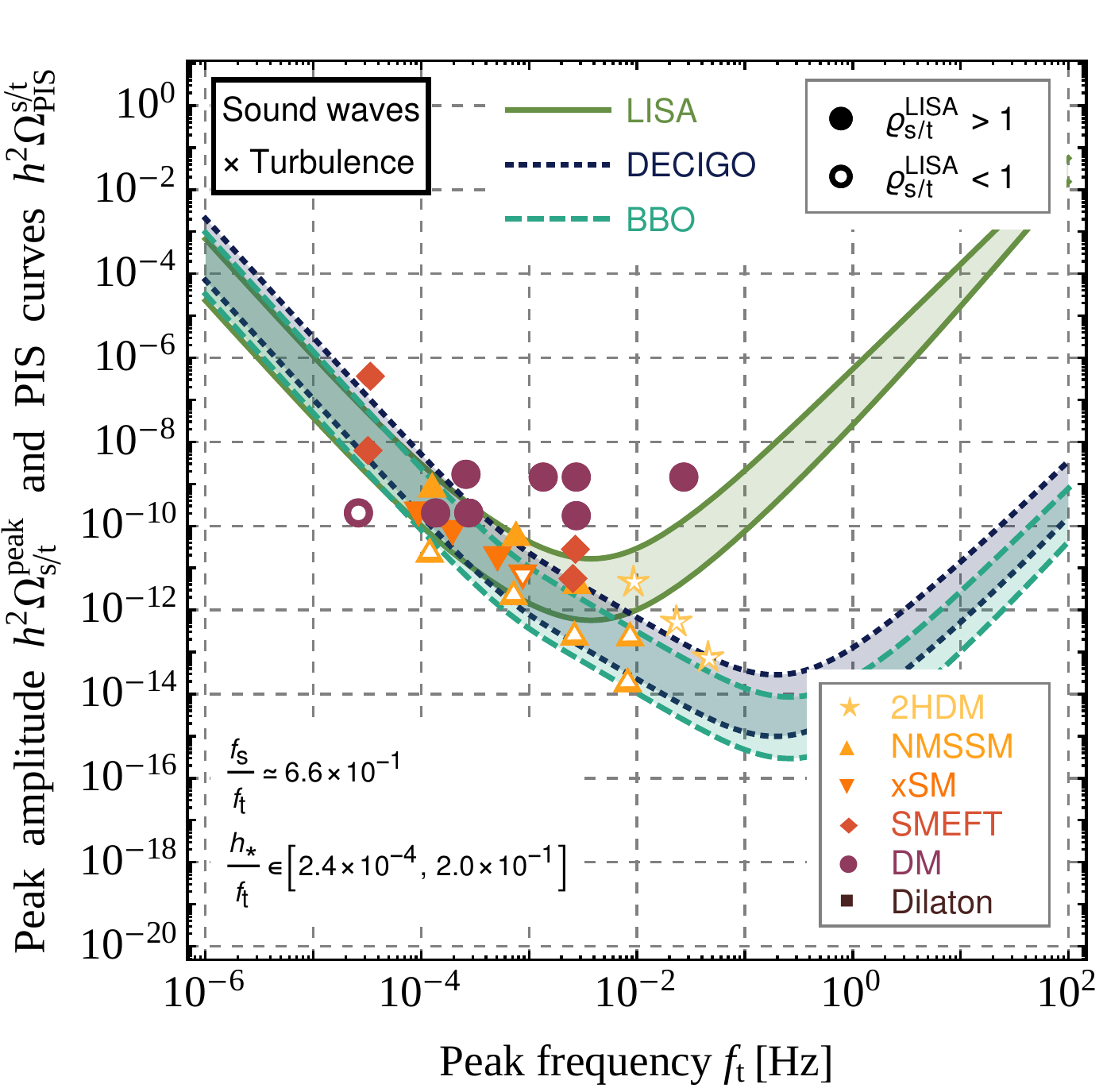}
\caption{PISCs for LISA, DECIGO, and BBO together with the predictions of all BPs. See text.}
\label{fig:bps}
\end{center}
\end{figure}


Our PISC method amounts, by contrast, to comparing apples and apples.
That is, our PISCs are constructed in exactly such a way that they represent the optimal sensitivity curves to be used in scatter plots of peak frequencies and peak amplitudes.
This is illustrated in Fig.~\ref{fig:bps}, where we now combine our PISCs for LISA, DECIGO, and BBO with the predictions of the benchmark points listed in Tab.~\ref{tab:bp}.
In contrast to Fig.~\ref{fig:spectra}, the plots in Fig.~\ref{fig:bps} are much easier to read, while at the same time, they still encode the full SNR information on the $y$-axis.
A slight difference between Fig.~\ref{fig:pis} and Fig.~\ref{fig:bps}, though, is that we are now no longer able to draw sensitivity \textit{curves} for all six signal channels; for all channels involving the signal from turbulence, we have to draw sensitivity \textit{bands}.
The reason for this is the dependence of the shape function $\mathcal{S}_{\rm t}$ on the Hubble frequency $h_*$ [see Eq.~\eqref{eq:S}], whose value is independent of the parameter combination $\beta/H_*/v_w$, unlike the values of $f_{\rm b}$, $f_{\rm s}$, and $f_{\rm t}$ [see Eqs.~\eqref{eq:hstar} and \eqref{eq:fbst}].
In Fig.~\ref{fig:bps}, the different parameter dependence of $h_*$ and the three peak frequencies is accounted for by the width of the PISBs, which reflects the variation of the ratio $h_*/f_t$ in our data set.
We also point out that, as a consequence of this finite width, it is not clear whether benchmark points \textit{inside} PISBs do or do not have a partial SNR larger than one.
To remedy this shortcoming, we distinguish between ``empty'' and ``filled'' points in Fig.~\ref{fig:bps}, which respectively correspond to partial SNRs for LISA, $\varrho_{\rm i/j}^{\rm LISA}$, smaller or larger than one.


As in the previous section, let us also compare our results to the standard approach of SNR scatter and contour plots.
In contrast to these standard plots, our PISC plots in Fig.~\ref{fig:bps} do not require contour lines or a color code to indicate the expected SNR.
This provides us with the freedom to use a color code for distinguishing between the predictions of different models.
Similarly, the fact that we do not rely on SNR contour lines in Fig.~\ref{fig:bps} allows us to plot and compare the PISCs and PISBs of three future experiments at the same time.
We argue that such a simultaneous comparison of different models and different experiments would be significantly more difficult if one were to work with SNR plots only.
Finally, we stress once more that it would be trivial to restrict oneself to individual signal channels in our analysis.
For instance, if one were interested in the signal from sound waves only, the entire relevant information would be readily contained in the middle left panel of Fig.~\ref{fig:bps}.


\subsection{Relaxing some of the underlying assumptions}
\label{subsec:assumptions}


The sensitivity curves and bands in Figs.~\ref{fig:pis} and \ref{fig:bps} are constructed in such a way that they have a natural interpretation in terms of the expected SNR.
Consider, \textit{e.g.}, a benchmark point that is separated from the PISC in the i/j-channel by a multiplicative factor $\Delta y$ along the $y$-axis.
Thanks to the definitions and conventions adopted in Sec.~\ref{subsec:definition}, this benchmark point predicts a partial SNR of exactly $\varrho_{\rm i/j} = \Delta y$.
Benchmark points that directly lie \textit{on} a PISC correspondingly predict $\varrho_{\rm i/j} = 1$.
Each of our PISCs is therefore canonically normalized to a unit threshold value for the corresponding partial SNR.
Similarly, all of our PISCs and PISBs are normalized to an observing time of $t_{\rm obs} = 1\,\textrm{yr}$, which we simply choose for convenience.
Our PISBs rely in addition on the explicit expressions for the Hubble and peak frequencies in Eqs.~\eqref{eq:hstar} and \eqref{eq:fbst}.
In this section, we shall now demonstrate how our plots can be generalized if one is interested in relaxing some of these assumptions, \textit{i.e.}, if one is interested in a different normalization or a different relation among the Hubble and peak frequencies.

\begin{figure}
\begin{center}

\includegraphics[width=0.45\textwidth]{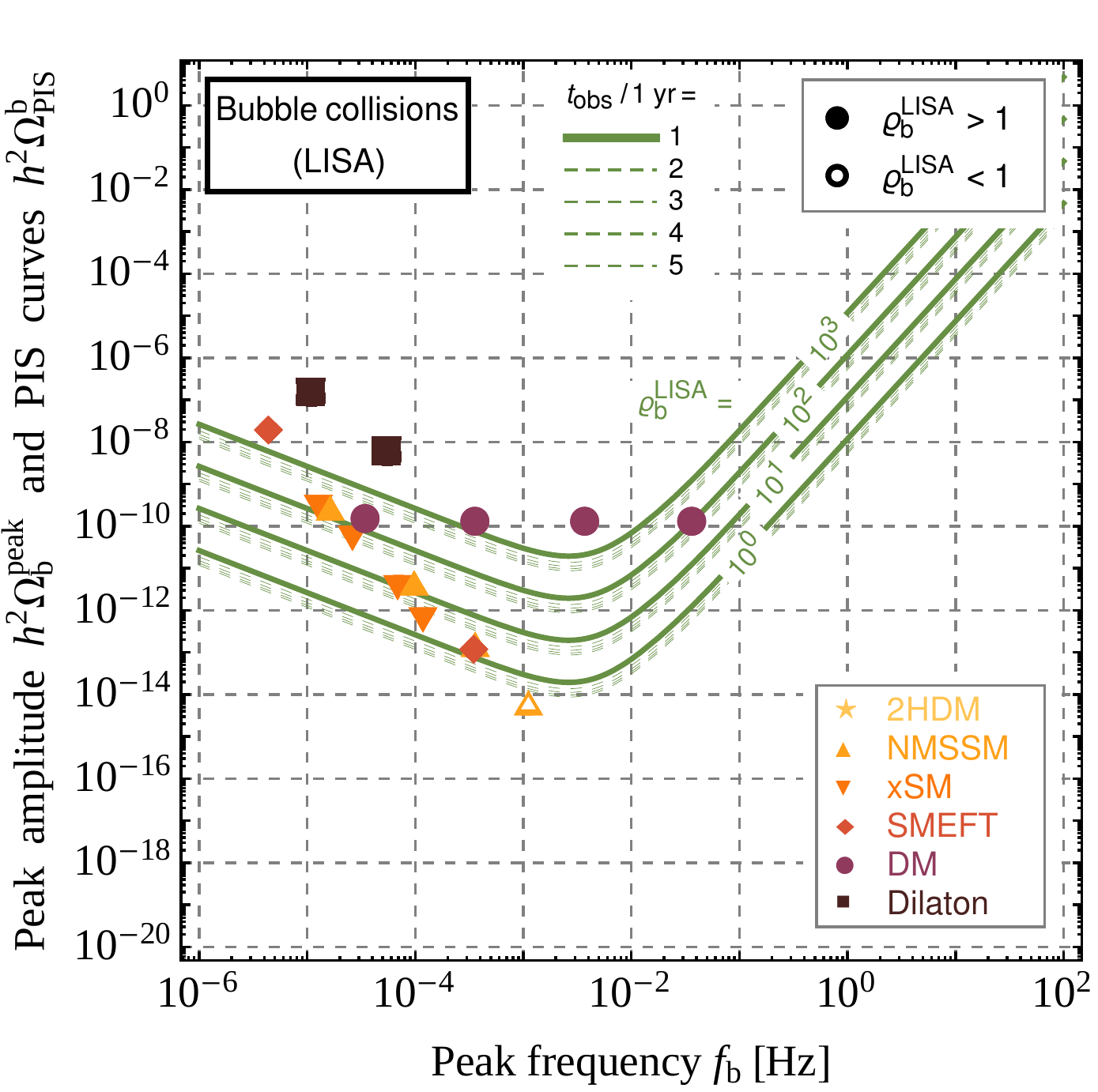}
\caption{PISC for LISA in the b-channel for different values of $\varrho_{\rm thr}$ and $t_{\rm obs}$.
See Eq.~\eqref{eq:OPISgen}.}
\label{fig:snr}
\end{center}
\end{figure}

In a first step, let us generalize Eqs.~\eqref{eq:rhoMaster}, \eqref{eq:rhoij}, and \eqref{eq:OPIS} to arbitrary $\varrho_{\rm thr}$ and $t_{\rm obs}$,
\begin{align}
\label{eq:OPISgen}
\varrho & = \sqrt{\varrho_{\rm b}^2 + \varrho_{\rm s}^2 + \varrho_{\rm t}^2 + \varrho_{\rm b/s}^2 + \varrho_{\rm b/t}^2 + \varrho_{\rm s/t}^2} \,, \\\nonumber
\varrho_{\rm i/j} & = \varrho_{\rm thr}\,\frac{\Omega_{\rm i/j}^{\rm peak}}{\Omega_{\rm PIS}^{\rm i/j}\left(\varrho_{\rm thr},t_{\rm obs}\right)} \,, \\\nonumber
\Omega_{\rm PIS}^{\rm i/j}\left(\varrho_{\rm thr},t_{\rm obs}\right) & = \varrho_{\rm thr}\left[\left(2 - \delta_{\rm ij}\right) n_{\rm det}\, t_{\rm obs} \int_{f_{\rm min}}^{f_{\rm max}} df\:\frac{\mathcal{S}_{\rm i}\left(f\right)\mathcal{S}_{\rm j}\left(f\right)}{\Omega_{\rm noise}^2\left(f\right)}\right]^{-1/2} \,,
\end{align}
where $\varrho_{\rm thr}$ now denotes the partial SNR threshold for a single channel.
The generalized peak-integrated sensitivity in the i/j-channel can thus be interpreted as the minimal peak amplitude that is necessary to reach, after an observing time $t_{\rm obs}$, a partial SNR of $\varrho_{\rm thr}$ in this channel.
A hypothetical scenario predicting peak amplitudes that just reach the respective threshold sensitivities in all six channels would therefore predict a total SNR of $\varrho = \sqrt{6}\,\varrho_{\rm thr}$.
As evident from Eq.~\eqref{eq:OPISgen}, the generalized sensitivities scale as follows with $\varrho_{\rm thr}$ and $t_{\rm obs}$,
\begin{align}
\label{eq:scaling}
\Omega_{\rm PIS}^{\rm i/j}\left(\varrho_{\rm thr},t_{\rm obs}\right) \propto \frac{\varrho_{\rm thr}}{\sqrt{t_{\rm obs}}} \,,
\end{align}
which is the same dependence as in the case of the usual power-law-integrated sensitivities [see Eq.~\eqref{eq:plis}].
We illustrate this scaling behavior in Fig.~\ref{fig:snr}, we where plot the generalized PISC for LISA in the b-channel for different values of $\varrho_{\rm thr}$ and $t_{\rm obs}$.
The main message of this plot is twofold:
First of all, it demonstrates that our method allows for an easy implementation of different SNR thresholds. 
Once the base PISCs for $\varrho_{\rm thr} = 1$ have been constructed, the sensitivity curves for larger or smaller SNR thresholds can be obtained by simply shifting these base curves up or down, \textit{i.e.}, by rescaling them by a factor $\varrho_{\rm thr}$.
This needs to be compared to the situation in the case of SNR scatter and contour plots, which typically do \textit{not} exhibit a universal relation between shifts in the expected SNR, $\Delta\varrho$, and shifts in one of the coordinate directions, $\Delta x$ or $\Delta y$.
The second message of Fig.~\ref{fig:snr} is that our PISC plots are reminiscent of plots that one often encounters in other fields of experimental physics, such as, \textit{e.g.}, the standard sensitivity plots for DM direct-detection experiments or the Brazil-band plots that were often shown by the experiments at the Large Hadron Collider prior to the discovery of the SM Higgs boson.
What we mean by this is that our PISC plots show, in a manner intuitive for particle physicists, how future GW experiments will approach from above and cut into the signal regions of specific SFOPT scenarios.
As in the case of the DM and Higgs plots, a better sensitivity reach of an experiment corresponds to a PISC extending to lower values along the $y$-axis in our plots.
Thanks to the scaling with the observing time $t_{\rm obs}$ in Eq.~\eqref{eq:scaling}, the expected experimental progress over the years can in particular be pictured as pushing our PISCs further and further down in the vertical direction.
Again, the situation in the case of SNR scatter and contour plots is quite different, as these plots typically do \textit{not} exhibit any simple scaling relation with respect to the observing time $t_{\rm obs}$.


\begin{figure}
\begin{center}

\includegraphics[width=0.32\textwidth]{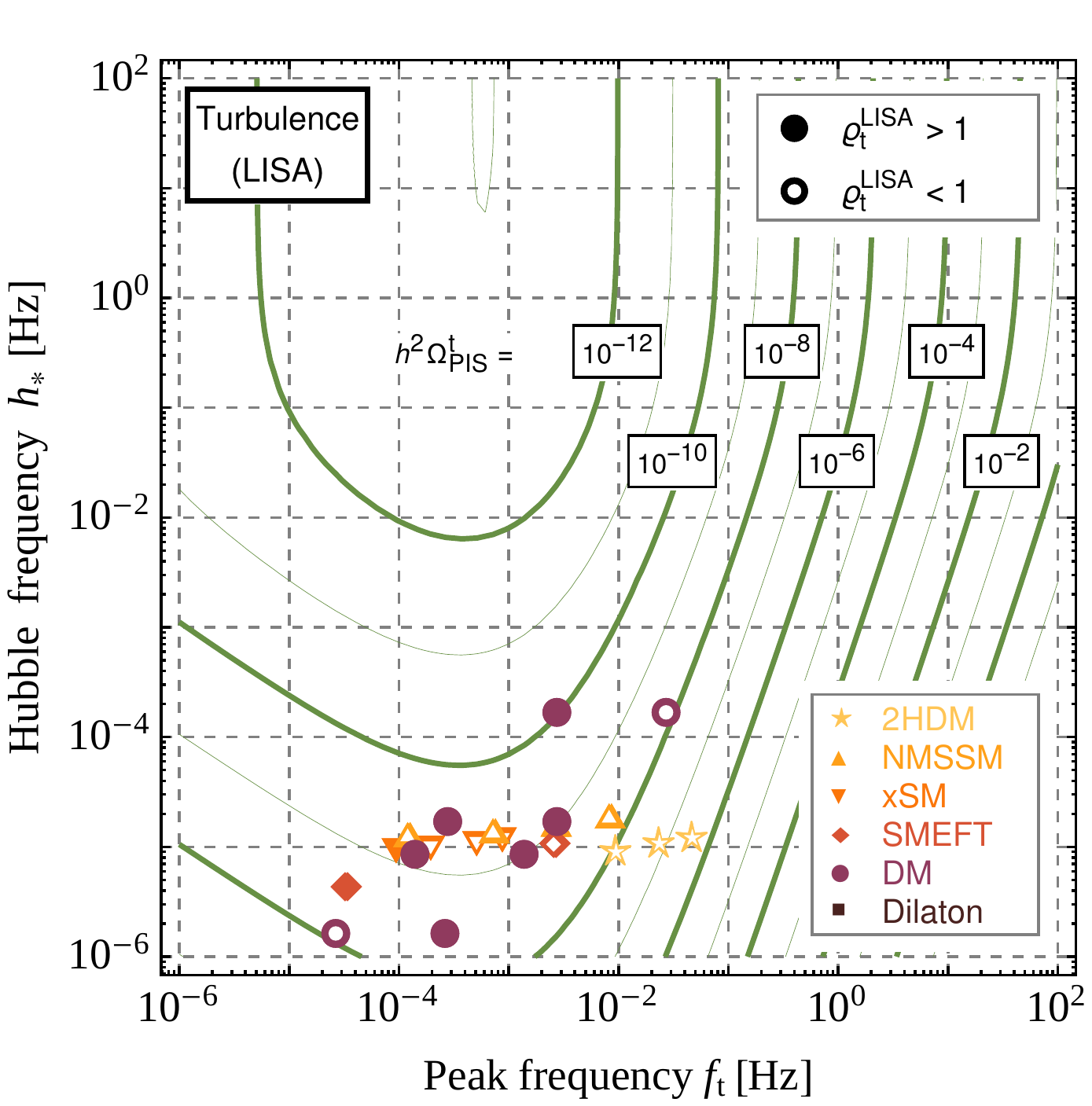}\,
\includegraphics[width=0.32\textwidth]{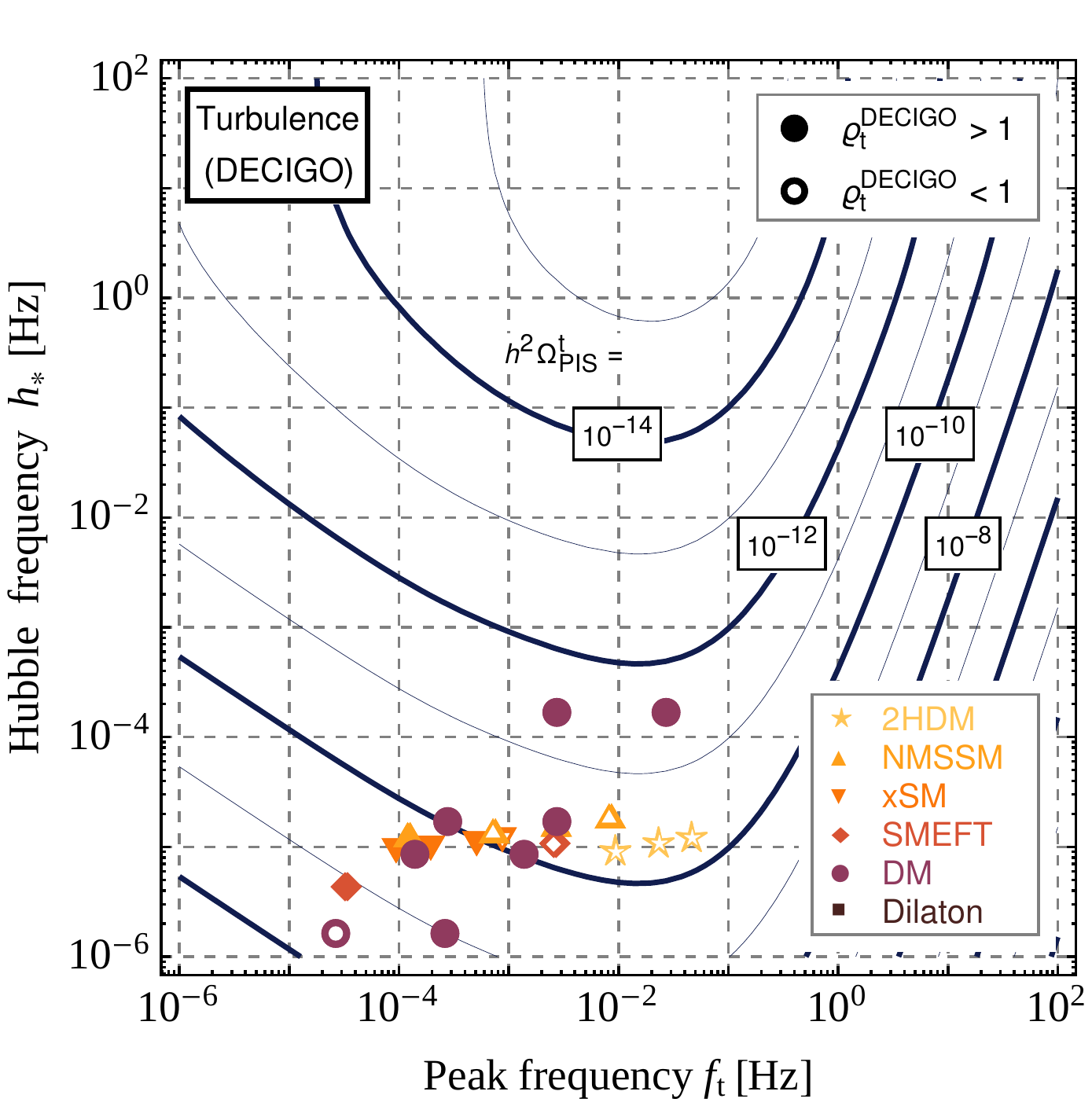}\,
\includegraphics[width=0.32\textwidth]{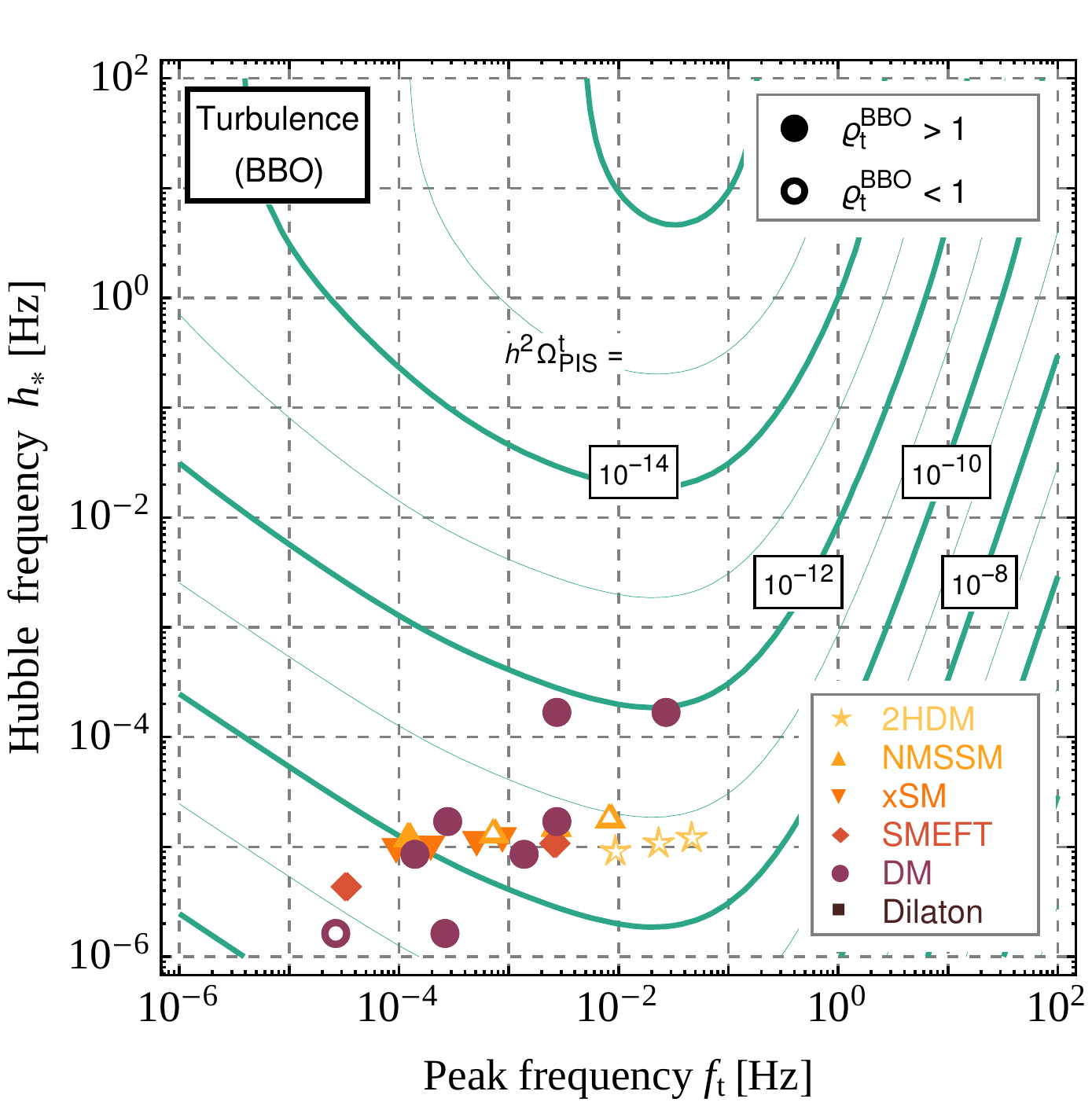}
\caption{PISCs for LISA, DECIGO, and BBO in the t-channel as functions of $f_{\rm t}$ and $h_*$.
See text.}
\label{fig:turbulence}
\end{center}
\end{figure}


Next, let us briefly discuss an alternative presentation of the sensitivity bands in Fig.~\ref{fig:bps}.
To this end, suppose that future progress will lead to a better theoretical understanding of the peak frequencies $f_{\rm b}$, $f_{\rm s}$, $f_{\rm t}$, in particular, to the realization that $f_{\rm t}$ is in fact not simply proportional to $h_*$ times one power of the parameter combination $\beta/H_*/v_w$.
In this case, it will be helpful to have a proper understanding of the full dependence of $\Omega_{\rm PIS}^{\rm t}$ on both frequencies, $f_{\rm t}$ and $h_*$ [see Eq.~\eqref{eq:OPISbst}].
Therefore, instead of constructing sensitivity bands as in Fig.~\ref{fig:bps}, one may as well work with contour plots of $\Omega_{\rm PIS}^{\rm t}$ in the $f_{\rm t}$\,--\,$h_*$ plane (see Fig.~\ref{fig:turbulence}).
Similar contour plots can also be drawn for the sensitivities in the b/t- and s/t-channels.
The advantage of this alternative presentation is that the Hubble frequency $h_*$ now no longer corresponds to a hidden parameter.
In Fig.~\ref{fig:bps}, it is unclear by construction with which curves in a PISB one should respectively compare the individual benchmark points.
The plots in Fig.~\ref{fig:turbulence} offer a trivial solution to this problem, as they explicitly feature $h_*$ on the $y$-axis.
The disadvantage of this method, however, is that it requires contour lines to indicate the experimental sensitivity.
In this sense, the partial SNR is now encoded in the $z$-direction, which means that one looses some of the attractive properties of our PISC plots.
In Fig.~\ref{fig:turbulence}, we use, \textit{e.g.}, empty and filled symbols to indicate which benchmark points predict a partial SNR smaller or larger than one.
However, beyond that, the plots in Fig.~\ref{fig:turbulence} contain no further information on the expected SNR.
In particular, it is unclear how far the individual points are separated from the threshold sensitivity at their respective locations in the $f_{\rm t}$\,--\,$h_*$ plane.


\subsection{Runaway phase transitions in vacuum}
\label{subsec:runaway}

\begin{figure}
\begin{center}

\includegraphics[width=0.45\textwidth]{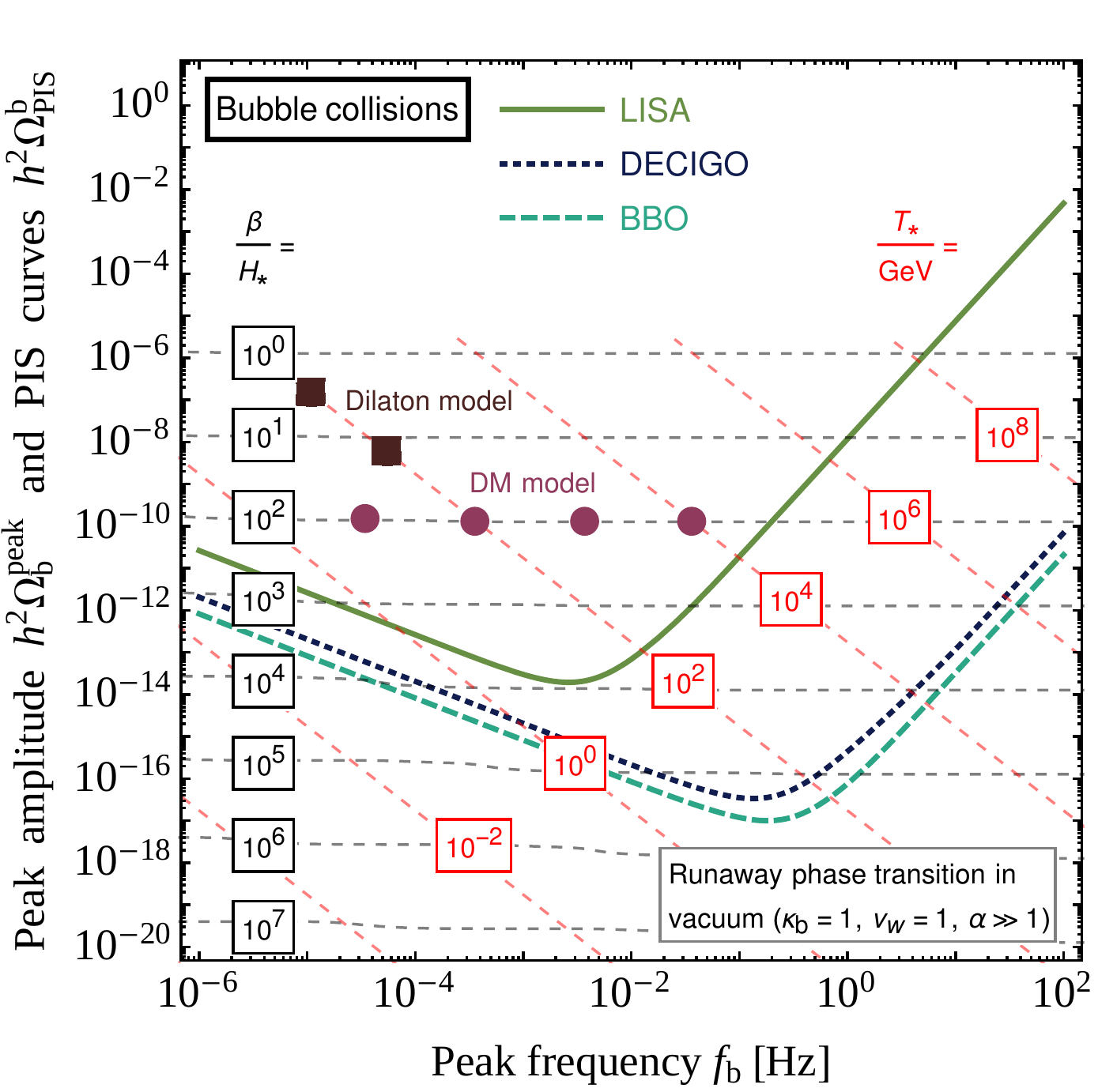}
\caption{Contours of constant $\beta/H_*$ and $T_*$ in the case of a runaway phase transition in vacuum.}
\label{fig:runaway}
\end{center}
\end{figure}

The only free parameters in the case of a RV phase transition are $\beta/H_*$ and $T_*$ (see Sec.~\ref{sec:signal}).
All other SFOPT parameters are fixed at the values listed in Eq.~\eqref{eq:rvpt}.
At the same time, the only relevant source of GWs during a RV phase transition are bubble collisions.
For this type of phase transition, there is hence a one-to-one correspondence between $\beta/H_*$ and $T_*$ on the one hand and the peak frequency and peak amplitude in the b-channel on the other hand.
This relation is shown in Fig.~\ref{fig:runaway}, where we overlay the PISC plot for the signal from bubble collisions with contours of constant $\beta/H_*$ and $T_*$.
As can be read off from this plot, LISA will be sensitive to $\beta/H_*$ values as large as $\beta/H_* \sim 10^4$ if $T_*$ is close to 10 GeV, while DECIGO and BBO will be able to probe RV phase transitions up to $\beta/H_* \sim 10^5$ for temperatures $T_*$ in the same range.
Our results for LISA in Fig.~\ref{fig:runaway} are consistent with Fig.~6 in Ref.~\cite{Caprini:2015zlo}, which shows LISA's sensitivity to the signal from bubble collisions in the $\beta/H_*$\,--\,$T_*$ plane.
The main difference between Fig.~\ref{fig:runaway} and Fig.~6 in Ref.~\cite{Caprini:2015zlo}, however, is that our PISC plot encodes the full SNR information on the $y$-axis.
Despite the fact that both plots are related by nothing but a simple parameter transformation, this is a distinct advantage of our Fig.~\ref{fig:runaway}.


The plot in Fig.~\ref{fig:runaway} represents a simple example that demonstrates how information on the underlying SFOPT parameters can be included in PISC plots.
Further, more sophisticated examples can be found in the companion paper~\cite{Alanne:2019bsm}, where we use our PISC method for a comprehensive analysis of the GW signal in the xSM.
In Ref.~\cite{Alanne:2019bsm}, we especially show how PISC plots can be used to investigate the dependence of the signal on the SFOPT parameters as well as on model parameters such as particle masses or coupling constants.
In addition, we use our PISC plots as a starting point for constructing distribution functions (\textit{i.e.}, histograms) in the space of peak frequencies and peak amplitudes.
These distribution functions provide a useful tool to characterize the GW phenomenology of the xSM, in particular, when combined with data on the underlying model parameters.
In future work, it would be interesting to repeat our analysis in Ref.~\cite{Alanne:2019bsm} for a broad class of BSM models.
Such a global analysis, including PISC plots and histograms such as those in Ref.~\cite{Alanne:2019bsm}, would allow for a comprehensive model comparison at both the qualitative and quantitative level.


\subsection{New signal predictions after theory updates}
\label{subsec:update}


In this paper, we apply our PISC approach to all models and benchmark points in Ref.~\cite{Caprini:2015zlo}, which allows us to demonstrate how to use our method if the total GW signal receives three individual contributions, $\Omega_{\rm b}$, $\Omega_{\rm s}$, and $\Omega_{\rm t}$ (see the discussion in Sec.~\ref{sec:signal}).
An equivalent analysis for the models and benchmark points in Ref.~\cite{Caprini:2019egz}, which only considers the GW signal from sound waves, can be found in Ref.~\cite{Schmitz:2020rag}.
In this section, we shall now demonstrate how the different modeling of the signal in Refs.~\cite{Caprini:2015zlo,Caprini:2019egz} is reflected in our PISC plots.
As we shall see, this highlights another advantage of our method:
Any update in the theoretical description of the signal that does not affect the spectral shape functions in Eq.~\eqref{eq:S} leaves our sensitivity curves invariant.
An improved theoretical understanding of the peak amplitudes and peak frequencies merely shifts the individual benchmark points in our plots.%
\footnote{In the context of a particular model, this feature could, \textit{e.g.}, be exploited to discuss the ``trajectories'' of individual benchmark points in our PISC plots that reflect their evolution in consequence of a theory update.}
This needs to be compared to the SNR scatter plots in Refs.~\cite{Caprini:2015zlo,Caprini:2019egz} as well as to the plots generated by the online tool \textsc{PTPlot}, which crucially depend on the precise modeling of the GW source.
At the same time, we emphasize once more that any future update of the spectral shape functions in Eq.~\eqref{eq:S} will require an update of the PISCs presented in this paper.


An important difference between Refs.~\cite{Caprini:2015zlo,Caprini:2019egz} is that Ref.~\cite{Caprini:2019egz} discards the GW signal from bubble collisions (as well as the GW signal from turbulence, which still requires a better theoretical understanding).
The reason for this is the realization~\cite{Bodeker:2017cim,Ellis:2019oqb,Ellis:2020nnr,Hoeche:2020rsg} that the efficiency factor $\kappa_{\rm b}$ can be significantly suppressed compared to earlier estimates [see Eq.~\eqref{eq:kappaRP}].
As pointed out in Ref.~\cite{Bodeker:2017cim}, soft particle emission by particles passing through the bubble walls (so-called transition radiation or splitting processes) leads to an additional pressure acting on the bubble walls in proportion to the wall Lorentz factor $\Delta P_{\rm split} \propto \gamma^n$.
While the next-to-leading-order calculation in Ref.~\cite{Bodeker:2017cim} found a scaling exponent $n=1$, the all-orders resummation in Ref.~\cite{Hoeche:2020rsg} recently demonstrated that, in reality, $n=2$.%
\footnote{Related analyses recently appeared in Refs.~\cite{Vanvlasselaer:2020niz,Balaji:2020yrx}.}
Based on this result, Ref.~\cite{Ellis:2020nnr} obtained the following updated expression for the efficiency factor $\kappa_{\rm b}$,
\begin{align}
\label{eq:kappab}
\kappa_{\rm b} = \left(1-\frac{\alpha_\infty}{\alpha}\right) \times
\begin{cases}
1-1/3\left(\tilde{\gamma}_*/\gamma_{\rm eq}\right)^2  & ; \quad \tilde{\gamma}_* < \gamma_{\rm eq} \\
2/3\,\gamma_{\rm eq}/\tilde{\gamma}_*  & ; \quad \tilde{\gamma}_* > \gamma_{\rm eq} 
\end{cases} \,,
\end{align}
where $\gamma_{\rm eq}$ denotes the Lorentz factor that is reached when all forces acting on the bubble walls have equilibrated, so that the walls no longer accelerate, and $\tilde{\gamma}_*$ is the Lorentz factor that the bubble walls would reach by the time of collision if there was no friction whatsoever,
\begin{align}
\gamma_{\rm eq} = \left(\frac{\Delta V - \Delta P}{P_{\rm split}/\gamma^2}\right)^{1/2} \,,\quad \tilde{\gamma}_* = \frac{2\,R_*}{3\,R_0} \,.
\end{align}
Here, $\Delta V$ is the change in the effective potential across the phase transition, $\Delta P$ stands for the leading-order friction term, and $R_0$ and $R_*$ denote the average bubble radius at the time of nucleation and collision, respectively.
In many models, one finds that\,---\,unless the phase transition is strongly supercooled or transition radiation significantly suppressed (\textit{e.g.}, because the symmetry-breaking field does not couple to gauge bosons)\,---\,$\gamma_{\rm eq}$ is typically much smaller than $\tilde{\gamma}_*$.
This means that many phase transitions that were originally believed to be of the RP type are actually NP phase transitions, where the bubble walls reach a terminal velocity, $\gamma_* = \min\left\{\tilde{\gamma}_*,\gamma_{\rm eq}\right\} = \gamma_{\rm eq}$, and the efficiency factor $\kappa_{\rm b}$ is strongly suppressed, $\kappa_{\rm b} \ll 1$.


The improved understanding of $\kappa_{\rm b}$ affects benchmark points \#08 to \#15, \#18, and \#19.
In principle, it would be desirable to reconsider all of these points, making use of the updated expression in Eq.~\eqref{eq:kappab}.
However, this is complicated by the fact that $\tilde{\gamma}_*$ depends on the initial bubble radius $R_0$, which requires knowledge of the initial profile and Euclidean action of the symmetry-breaking scalar field at the time of nucleation~\cite{Ellis:2019oqb}. 
In the companion paper~\cite{Alanne:2019bsm}, we perform such an analysis for the xSM, which confirms that $\kappa_{\rm b} \ll 1$ across the entire parameter space.
Therefore, instead of explicitly computing $R_0$ and reevaluating $\kappa_{\rm b}$ for points \#08 to \#15, \#18, and \#19, we simply conclude that all RP phase transitions in Tab.~\ref{tab:bp} should either be ignored or replaced by an equivalent NP phase transition.


A second difference between Refs.~\cite{Caprini:2015zlo,Caprini:2019egz} is that Ref.~\cite{Caprini:2019egz} accounts for the formation of shocks at some time $\tau_{\rm sh}$ after the phase transition.
If this happens within a Hubble time, $H_*\tau_{\rm sh} < 1$, the GW signal from sound waves picks up an extra suppression factor~\cite{Ellis:2018mja,Ellis:2020awk},%
\footnote{For a generalization of this factor that also accounts for the expansion of the background, see Ref.~\cite{Guo:2020grp}.}
\begin{align}
\label{eq:taush}
\Omega_{\rm s}^{\rm peak} \rightarrow \min\left\{1,H_*\tau_{\rm sh}\right\} \times \Omega_{\rm s}^{\rm peak} \,,
\end{align}
where $H_*\tau_{\rm sh}$ can be computed in terms of the enthalpy-weighted root-mean-square of the plasma velocity, $\bar{U}_f$, which follows from the kinetic-energy fraction of the bulk plasma, $K$, 
\begin{align}
H_*\tau_{\rm sh} = \left(8\pi\right)^{1/3} \frac{v_w}{\beta/H_*}\frac{1}{\bar{U}_f} \,,\quad \bar{U}_f = \left(\frac{3}{4}\,K\right)^{1/2} \,,\quad K = \frac{\kappa_{\rm s}\,\alpha}{1+\alpha} \,.
\end{align}


\begin{figure}
\begin{center}

\includegraphics[width=0.45\textwidth]{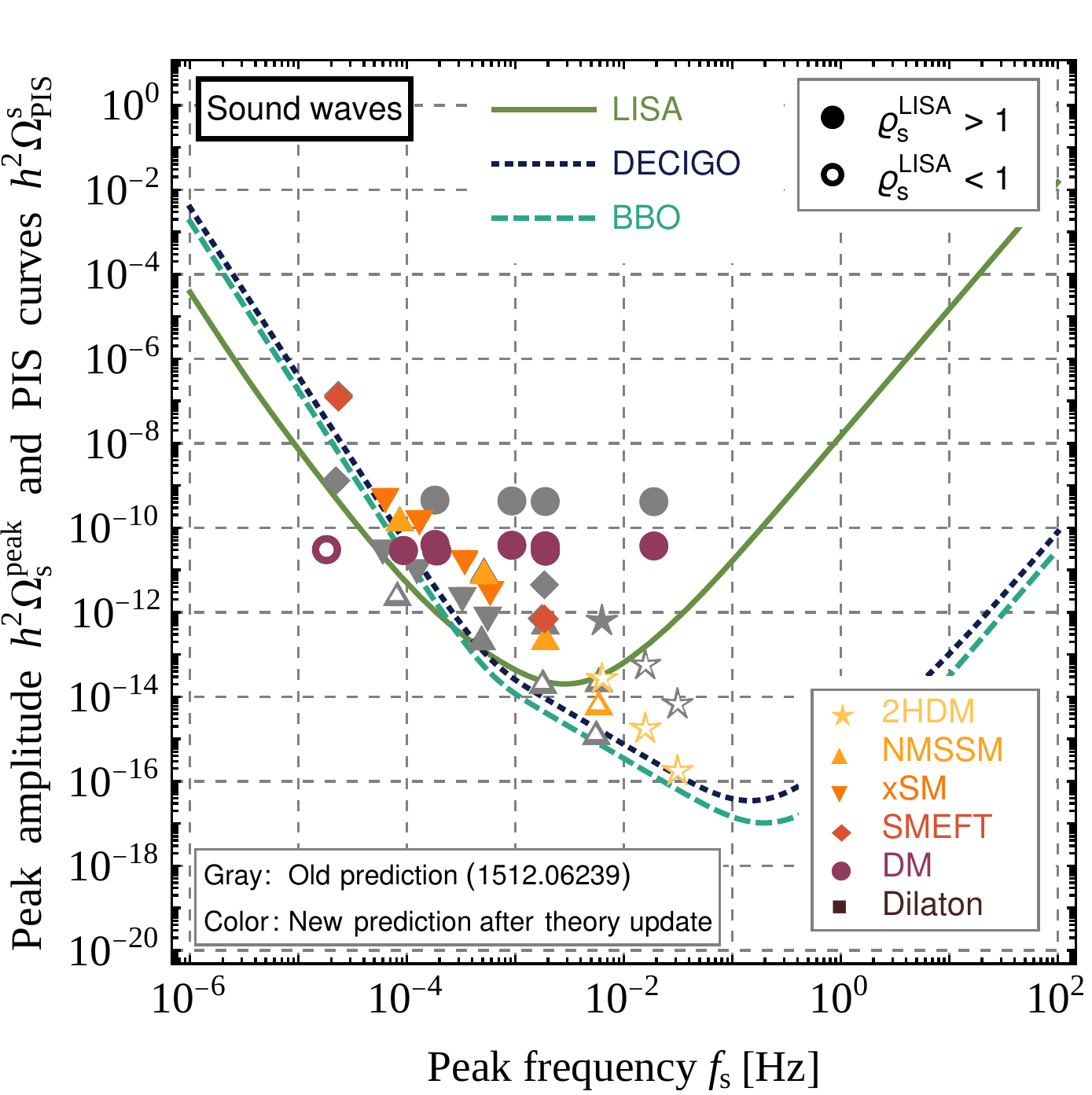}
\caption{Update of the benchmark points in Ref.~\cite{Caprini:2015zlo} according to Refs.~\cite{Ellis:2018mja,Ellis:2020nnr,Hoeche:2020rsg}.
See text.}
\label{fig:update}
\end{center}
\end{figure}


In Fig.~\ref{fig:update}, we summarize how the different treatment of $\kappa_{\rm b}$ and $\Omega_{\rm s}^{\rm peak}$ in Ref.~\cite{Caprini:2019egz} compared to Ref.~\cite{Caprini:2015zlo} affects our PISC plot in the s-channel.
In this figure, we no longer show the predictions of the RP phase transitions in the NMSSM and SMEFT; we replace the RP phase transitions in the xSM by equivalent NP phase transitions; and we rescale all peak amplitudes by the suppression factor in Eq.~\eqref{eq:taush}.
These three steps remove and shift some of our benchmark points.
However, for the purposes of this paper, the main message of Fig.~\ref{fig:update} is that the three PISCs simply remain the same as in the middle left panel of Fig.~\ref{fig:pis}, despite the comprehensive theory update.
Of course, an update of the spectral shape functions would require a revision of the sensitivity curves.
A more extensive version of Fig.~\ref{fig:update}, including almost 4000 benchmark points for ten different particle physics models, can be found in Ref.~\cite{Schmitz:2020rag}.
Both plots will continue to serve as useful resources in the future when new theoretical predictions for $f_{\rm s}$ and $\Omega_{\rm s}^{\rm peak}$ should become available (see, \textit{e.g.}, Refs.~\cite{Giese:2020rtr,Giese:2020znk}).


\subsection{Semianalytical fit functions}
\label{subsec:fits}

\begin{figure}
\begin{center}

\includegraphics[width=0.32\textwidth]{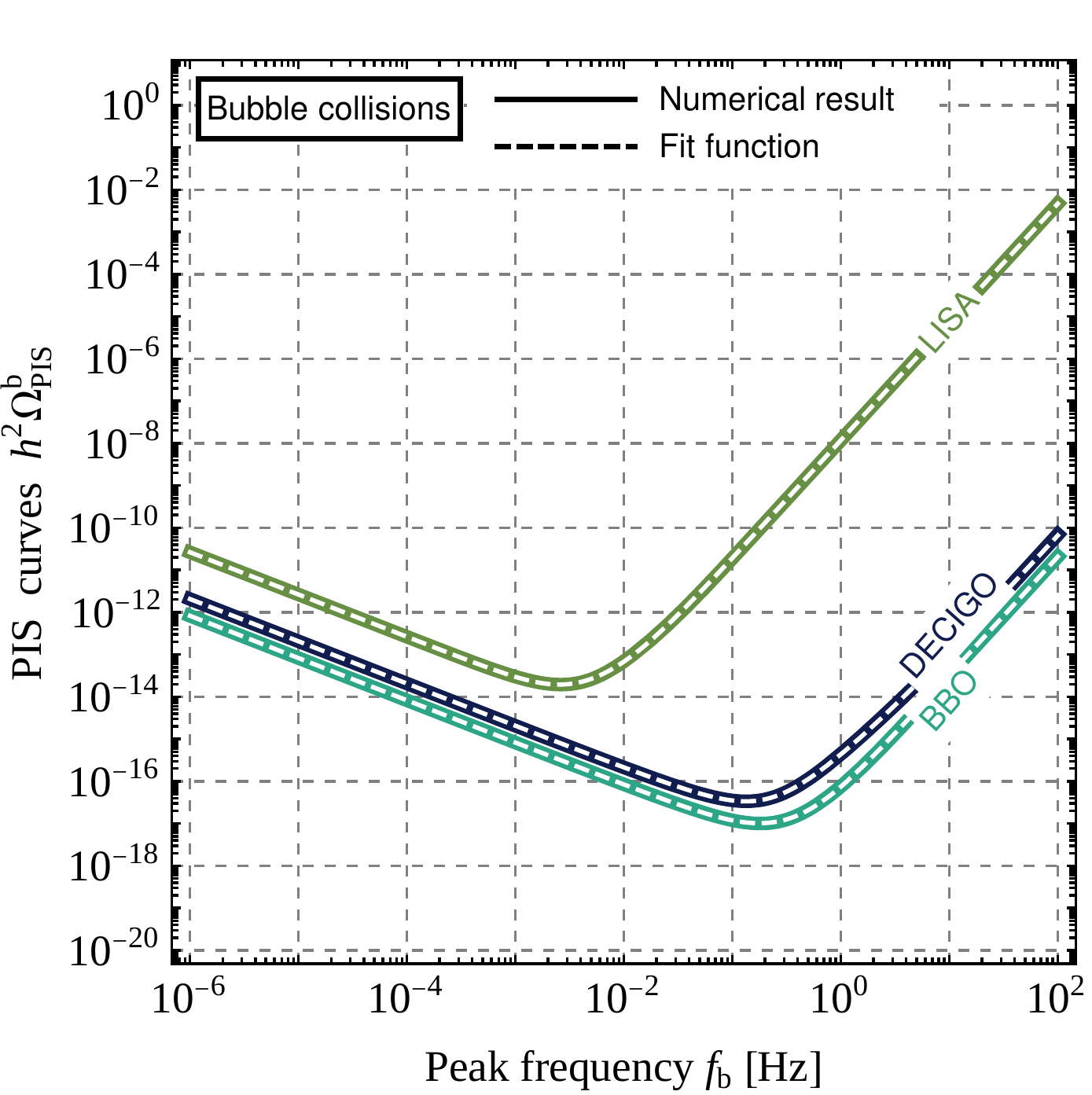}\,
\includegraphics[width=0.32\textwidth]{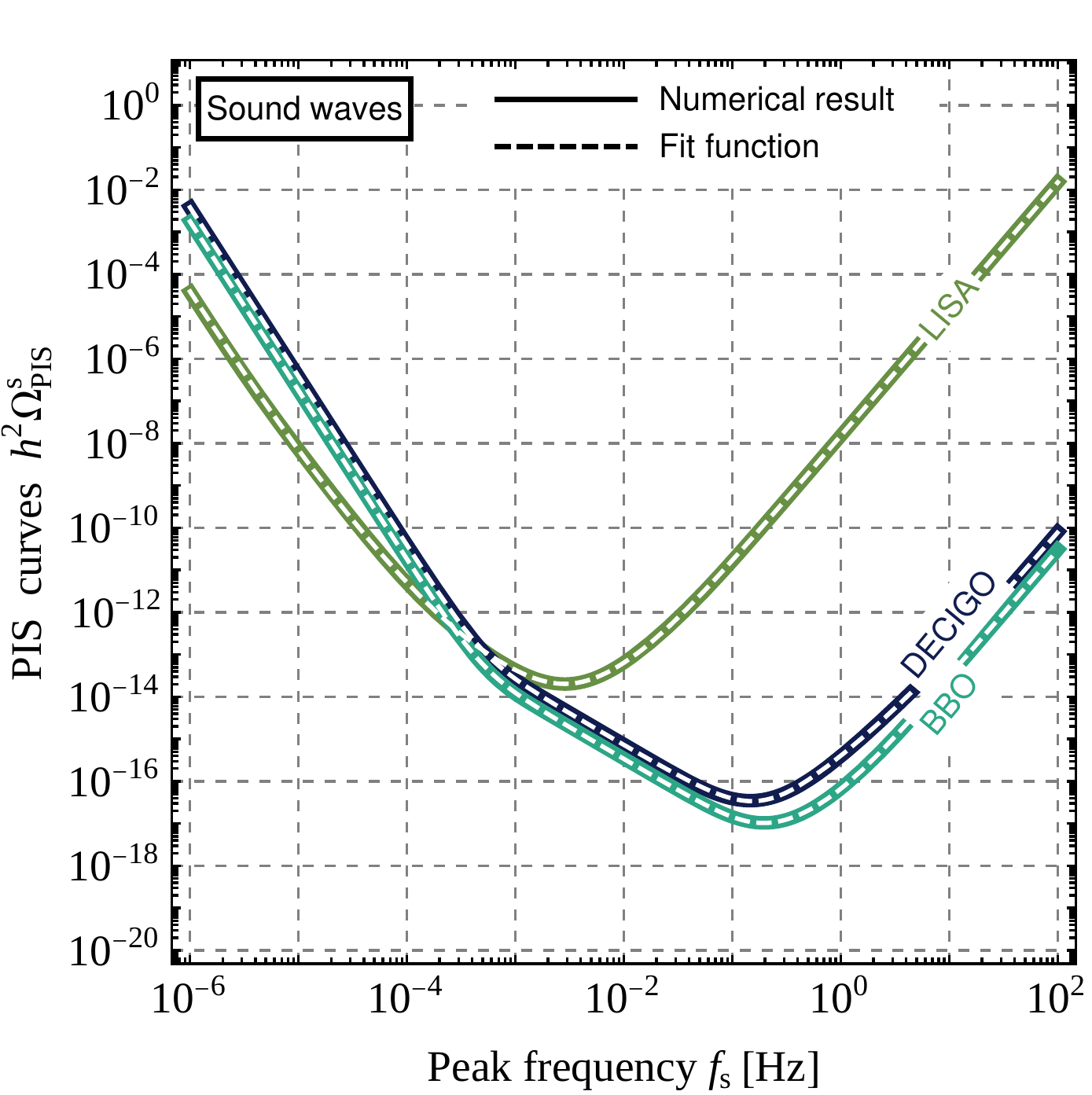}\,
\includegraphics[width=0.32\textwidth]{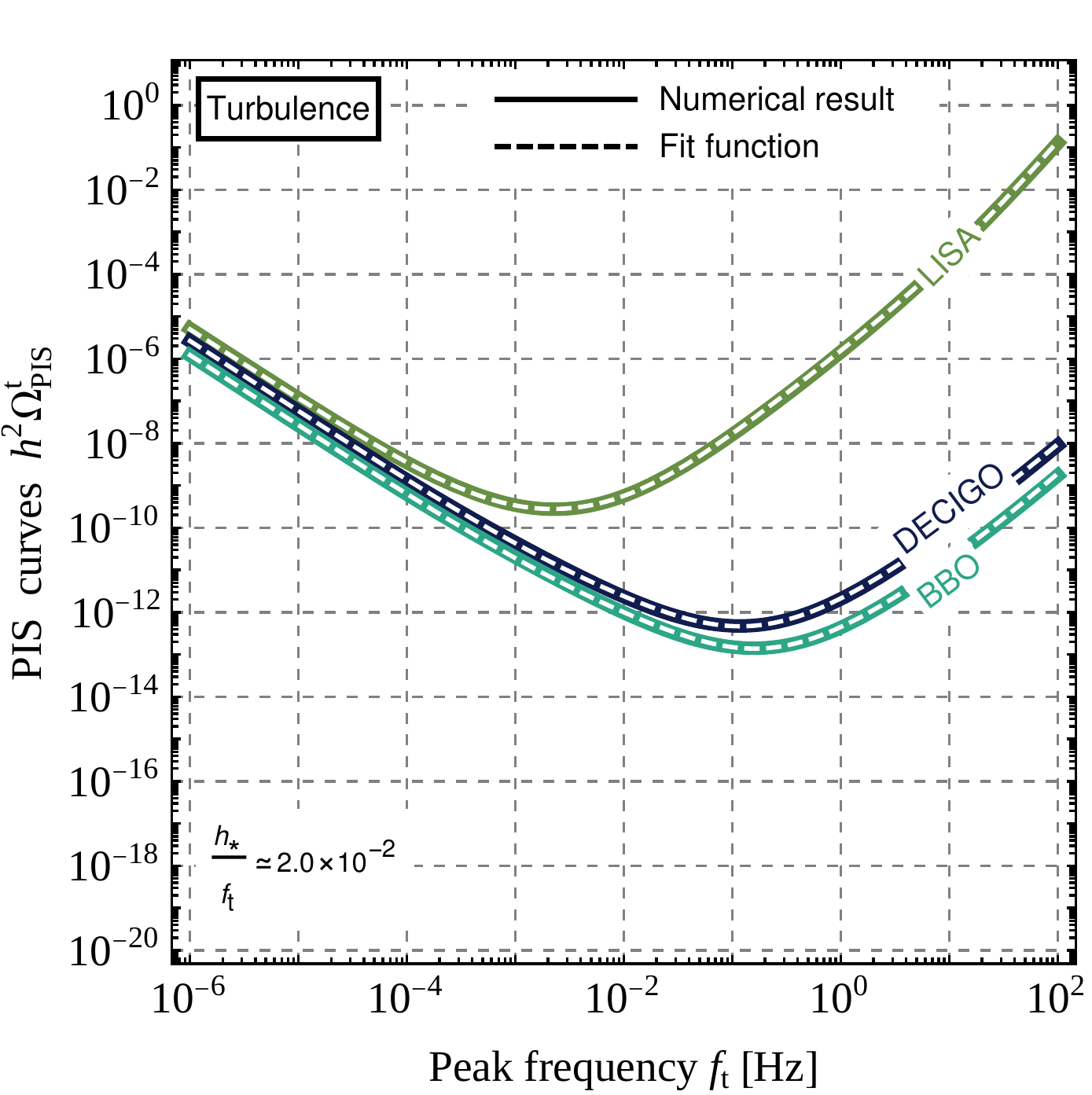}
\caption{Numerical results and fit functions for the PISCs in the b-, s-, and t-channel.}
\label{fig:fit}
\end{center}
\end{figure}

In the previous sections, we studied 18 different sensitivities:
We considered three different experiments (LISA, DECIGO, and BBO), and for each experiment, we constructed PISCs in six different channels (b, s, t, b/s, b/t, s/t). 
We shall now conclude our analysis by providing semianalytical fit functions for all of these 18 sensitivities.
In doing so, let us assume that all sensitivities can be reasonably well approximated by power series of the following form,
\begin{align}
\Omega_{\rm PIS}^{\rm i/j}\left(f_{\rm b},f_{\rm s},f_{\rm t},h_*\right) \simeq \sum_{a,b,c,d} c_{\left(a,b,c,d\right)}\,f_{\rm b}^a\,f_{\rm s}^b\,f_{\rm t}^c\,h_*^d \,,
\end{align}
for an appropriate set of 4-tuples $\left(a,b,c,d\right) \in \mathbb{R}^4$.
Based on this ansatz, we are then able to fit our numerical data and determine the coefficients $c_{\left(a,b,c,d\right)}$ for each of our 18 sensitivities.
Below, we present our results for LISA, DECIGO, and BBO using the following notation,
\begin{align}
x_{\rm b}   = \frac{f_{\rm b}}{1\,\textrm{mHz}} \,, \:\:\:\:
x_{\rm s}   = \frac{f_{\rm s}}{1\,\textrm{mHz}} \,, \:\:\:\:
x_{\rm t}   = \frac{f_{\rm t}}{1\,\textrm{mHz}} \,, \:\:\:\:
x_{\rm b/s} = \frac{f_{\rm b}}{f_{\rm s}}       \,, \:\:\:\:
x_{\rm b/t} = \frac{f_{\rm b}}{f_{\rm t}}       \,, \:\:\:\:
x_{\rm t/h} = \frac{f_{\rm t}}{h_*}             \,.
\end{align}
Our fit functions are constructed such that they reproduce our numerical results to high precision across the entire range of relevant frequencies.
This is, \textit{e.g.}, illustrated in Fig.~\ref{fig:fit}, where we compare our numerical results and fit functions in the b-, s-, and t-channels  to each other.
In the other three channels, our fit functions are of an equally high quality.


\subsubsection*{LISA}


\begin{small}
\begin{align}
\label{eq:LISAb}
\frac{h^2\Omega_{\rm PIS}^{\rm b}}{10^{-14}} & \simeq
2.63 \times 10^0    \,x_{\rm b}^{-1}  +
3.26 \times 10^{-1} \,x_{\rm b}^1     + 
3.29 \times 10^{-3} \,x_{\rm b}^2     +
4.67 \times 10^{-3} \,x_{\rm b}^{2.8} \,,
\\
\label{eq:LISAs}
\frac{h^2\Omega_{\rm PIS}^{\rm s}}{10^{-14}} & \simeq
3.58 \times 10^{-3} \,x_{\rm s}^{-4} +
3.26 \times 10^{-1} \,x_{\rm s}^{-3} +
1.20 \times 10^0    \,x_{\rm s}^{-2} +
2.48 \times 10^0    \,x_{\rm s}^{-1} \\\nonumber & +
2.85 \times 10^{-1} \,x_{\rm s}^1    + 
1.81 \times 10^{-2} \,x_{\rm s}^2    + 
1.50 \times 10^{-3} \,x_{\rm s}^3    \,,
\\
\label{eq:LISAt}
\frac{h^2\Omega_{\rm PIS}^{\rm t}}{10^{-12}} & \simeq
1.07 \times 10^0    \,x_{\rm t/h}^{0.98}\,x_{\rm t}^{-5/3} +
1.96 \times 10^0    \,x_{\rm t/h}^{1.04}\,x_{\rm t}^{-1}   +
3.50 \times 10^0    \,x_{\rm t/h}^{0.96}\,x_{\rm t}^0      \\\nonumber & +
4.77 \times 10^{-1} \,x_{\rm t/h}^{1.03}\,x_{\rm t}^1      + 
3.32 \times 10^{-2} \,x_{\rm t/h}^{0.96}\,x_{\rm t}^2      + 
1.05 \times 10^{-4} \,x_{\rm t/h}^{0.05}\,x_{\rm t}^3      \,,
\\ 
\frac{h^2\Omega_{\rm PIS}^{\rm b/s}}{10^{-14}} & \simeq
3.79 \times 10^{-1} \,x_{\rm b/s}^{-0.48}\,x_{\rm s}^{-2.5} +
1.27 \times 10^{-1} \,x_{\rm b/s}^{-0.62}\,x_{\rm s}^{-2}   +
1.82 \times 10^0    \,x_{\rm b/s}^{-0.48}\,x_{\rm s}^{-1}   \\\nonumber & +
1.39 \times 10^{-1} \,x_{\rm b/s}^{-0.51}\,x_{\rm s}^0      + 
1.02 \times 10^{-1} \,x_{\rm b/s}^{-0.51}\,x_{\rm s}^1      + 
3.77 \times 10^{-3} \,x_{\rm b/s}^{ 0.58}\,x_{\rm s}^2      \\\nonumber & +
1.87 \times 10^{-3} \,x_{\rm b/s}^{ 1.40}\,x_{\rm s}^{2.9}  \,,
\\ 
\frac{h^2\Omega_{\rm PIS}^{\rm b/t}}{10^{-13}} & \simeq
1.18 \times 10^0    \,x_{\rm b/t}^{-0.57}\,x_{\rm t/h}^{0.45}\,x_{\rm t}^{-4/3} +
5.27 \times 10^{-1} \,x_{\rm b/t}^{-0.27}\,x_{\rm t/h}^{0.62}\,x_{\rm t}^{-1}   \\\nonumber & +
1.32 \times 10^0    \,x_{\rm b/t}^{-0.59}\,x_{\rm t/h}^{0.46}\,x_{\rm t}^0      + 
4.26 \times 10^{-2} \,x_{\rm b/t}^{ 1.20}\,x_{\rm t/h}^{0.55}\,x_{\rm t}^2      + 
3.68 \times 10^{-4} \,x_{\rm b/t}^{ 1.47}\,x_{\rm t/h}^{0.23}\,x_{\rm t}^{2.9}  \,,
\\ 
\frac{h^2\Omega_{\rm PIS}^{\rm s/t}}{10^{-13}} & \simeq
3.19 \times 10^{-1} \,x_{\rm t/h}^{ 0.50}\,x_{\rm t}^{-17/6} +
2.06 \times 10^0    \,x_{\rm t/h}^{ 0.50}\,x_{\rm t}^{-2}    +
3.25 \times 10^0    \,x_{\rm t/h}^{ 0.51}\,x_{\rm t}^{-1}    \\\nonumber & +
7.77 \times 10^{-1} \,x_{\rm t/h}^{ 0.39}\,x_{\rm t}^0       + 
1.61 \times 10^{-1} \,x_{\rm t/h}^{ 0.56}\,x_{\rm t}^1       + 
1.74 \times 10^{-2} \,x_{\rm t/h}^{ 0.34}\,x_{\rm t}^2       \\\nonumber & +
9.58 \times 10^{-4} \,x_{\rm t/h}^{ 0.63}\,x_{\rm t}^{2.5}   +
1.76 \times 10^{-4} \,x_{\rm t/h}^{-0.13}\,x_{\rm t}^3       \,.
\end{align}
\end{small}


\subsubsection*{DECIGO}


\begin{small}
\begin{align}
\frac{h^2\Omega_{\rm PIS}^{\rm b}}{10^{-15}} & \simeq
2.06 \times 10^0     \,x_{\rm b}^{-1}  +
7.61 \times 10^{-3}  \,x_{\rm b}^0     + 
4.58 \times 10^{-5}  \,x_{\rm b}^1     + 
2.19 \times 10^{-7}  \,x_{\rm b}^2     \\\nonumber & +
6.35 \times 10^{-10} \,x_{\rm b}^{2.8} \,,
\\ 
\frac{h^2\Omega_{\rm PIS}^{\rm s}}{10^{-14}} & \simeq
3.82 \times 10^{-1}  \,x_{\rm s}^{-4}   +
2.26 \times 10^0     \,x_{\rm s}^{-1.5} +
1.10 \times 10^{-3}  \,x_{\rm s}^0      +
2.56 \times 10^{-6}  \,x_{\rm s}^1      \\\nonumber & +
2.91 \times 10^{-8}  \,x_{\rm s}^2      + 
7.54 \times 10^{-12} \,x_{\rm s}^3      \,,
\\ 
\frac{h^2\Omega_{\rm PIS}^{\rm t}}{10^{-13}} & \simeq
5.31 \times 10^0     \,x_{\rm t/h}^{1.00}\,x_{\rm t}^{-5/3} +
3.34 \times 10^0     \,x_{\rm t/h}^{1.00}\,x_{\rm t}^{-1}   +
3.75 \times 10^{-2}  \,x_{\rm t/h}^{0.99}\,x_{\rm t}^0      \\\nonumber & +
2.16 \times 10^{-4}  \,x_{\rm t/h}^{1.01}\,x_{\rm t}^1      + 
1.92 \times 10^{-7}  \,x_{\rm t/h}^{0.96}\,x_{\rm t}^2      + 
9.33 \times 10^{-12} \,x_{\rm t/h}^{0.07}\,x_{\rm t}^3      \,,
\\ 
\frac{h^2\Omega_{\rm PIS}^{\rm b/s}}{10^{-15}} & \simeq
6.27 \times 10^0     \,x_{\rm b/s}^{-0.49}\,x_{\rm s}^{-2.5} +
6.17 \times 10^0     \,x_{\rm b/s}^{-0.53}\,x_{\rm s}^{-1.5} +
1.29 \times 10^0     \,x_{\rm b/s}^{-0.44}\,x_{\rm s}^{-1}   \\\nonumber & +
2.07 \times 10^{-3}  \,x_{\rm b/s}^{-0.66}\,x_{\rm s}^0      + 
2.18 \times 10^{-5}  \,x_{\rm b/s}^{-0.63}\,x_{\rm s}^1      + 
1.76 \times 10^{-7}  \,x_{\rm b/s}^{ 0.85}\,x_{\rm s}^2      \\\nonumber & +
1.23 \times 10^{-10} \,x_{\rm b/s}^{ 1.24}\,x_{\rm s}^{2.9}  \,,
\\ 
\frac{h^2\Omega_{\rm PIS}^{\rm b/t}}{10^{-14}} & \simeq
2.75 \times 10^0     \,x_{\rm b/t}^{-0.51}\,x_{\rm t/h}^{0.49}\,x_{\rm t}^{-4/3} +
6.57 \times 10^{-1}  \,x_{\rm b/t}^{-0.46}\,x_{\rm t/h}^{0.53}\,x_{\rm t}^{-1}   \\\nonumber & +
2.45 \times 10^{-2}  \,x_{\rm b/t}^{-0.52}\,x_{\rm t/h}^{0.49}\,x_{\rm t}^0      + 
2.65 \times 10^{-7}  \,x_{\rm b/t}^{ 1.10}\,x_{\rm t/h}^{0.49}\,x_{\rm t}^2      + 
6.04 \times 10^{-11} \,x_{\rm b/t}^{ 1.84}\,x_{\rm t/h}^{0.38}\,x_{\rm t}^{2.9}  \,,
\\ 
\frac{h^2\Omega_{\rm PIS}^{\rm s/t}}{10^{-13}} & \simeq
1.01 \times 10^0     \,x_{\rm t/h}^{0.50}\,x_{\rm t}^{-17/6} +
2.56 \times 10^0     \,x_{\rm t/h}^{0.50}\,x_{\rm t}^{-1.5}  +
1.55 \times 10^{-1}  \,x_{\rm t/h}^{0.49}\,x_{\rm t}^{-1}    \\\nonumber & +
7.02 \times 10^{-4}  \,x_{\rm t/h}^{0.50}\,x_{\rm t}^0       + 
8.09 \times 10^{-6}  \,x_{\rm t/h}^{0.50}\,x_{\rm t}^1       + 
6.91 \times 10^{-9}  \,x_{\rm t/h}^{0.49}\,x_{\rm t}^2       \\\nonumber & +
1.32 \times 10^{-10} \,x_{\rm t/h}^{0.51}\,x_{\rm t}^{2.5}   +
3.56 \times 10^{-13} \,x_{\rm t/h}^{0.13}\,x_{\rm t}^3       \,.
\end{align}
\end{small}


\subsubsection*{BBO}


\begin{small}
\begin{align}
\frac{h^2\Omega_{\rm PIS}^{\rm b}}{10^{-16}} & \simeq
8.24 \times 10^0    \,x_{\rm b}^{-1}  +
1.61 \times 10^{-2} \,x_{\rm b}^0     + 
1.98 \times 10^{-4} \,x_{\rm b}^1     + 
4.24 \times 10^{-9} \,x_{\rm b}^2     \\\nonumber & +
2.06 \times 10^{-9} \,x_{\rm b}^{2.8} \,,
\\ 
\frac{h^2\Omega_{\rm PIS}^{\rm s}}{10^{-14}} & \simeq
1.77 \times 10^{-1}  \,x_{\rm s}^{-4}   +
1.06 \times 10^0     \,x_{\rm s}^{-1.5} +
1.35 \times 10^{-4}  \,x_{\rm s}^0      +
2.23 \times 10^{-6}  \,x_{\rm s}^1      \\\nonumber & +
1.29 \times 10^{-9}  \,x_{\rm s}^2      + 
2.99 \times 10^{-12} \,x_{\rm s}^3      \,,
\\ 
\frac{h^2\Omega_{\rm PIS}^{\rm t}}{10^{-13}} & \simeq
2.44 \times 10^0     \,x_{\rm t/h}^{1.00}\,x_{\rm t}^{-5/3} +
1.41 \times 10^0     \,x_{\rm t/h}^{1.00}\,x_{\rm t}^{-1}   +
1.00 \times 10^{-2}  \,x_{\rm t/h}^{1.00}\,x_{\rm t}^0      \\\nonumber & +
5.06 \times 10^{-5}  \,x_{\rm t/h}^{1.00}\,x_{\rm t}^1      + 
3.19 \times 10^{-8}  \,x_{\rm t/h}^{0.98}\,x_{\rm t}^2      + 
1.37 \times 10^{-12} \,x_{\rm t/h}^{0.31}\,x_{\rm t}^3      \,,
\\ 
\frac{h^2\Omega_{\rm PIS}^{\rm b/s}}{10^{-15}} & \simeq
2.94 \times 10^0     \,x_{\rm b/s}^{-0.49}\,x_{\rm s}^{-2.5} +
3.07 \times 10^0     \,x_{\rm b/s}^{-0.51}\,x_{\rm s}^{-1.5} +
4.98 \times 10^{-1}  \,x_{\rm b/s}^{-0.47}\,x_{\rm s}^{-1}   \\\nonumber & +
2.62 \times 10^{-4}  \,x_{\rm b/s}^{-0.64}\,x_{\rm s}^0      + 
8.53 \times 10^{-6}  \,x_{\rm b/s}^{-0.50}\,x_{\rm s}^1      + 
6.33 \times 10^{-11} \,x_{\rm b/s}^{ 1.44}\,x_{\rm s}^{2.9}  \,,
\\ 
\frac{h^2\Omega_{\rm PIS}^{\rm b/t}}{10^{-14}} & \simeq
1.20 \times 10^0     \,x_{\rm b/t}^{-0.51}\,x_{\rm t/h}^{0.49}\,x_{\rm t}^{-4/3} +
2.72 \times 10^{-1}  \,x_{\rm b/t}^{-0.45}\,x_{\rm t/h}^{0.53}\,x_{\rm t}^{-1}   \\\nonumber & +
7.18 \times 10^{-3}  \,x_{\rm b/t}^{-0.53}\,x_{\rm t/h}^{0.49}\,x_{\rm t}^0      + 
4.28 \times 10^{-8}  \,x_{\rm b/t}^{ 1.12}\,x_{\rm t/h}^{0.50}\,x_{\rm t}^2      + 
1.39 \times 10^{-11} \,x_{\rm b/t}^{ 1.73}\,x_{\rm t/h}^{0.38}\,x_{\rm t}^{2.9}  \,,
\\ 
\frac{h^2\Omega_{\rm PIS}^{\rm s/t}}{10^{-13}} & \simeq
4.67 \times 10^{-1}  \,x_{\rm t/h}^{ 0.50}\,x_{\rm t}^{-17/6} +
1.21 \times 10^0     \,x_{\rm t/h}^{ 0.50}\,x_{\rm t}^{-1.5}  +
6.27 \times 10^{-2}  \,x_{\rm t/h}^{ 0.49}\,x_{\rm t}^{-1}    \\\nonumber & +
1.53 \times 10^{-4}  \,x_{\rm t/h}^{ 0.50}\,x_{\rm t}^0       + 
2.07 \times 10^{-6}  \,x_{\rm t/h}^{ 0.50}\,x_{\rm t}^1       + 
5.91 \times 10^{-10} \,x_{\rm t/h}^{ 0.40}\,x_{\rm t}^2       \\\nonumber & +
2.71 \times 10^{-11} \,x_{\rm t/h}^{ 0.55}\,x_{\rm t}^{2.5}   +
2.90 \times 10^{-13} \,x_{\rm t/h}^{-0.29}\,x_{\rm t}^3       \,.
\end{align}
\end{small}


The above fit functions allow us to write down a quasianalytic expression for the SNR,
\begin{align}
\label{eq:solution}
\varrho = \left[\frac{t_{\rm obs}}{1\,\textrm{yr}}\sum_{\rm i/j}\left(\frac{\Omega_{\rm i/j}^{\rm peak}}{\Omega_{\rm PIS}^{\rm i/j}\left(f_{\rm b},f_{\rm s},f_{\rm t},h_*\right)}\right)^2\right]^{1/2}
\end{align}
for LISA, DECIGO, and BBO, respectively.
In this sense, the concept of PISCs in combination with the fit functions presented in this section amount to a quasianalytic solution to the problem of computing the SNR for the GW signal from a cosmological phase transition.
As evident from Eq.~\eqref{eq:solution}, this analytic solution only depends on the SFOPT parameters $\alpha$, $\beta/H_*$, $T_*$, $v_w$, $\kappa_{\rm b}$, $\kappa_{\rm s}$, and $\kappa_{\rm t}$; the frequency dependence of the signal as well as the experimental noise spectra are already take care of by our fit functions.
Our result in Eq.~\eqref{eq:solution} therefore renders any further numerical step (\textit{i.e.}, integration) in the computation of $\varrho$ obsolete.
With the results in this section at hand, the SNR can be evaluated analytically.
This is in particular also true if one is only interested in the signal from a single source.
Suppose, \textit{e.g.}, we were only interested in LISA's sensitivity to the signal from sound waves.
In this case, the full information on the expected SNR will be contained in the following expression,
\begin{align}
\label{eq:rhoLISAs}
\varrho_{\rm s}^{\rm LISA} = \left(\frac{t_{\rm obs}}{1\,\textrm{yr}}\right)^{1/2}\frac{\Omega_{\rm s}^{\rm peak}}{
c_{-4}\,x_{\rm s}^{-4} +
c_{-3}\,x_{\rm s}^{-3} +
c_{-2}\,x_{\rm s}^{-2} +
c_{-1}\,x_{\rm s}^{-1} +
c_1\,x_{\rm s}^1    + 
c_2\,x_{\rm s}^2    + 
c_3\,x_{\rm s}^3} \,,
\end{align}
where the numerical values of the coefficients $c_{-4}$, $c_{-3}$, etc.\ can be read off from Eq.~\eqref{eq:LISAs}.
This is an important result of our analysis.
We stress that it is independent of the explicit form of $\Omega_{\rm s}^{\rm peak}$, such that it can be compared to the results in both Ref.~\cite{Caprini:2015zlo} and Ref.~\cite{Caprini:2019egz}.


\section{Conclusions and outlook}
\label{sec:conclusions}


For a DM direct-detection experiment, it is typically straightforward to answer the following two questions:
(i) \textit{``What is the experiment's sensitivity to the DM cross section $\mathit{\sigma_{\textrm{\textit{DM}}}}$ as a function of the DM mass $\mathit{m_{\textrm{\textit{DM}}}}$?''}
(ii) \textit{``To what extend will the experiment explore the parameter regions preferred by specific DM scenarios?''}
This situation in the field of DM experiments needs to be compared to searches for GWs from a SFOPT in the early Universe.
In this case, one can likewise ask:
(i) \textit{``What is an experiment's sensitivity to the GW signal $\mathit{\Omega_{\textrm{\textit{SFOPT}}}}$ as a function of the GW frequency $\mathit{f}$?''}
(ii) \textit{``To what extend will it explore the parameter regions preferred by specific SFOPT scenarios?''}
In contrast to DM searches, there are at present no commonly accepted answers to these questions in the GW community.
Existing approaches, such as PLISC and SNR plots, certainly do convey a useful impression of the sensitivities of current and future experiments, but still suffer from a number of shortcomings.
Graphical analyses based on PLISCs no longer contain information on the expected SNR and are only meaningful as long as the expected signal does not deviate too much from a pure power law.
SNR plots, on the other hand, present the reach of GW experiments in terms of auxiliary parameters instead of observable quantities, are often times restricted to two-dimensional slices through the higher-dimensional parameter space, and require additional elements such as a color code or contour lines to indicate the expected SNR.
In this sense, neither of these approaches is truly on par with the sensitivity curves for DM experiments.
This observation is the basis for our PISC proposal.
Suppose, \textit{e.g.}, one asked: \textit{``What is LISA's projected sensitivity to GWs from sound waves?''}
We argue that the best possible answer to this question would be LISA's $\Omega_{\rm PIS}^{\rm s}$ curve as a function of $f_{\rm s}$ [see Fig.~\ref{fig:update} as well as Eqs.~\eqref{eq:LISAs} and \eqref{eq:rhoLISAs}].


The PISCs constructed in this paper exhibit twelve characteristic features (see also Sec.~6 of Ref.~\cite{Alanne:2019bsm}).
They
(i) retain the full information on the SNR, encoding it on the $y$-axis of the PISC plots;
(ii) do not require extra graphical elements such as a color code or contour lines to indicate the expected SNR;
(iii) do not require one to slice the parameter space into hypersurfaces;
(iv) only depend on the experimental noise spectra and spectral shape functions in Eq.~\eqref{eq:S}, which renders them insensitive to the theoretical uncertainties in calculating the peak amplitudes in Eq.~\eqref{eq:Opeak};
(v) can be generalized to arbitrary shape functions $\mathcal{S}$;
(vi) indicate sensitivities in terms of observables that will play an important role in the experimental data analysis rather than auxiliary quantities;
(vii) directly illustrate how GW experiments will approach from above and cut into the signal regions of specific models;
(viii) set the stage for the systematic analysis of underlying model-parameter dependences (see the example study in Ref.~\cite{Alanne:2019bsm});
(ix) allow for an easy comparison of the projected sensitivities of different experiments;
(x) allow one to combine and compare different signal contributions at one's convenience; and
(xi) close the gap between experiment and theory by eliminating the model-independent and redundant frequency integration in Eq.~\eqref{eq:rho}.
It is also easy to (xii) approximate our PISCs by fit functions, such that the SNR can be written as a function of the SFOPT parameters only.
In this sense, our PISC method provides a quasianalytical solution to the problem of computing the SNR for the GW signal from a SFOPT.
At the same time, we stress the important caveat that the PISCs presented in this paper are always only as good as our knowledge of the spectral shape functions in Eq.~\eqref{eq:S}.
Each update of these functions will require an update of our sensitivity curves.
This opens up the possibility to use PISC plots as a bookkeeping tool to keep track of the theoretical progress in the field.
Similarly, the sensitivity \textit{curves} constructed in this paper can be generalized to sensitivity \textit{bands} indicating the theoretical uncertainty in the expected shape of the GW spectrum~\cite{Schmitz:2020rag}.


There are several natural directions in which our analysis in this paper could be extended.
An obvious extension would, \textit{e.g.}, consist in applying our method to further experiments.
To facilitate such an analysis, we review the strain noise spectra of a number of interferometer and PTA experiments in Appendix~\ref{app:review}.
Beyond that, our method could also be extended to experiments that we do not consider in Appendix~\ref{app:review}, such as, \textit{e.g.}, AEDGE~\cite{Bertoldi:2019tck}, AIGSO~\cite{Gao:2017rgh,Wang:2019oeu}, AION~\cite{Badurina:2019hst}, AMIGO~\cite{Ni:2019nau}, Taiji~\cite{Hu:2017mde}, TianGO~\cite{Kuns:2019upi}, TianQin~\cite{Luo:2015ght,Hu:2018yqb}, etc.
Furthermore, it would be desirable to repeat our xSM analysis in the companion paper~\cite{Alanne:2019bsm} for as many BSM models as possible.
The ultimate goal of such an effort would be a comprehensive database providing the necessary means for a systematic and quantitative model comparison.
Similarly as in the case of DM experiments, such a database would allow one to construct the signal regions for various models and to illustrate, by means of our PISC plots, how these signal regions are going to be probed by future experiments.
The plots in Fig.~\ref{fig:bps} provide a glimpse of how such a simultaneous comparison of different models and different experiments could eventually look like.
However, to be able to make stronger and more quantitative statements, it will be necessary to consider a significantly larger number of benchmark points for each model.
In Ref.~\cite{Alanne:2019bsm}, \textit{e.g.}, we study roughly 6000 points, which is necessary to fully chart and trace out the signal region of the xSM in our PISC plots.
Finally, we point out that our PISC method could also be extended to any SGWB signal whose spectral shape is described by a clearly defined shape function $\mathcal{S}$.
In this case, one would be able to construct \textit{shape-integrated sensitivity curves} (SISCs) in analogy to our PISCs.
We leave a more detailed discussion of this possibility for future work.
Instead, we conclude by stressing that the novel concept of peak-integrated sensitivity curves bears the potential to develop into a new useful standard tool for model builders, phenomenologists, and experimentalists that are interested in the GW signal from a phase transition in the early Universe.


\section*{Acknowledgments}


I would like to thank Marco Peloso for comments and encouragement at the early stages of this project, in particular, for the suggestion to work out semianalytical fit functions for my numerical results.
In addition, I would like to thank Sachiko Kuroyanagi for sharing with me her numerical results on the DECIGO overlap reduction function and Hirotaka Yuzurihara for useful comments on the KAGRA sensitivity curve.
I am also grateful to Tommi Alanne, Moritz Platscher, and Thomas Hugle for the fruitful collaboration on the companion paper~\cite{Alanne:2019bsm}.
This project has received funding from the European Union's Horizon 2020 Research and Innovation Programme under grant agreement number 796961, ``AxiBAU''.


\appendix


\section{Review: Noise spectra of interferometer and pulsar timing experiments}
\label{app:review}


The construction of both power-law- and peak-integrated sensitivity curves requires knowledge of the experimental strain noise power spectra.
In this appendix, we shall therefore review the strain noise spectra of LISA, DECIGO, and BBO, and in addition, several other interferometer and PTA experiments.
Specifically, we are going to consider: aLIGO, aVirgo, KAGRA, CE, ET, LISA, DECIGO, BBO, NANOGrav, PPTA, EPTA, IPTA, and SKA.
As for the second-generation ground-based interferometers (aLIGO, aVirgo, and KAGRA), we will derive the effective strain noise spectra of three different detector networks: HL, HLV, and HLVK (see Sec.~\ref{subsec:example}).
Our analysis in this appendix is supposed to facilitate the generalization of our PISC method to experiments beyond LISA, DECIGO, and BBO.
In addition, we hope that it will serve as a useful resource for a broader range of applications.


In Sec.~\ref{subsec:formalism}, we will first introduce the necessary formalism and fix our conventions.
In Sec.~\ref{subsec:transfer}, we will then present all relevant transfer functions (\textit{i.e.}, signal response and overlap reduction functions), before turning to the individual detector noise spectra in Sec.~\ref{subsec:detector}.
In Sec.~\ref{subsec:strain}, we will finally put everything together and construct the strain noise spectra.
For an overview of all quantities playing a role in the following discussion, see Tab.~\ref{tab:quantities}.


\subsection{Formalism}
\label{subsec:formalism}


\begin{table}
\begin{center}
\renewcommand{\arraystretch}{1.22}
\caption{Overview of different quantities playing a role in this appendix.}
\label{tab:quantities}
\bigskip
\begin{tabular}{|l||l|}
\hline
Quantity & Definition \\
\hline\hline
$h_{ij}$                  & GW in position space (tensor perturbation of the spacetime metric) \\
$h_{\mathbf{n}}^p$        & GW mode in Fourier space describing a sinusoidal plane wave \\
$S_{\rm signal}$          & GW strain power spectrum \\
$\Omega_{\rm signal}$     & GW energy density spectrum \\
$\Omega_{\rm gw}$         & GW energy density integrated over all frequencies and normalized to $\rho_{\rm c}$ \\
\hline
$d_I$                     & Time series data recorded by detector $I$\\
$\tilde{d}_I$             & Fourier transform of $d_I$ (data mode in Fourier space) \\
$S_{IJ}$                  & Filtered cross-correlation signal of a pair of detectors $IJ$ \\
$Q_{IJ}$                  & Optimal filter function for a pair of detectors $IJ$ \\
$\widetilde{Q}_{IJ}$      & Fourier transform of $Q_{IJ}$ (optimal filter function in Fourier space) \\
\hline
$s_I$                     & Signal response of detector $I$ (signal contribution to the time series data) \\
$\tilde{s}_I$             & Fourier transform of $s_I$ (signal mode in Fourier space) \\
$C_{IJ}$                  & Covariance matrix for the signal responses in a detector network\\
$\widetilde{C}_{IJ}$      & Signal response cross power spectrum of a pair of detectors $IJ$ \\
\hline
$n_I$                     & Noise of detector $I$ (noise contribution to the time series data) \\
$\tilde{n}_I$             & Fourier transform of $n_I$ (noise mode in Fourier space) \\
$D_{\rm noise}^I$         & Detector noise auto power spectrum of detector $I$ \\
$S_{\rm noise}^I$         & Strain noise auto power spectrum of detector $I$ \\
$S_{\rm noise}^{\rm eff}$ & Effective strain noise power spectrum of a detector network \\
$\Omega_{\rm noise}$      & $S_{\rm noise}^{\rm eff}$ expressed in terms of a GW energy density spectrum \\
\hline
$R_I^{ij}$                & Impulse response of detector $I$ \\
$R_{\mathbf{n},I}^p$      & Response function of detector $I$ (absolute value defines antenna pattern)\\
$\mathcal{R}_I$           & Signal response function (detector transfer function) of detector $I$ \\
$\Gamma_{IJ}$             & Overlap reduction function of a pair of detectors $IJ$ \\
$\gamma_{IJ}$             & Normalized overlap reduction function of a pair of detectors $IJ$ \\
\hline
\end{tabular}
\end{center}
\end{table}


We are interested in experimental searches for a stochastic, Gaussian, stationary, isotropic, and unpolarized GW background.
A detailed review of the formalism to describe such searches can be found in Ref.~\cite{Romano:2016dpx}.
In the following, we will adopt the conventions of Ref.~\cite{Romano:2016dpx}, but restrict ourselves to a briefer exposition and customize our notation.
Most SGWB searches aim at measuring a nonzero cross-correlation signal in the outputs of two detectors whose intrinsic noise spectra are uncorrelated~\cite{Allen:1996vm,Allen:1997ad,Maggiore:1999vm}.
Let us now outline the main quantities entering the description of such a measurement and derive the expected SNR.


The raw data $d_I$ of a single detector $I$ amounts to a time series output of the form
\begin{align}
d_I\left(t\right) = s_I\left(t\right) + n_I\left(t\right) \,,
\end{align}
which receives a signal contribution $s_I$ and a noise contribution $n_I$.
Here, the signal contribution represents the detector response to the incoming GWs, which depends on both the properties of the GWs and the geometry of the detector.
In the following, large parts of our discussion will refer to the frequency domain, where $d_I$, $s_I$, and $n_I$ are replaced by their Fourier transforms $\tilde{d}_I$, $\tilde{s}_I$, and $\tilde{n}_I$.
In our convention, these Fourier modes are defined via
\begin{align}
F_I\left(t\right) = \int_{-\infty}^\infty df\:\widetilde{F}_I\left(f\right) e^{2\pi ift} \,, \quad F \in \left\{d,s,n\right\} \,.
\end{align}
Both the signal and the noise modes are assumed to correspond to Gaussian random variables.
Without loss of generality, we can set the expectation values of all modes to zero, $\big<\widetilde{F}_I\big> = 0$.
The entire available information on the statistical properties of both the signal and the noise is thus contained in quadratic expectation values of the form $\big<\widetilde{F}_I^{\vphantom{*}}\widetilde{F}_J^*\big>$, where $J = I$ or $J \neq I$.
For a single detector $I$, $\left<\tilde{n}_I^{\vphantom{*}}\tilde{n}_I^*\right>$ defines, \textit{e.g.}, the detector noise auto power spectrum $D_{\rm noise}^I$,
\begin{align}
\label{eq:Enn}
\left<\tilde{n}_I^{\vphantom{*}}\left(f^{\vphantom{\prime}}\right)\tilde{n}_I^*\left(f'\right)\right> = \frac{1}{2}\,\delta^{(1)}\left(f-f'\right)\,D_{\rm noise}^I\left(f\right) \,.
\end{align}
Here, the factor $1/2$ reflects the fact that, in our convention, all power spectra are defined to be single-sided.
The variance of the detector noise, $\sigma_I^2$, can therefore be written as follows,
\begin{align}
\label{eq:sigmaI2}
\sigma_I^2 = \left<n_I^2\right> - \left<n_I^{\vphantom{2}}\right>^2 = \left<n_I^2\right> = \int_0^\infty df\:D_{\rm noise}^I\left(f\right) \,,
\end{align}
with $D_{\rm noise}^I$ being integrated over the frequency range $\left[0,+\infty\right)$ rather than $\left(-\infty,+\infty\right)$.


Next, let us consider a network of detectors, $I,J = 1,2,\cdots$, and derive a similar integral representation for its response to the signal.
In this case, we now have to compute the covariance matrix $C_{IJ} = \left<s_I s_J\right> - \left<s_I\right>\left<s_J\right> = \left<s_I s_J\right>$ instead of just a single variance.
The two-point correlation function $\left<s_I s_J\right>$ accounts again for both the properties of the SGWB and the geometry of the detector network.
The incoming GWs are described by tensor perturbations of the spacetime metric $g_{\mu\nu}$.
In transverse-traceless gauge, we can write
\begin{align}
g_{\mu\nu}\left(t,\mathbf{x}\right)dx^\mu dx^\nu = -dt^2 + \left(\delta_{ij} + h_{ij}\left(t,\mathbf{x}\right)\right)dx^i dx^j \,,
\end{align}
where the tensor perturbations $h_{ij}$ can be decomposed into plane waves as follows,
\begin{align}
\label{eq:hij}
h_{ij}\left(t,\mathbf{x}\right) = \sum_{p = +,\times}\int_{-\infty}^\infty df \int d^2\mathbf{n}\:h_{\mathbf{n}}^p\left(f\right) \left(e_{\mathbf{n}}^p\right)_{ij} e^{2\pi if\left(t - \mathbf{n}\mathbf{x}\right)} \,.
\end{align}
Here, $h_{\mathbf{n}}^p\left(f\right)$ denotes the amplitude of a sinusoidal plane GW with frequency $f$, polarization $p$, and propagation direction $\mathbf{n}$, while $\left(e_{\mathbf{n}}^p\right)_{ij}$ is the corresponding polarization tensor,
\begin{align}
\label{eq:eij}
(e_{\mathbf{n}}^p)_{ij} =
(e_{\mathbf{n}}^p)_{ji} \,, \quad
(e_{\mathbf{n}}^p)_{ii} = 0 \,, \quad
n_i\,(e_{\mathbf{n}}^p)_{ij} = 0 \,, \quad
(e_{\mathbf{n}}^p)_{ij}\,(e_{\mathbf{n}}^{p'})_{ij}^* = 2\,\delta^{pp'} \,.
\end{align}
Analogous to Eq.~\eqref{eq:Enn}, the quadratic expectation value of the Fourier modes $h_{\mathbf{n}}^p$ reads
\begin{align}
\label{eq:Ehh}
\big<h_{\mathbf{n}^{\vphantom{\prime}}}^{p^{\vphantom{\prime}}}\left(f^{\vphantom{\prime}}\right)h_{\mathbf{n}'}^{p'*}\left(f'\right)\big> = \frac{1}{16\pi}\,\delta^{pp'}\,\delta^{(2)}\left(\mathbf{n}-\mathbf{n}'\right)\delta^{(1)}\left(f-f'\right) S_{\rm signal}\left(f\right) \,,
\end{align}
where the Kronecker delta and Dirac delta functions account for the fact that we assume the SGWB to be unpolarized, isotropic, and stationary.
The nontrivial information on the RHS of Eq.~\eqref{eq:Ehh} is encoded in the GW strain power spectrum $S_{\rm signal}$, which describes the total strain power of the SGWB summed over both polarization states, $p = +, \times$, and integrated over the entire sky as a function of frequency.
Analogous to Eq.~\eqref{eq:sigmaI2}, the strain power spectrum can be used to write down an integral representation of the strain variance,%
\footnote{The factor of $2$ on the RHS of this relation follows from the normalization of the polarization tensors in Eq.~\eqref{eq:eij} and is therefore nothing but a matter of convention.
It would be straightforward to avoid this factor by performing the rescaling: $1/\sqrt{2}\,(e_{\mathbf{n}}^p)_{ij} \rightarrow (e_{\mathbf{n}}^p)_{ij}$, $\sqrt{2}\,h_{\mathbf{n}}^p \rightarrow h_{\mathbf{n}}^p$, and $2\,S_{\rm signal} \rightarrow S_{\rm signal}$.
In passing, we also mention that this factor of 2 is sometimes ascribed to the fact that the metric perturbations $h_{ij}$ receive contributions from two different polarizations (see, \textit{e.g.}, Refs.~\cite{Maggiore:1999vm,Caprini:2018mtu}).
However, this statement is slightly misleading since the strain power spectrum is already summed over both polarization states~\cite{Romano:2016dpx}.
That is, if we wrote $S_{\rm signal} = S_{\rm signal}^+ + S_{\rm signal}^\times$, the contributions for both polarizations would still feature a factor of $2$.}
\begin{align}
\label{eq:sigmah}
\sigma_h^2 = \left<h_{ij}^{\vphantom{*}}h_{ij}^*\right> - \left<h_{ij}^{\vphantom{*}}\right>\left<h_{ij}^*\right> = \left<h_{ij}^{\vphantom{*}}h_{ij}^*\right> = 2\,\int_0^\infty df\:S_{\rm signal}\left(f\right) \,.
\end{align}


The signal response $s_I$ of detector $I$ follows from convoluting the tensor perturbation $h_{ij}$ with the detector's impulse response $R_I^{ij}$.
In the frequency domain, this results in
\begin{align}
\label{eq:stildeI}
\tilde{s}_I\left(f\right) = \sum_{p = +,\times} \int d^2\mathbf{n}\:R_{\mathbf{n},I}^p\left(f\right) h_{\mathbf{n}}^p\left(f\right) \,,
\end{align}
where the response function $R_{\mathbf{n},I}^p\left(f\right)$ represents the signal response of detector $I$ to (\textit{i.e.}, the convolution of its impulse response with) a sinusoidal plane GW with frequency $f$, polarization $p$, and propagation direction $\mathbf{n}$.
The graph of $\big|R_{\mathbf{n},I}^p\big|$ as a function of $\mathbf{n}$ describes the antenna pattern of the detector for GWs with frequency $f$ and polarization $p$.
More details on the impulse response and the response function can be found in Ref.~\cite{Romano:2016dpx}.
For our purposes, the important message from Eq.~\eqref{eq:stildeI} is that it allows us to write the quadratic expectation value of the signal modes $\tilde{s}_I$ in a similar way as the expectation value in Eq.~\eqref{eq:Enn},
\begin{align}
\label{eq:Ess}
\left<\tilde{s}_I^{\vphantom{*}}\left(f^{\vphantom{\prime}}\right)\tilde{s}_J^*\left(f'\right)\right> = \frac{1}{2}\,\delta^{(1)}\left(f-f'\right)\,\widetilde{C}_{IJ}\left(f\right) = \frac{1}{2}\,\delta^{(1)}\left(f-f'\right)\,\Gamma_{IJ}\left(f\right)\,S_{\rm signal}\left(f\right) \,,
\end{align}
where we introduced the so-called overlap reduction function $\Gamma_{IJ}$ of the detector pair $IJ$,
\begin{align}
\Gamma_{IJ}\left(f\right) = \frac{1}{2}  \sum_{p = +,\times} \int \frac{d^2\mathbf{n}}{4\pi}\: R_{\mathbf{n},I}^p\left(f\right) R_{\mathbf{n},J}^{p*}\left(f\right) \,.
\end{align}
As evident from Eq.~\eqref{eq:Ess}, the overlap reduction function acts as the transfer function between between the GW strain power spectrum and the signal response cross power spectrum of the $IJ$ detector pair, $\widetilde{C}_{IJ} = \Gamma_{IJ}\,S_{\rm signal}$.
In line with our assumption of an isotropic and unpolarized SGWB, $\Gamma_{IJ}$ is defined as the sky- and polarization-averaged product of the response functions for the detectors $I$ and $J$.
In the special case of a single detector, $J = I$, it reduces in particular to the sky- and polarization-averaged square of the antenna pattern,
\begin{align}
\mathcal{R}_I\left(f\right) = \Gamma_{II}\left(f\right) = \frac{1}{2}  \sum_{p = +,\times} \int \frac{d^2\mathbf{n}}{4\pi}\,\left|R_{\mathbf{n},I}^p\left(f\right)\right|^2 \,,
\end{align}
which is also known as the detector transfer function or simply signal response function $\mathcal{R}_I$.
The function $\mathcal{R}_I$ relates the GW strain power spectrum $S_{\rm signal}$ to the signal response auto power spectrum $D_{\rm signal}^I$ and can be used to define the strain noise auto power spectrum $S_{\rm noise}^I$,
\begin{align}
\label{eq:DRS}
D_{\rm signal}^I = \mathcal{R}_I\,S_{\rm signal} \,,\quad D_{\rm noise}^I = \mathcal{R}_I\,S_{\rm noise}^I \,.
\end{align}


Often times, one also works with the normalized overlap reduction function
\begin{align}
\label{eq:gammaIJ}
\gamma_{IJ}\left(f\right) = \frac{5}{\sin^2\delta}\:\Gamma_{IJ}\left(f\right) \,,
\end{align}
which is normalized to $\gamma_{IJ}\left(f = 0\right) = 1$ for a pair of identical, co-located, and co-aligned interferometers with an opening angle $\delta$ between the two arms.
Below, we will be interested in the cases $\delta = \pi/2$ (aLIGO, aVirgo, KAGRA, CE) and $\delta = \pi/3$ (ET, LISA, DECIGO, BBO), for which the conversion factor between $\gamma_{IJ}$ and $\Gamma_{IJ}$ amounts to $1/5$ and $3/20$, respectively.
The relation in Eq.~\eqref{eq:Ess} finally allows us to write down the covariance matrix,
\begin{align}
C_{IJ} = \left<s_I s_J\right> - \left<s_I\right>\left<s_J\right> = \left<s_I s_J\right> = \int_0^\infty df\:\widetilde{C}_{IJ}\left(f\right) = \int_0^\infty df\:\Gamma_{IJ}\left(f\right)\,S_{\rm signal}\left(f\right) \,.
\end{align}
Again, we only integrate over positive frequencies because all power spectra are single-sided.


We are now all set to derive the expected SNR for a cross-correlation measurement of the SGWB.
The basic idea is to apply a filter function $Q_{IJ}$ to the cross-correlation signal $S_{IJ}$ that can be constructed from the data streams of the two detectors $I$ and $J$,
\begin{align}
S_{IJ} = \int_{-t_{\rm obs}/2}^{+t_{\rm obs}/2} dt \int_{-t_{\rm obs}/2}^{+t_{\rm obs}/2} dt'\, d_I\left(t^{\vphantom{\prime}}\right) Q_{IJ}\left(t-t'\right) d_J\left(t'\right) \,,
\end{align}
and to choose (\textit{i.e.}, match) this filter function so as to maximize the corresponding SNR,
\begin{align}
\label{eq:rhoSIJ}
\varrho_{IJ} = \frac{\left<S_{IJ}^{\vphantom{2}}\right>}{\sqrt{\left<S_{IJ}^2\right> - \left<S_{IJ}^{\vphantom{2}}\right>^2}} \,.
\end{align}
The solution of this optimization problem has been worked out in Ref.~\cite{Allen:1996vm,Allen:1997ad}; here, we will only state the final result.
In Fourier space, the optimal filter function $\widetilde{Q}_{IJ}$ turns out to be
\begin{align}
\widetilde{Q}_{IJ}\left(f\right) = \mathcal{N}\,\frac{\Gamma_{IJ}\left(f\right)\,S_{\rm signal}\left(f\right)}{D_{\rm noise}^I\left(f\right) D_{\rm noise}^I\left(f\right)} \,,
\end{align}
where $\mathcal{N}$ is an irrelevant normalization constant that cancels in the expression for the SNR in Eq.~\eqref{eq:rhoSIJ}.
Based on this result, the optimal SNR ends up acquiring the following form,
\begin{align}
\label{eq:rhoIJ}
\varrho_{IJ} = \left(n_{\rm det}\,t_{\rm obs} \int_{f_{\rm min}}^{f_{\rm max}}df \frac{\Gamma_{IJ}^2\left(f\right)S_{\rm signal}^2\left(f\right)}{D_{\rm noise}^I\left(f\right)D_{\rm noise}^J\left(f\right)}\right)^{1/2} \,,
\end{align}
where $n_{\rm det} = 2$ counts the number of detectors involved in the cross-correlation measurement and the frequency interval $\left[f_{\rm min},f_{\rm max}\right]$ defines the bandwidth of the $IJ$ detector pair.
This result is valid in the weak-signal regime, which assumes that the integrand of the frequency integral in Eq.~\eqref{eq:rhoIJ} is smaller than unity for all frequencies, $\Gamma_{IJ}^2 S_{\rm signal}^2 \ll D_{\rm noise}^I D_{\rm noise}^J$.


In view of Eq.~\eqref{eq:rhoIJ}, several comments are in order.
First of all, note that both the optimal filter and the optimal SNR depend on the strain power spectrum of the signal, $S_{\rm signal}$.
In principle, one would therefore need to know the exact shape of the signal that one intends to measure if one really wanted to identify a cross-correlation signal $S_{IJ}$ whose SNR matches the one in Eq.~\eqref{eq:rhoIJ}.
This is of course impossible, which is why, in practice, one has to resort to a library of template spectra.
Among these template spectra, the best approximation of the true signal will then result in the SNR value closest to the optimal SNR.
A second comment is that Eq.~\eqref{eq:rhoIJ} remains in fact valid if we consider an idealized auto-correlation measurement in a single detector rather than a cross-correlation measurement using a pair of detectors.
In the case of LISA, one will, \textit{e.g.}, be able to monitor the detector noise in real time.
For an auto-correlation measurement, the optimal SNR after perfect noise subtraction is then again given by Eq.~\eqref{eq:rhoIJ}, however, with $n_{\rm det}$ set to $n_{\rm det} = 1$ (see Ref.~\cite{Thrane:2013oya} and references therein).
A third comment finally is that it is straightforward forward to generalize Eq.~\eqref{eq:rhoIJ} to an entire network of detectors, $I,J = 1,2,\cdots$.
In this case, one simply has to compute the partial SNRs for all possible pairs of detectors and add them in quadrature,
\begin{align}
\varrho = \left(\sum_{J>I} \varrho_{IJ}^2\right)^{1/2} = \left(n_{\rm det}\,t_{\rm obs} \int_{f_{\rm min}}^{f_{\rm max}}df \sum_{J>I} \frac{\Gamma_{IJ}^2\left(f\right)S_{\rm signal}^2\left(f\right)}{D_{\rm noise}^I\left(f\right)D_{\rm noise}^J\left(f\right)}\right)^{1/2} \,.
\end{align}
Given this expression for $\varrho$, it is convenient to define an effective strain noise power spectrum 
\begin{align}
\label{eq:Seff}
S_{\rm noise}^{\rm eff}\left(f\right) = \left(\sum_{J > I}\frac{\Gamma_{IJ}^2\left(f\right)}{D_{\rm noise}^I\left(f\right)D_{\rm noise}^J\left(f\right)}\right)^{-1/2} \,,
\end{align}
which generalizes the idea of the strain noise power spectrum $S_{\rm noise}^I$ to a detector network.
With this definition, the SNR can now be written as follows,
\begin{align}
\varrho = \left[n_{\rm det}\,t_{\rm obs} \int_{f_{\rm min}}^{f_{\rm max}}df\left(\frac{S_{\rm signal}\left(f\right)}{S_{\rm noise}^{\rm eff}\left(f\right)}\right)^2\right]^{1/2} \,.
\end{align}
Finally, both the strain power spectrum of the signal, $S_{\rm signal}$, and the noise power spectrum of the detector network, $S_{\rm noise}^{\rm eff}$, can be expressed in terms of GW energy density spectra,
\begin{align}
\label{eq:Osignalnoise}
\Omega_{\rm signal}\left(f\right) = \frac{2\pi^2}{3H_0^2}\,f^3 S_{\rm signal}\left(f\right) \,, \quad \Omega_{\rm noise}\left(f\right) = \frac{2\pi^2}{3H_0^2}\,f^3 S_{\rm noise}^{\rm eff}\left(f\right) \,.
\end{align}
Here, the GW energy density spectrum $\Omega_{\rm signal}$ is defined as the energy density contained in GWs per logarithmic frequency interval and normalized to the critical energy density $\rho_{\rm c}$,
\begin{align}
\Omega_{\rm gw} = \frac{\rho_{\rm gw}}{\rho_{\rm c}} = \frac{1}{\rho_{\rm c}}\int_0^\infty d\ln f\:\frac{d\rho_{\rm gw}}{d\ln f} = \frac{1}{\rho_{\rm c}}\int_0^\infty d\ln f\:\Omega_{\rm signal}\left(f\right) \,.
\end{align}
Making use of Eq.~\eqref{eq:Osignalnoise}, we obtain our final result for the optimal SNR [see Eq.~\eqref{eq:rho}],
\begin{align}
\label{eq:rhofinal}
\varrho = \left[n_{\rm det}\,t_{\rm obs} \int_{f_{\rm min}}^{f_{\rm max}}df\left(\frac{\Omega_{\rm signal}\left(f\right)}{\Omega_{\rm noise}\left(f\right)}\right)^2\right]^{1/2} \,.
\end{align}


Eq.~\eqref{eq:rhofinal} is the starting point for the construction of both power-law- and peak-integrated sensitivity curves.
The construction of PISCs is discussed in detail in the main text (see Sec.~\ref{sec:pisc}); here, we shall now review the construction of PLISCs.
The main assumption behind the PLISC approach is that the signal can be described by a pure power law in the relevant frequency range.
For an arbitrary reference frequency $f_{\rm ref}$, we may thus write
\begin{align}
\label{eq:ansatz}
\Omega_{\rm signal}\left(f\right) = \Omega_p\left(\frac{f}{f_{\rm ref}}\right)^p \,.
\end{align}
For each power $p$, one can now determine the corresponding value of the amplitude $\Omega_p$ that results in a specific value of the SNR, typically, $\varrho_{\rm thr} = 1$.
The solutions for $\Omega_p$ are of the form
\begin{align}
\label{eq:OPLISp}
\Omega_p = \Omega_{\rm PLIS}^{(p)}\left(\varrho_{\rm thr},t_{\rm obs}\right) = \varrho_{\rm thr} \left[n_{\rm det}\,t_{\rm obs} \int_{f_{\rm min}}^{f_{\rm max}}df\left(\frac{\left(f/f_{\rm ref}\right)^p}{\Omega_{\rm noise}\left(f\right)}\right)^2\right]^{-1/2} \,.
\end{align}
Plugging these $\Omega_p$ values back into Eq.~\eqref{eq:ansatz}, one finds a set of power-law curves whose envelope (typically in a log-log plot of the frequency\,--\,amplitude plane) defines the PLISC,
\begin{align}
\label{eq:plis}
\Omega_{\rm PLIS}\left(f\right) = \max_p \left\{\Omega_{\rm PLIS}^{(p)}\left(\varrho_{\rm thr},t_{\rm obs}\right)\left(\frac{f}{f_{\rm ref}}\right)^p\right\} \,.
\end{align}
Note that $f_{\rm ref}$ cancels in this definition, which renders $\Omega_{\rm PLIS}$ insensitive to the exact choice of this auxiliary quantity.
The interpretation of the \textit{power-law-integrated sensitivity} (PLIS) is as follows:
Any power-law signal that intersects the PLISC, such that $\Omega_{\rm signal}\left(f\right) > \Omega_{\rm PLIS}\left(f\right)$ for at least some frequency $f$, results in an SNR above threshold; all curves tangential to the PLISC result in an SNR of exactly $\varrho_{\rm thr}$; and all curves that always stay below the PLISC have subthreshold SNR.
The factor by which the signal curve needs to be rescaled in order to align it with a tangent of the PLISC can thus be interpreted as the expected SNR.


It is also interesting to compare the power-law-integrated sensitivity in Eq.~\eqref{eq:OPLISp} to the peak-integrated sensitivity in Eq.~\eqref{eq:OPISgen}.
Obviously, we can reproduce the expression in Eq.~\eqref{eq:OPLISp} by setting $i=j$ and $\mathcal{S}_i = \mathcal{S}_j = \left(f/f_{\rm ref}\right)^p$ in Eq.~\eqref{eq:OPISgen}, \textit{i.e.}, if we assume a spectral shape function that is described by a pure power law.
In the case of GWs from a cosmological phase transition, for which better estimates of the spectral shape exist, this is certainly not the best choice.
Among other things, this is an important motivation for our PISC method.

 
\subsection{Transfer functions}
\label{subsec:transfer}


\subsubsection*{Signal response function for a single detector}


\begin{figure}
\begin{center}

\includegraphics[width=0.45\textwidth]{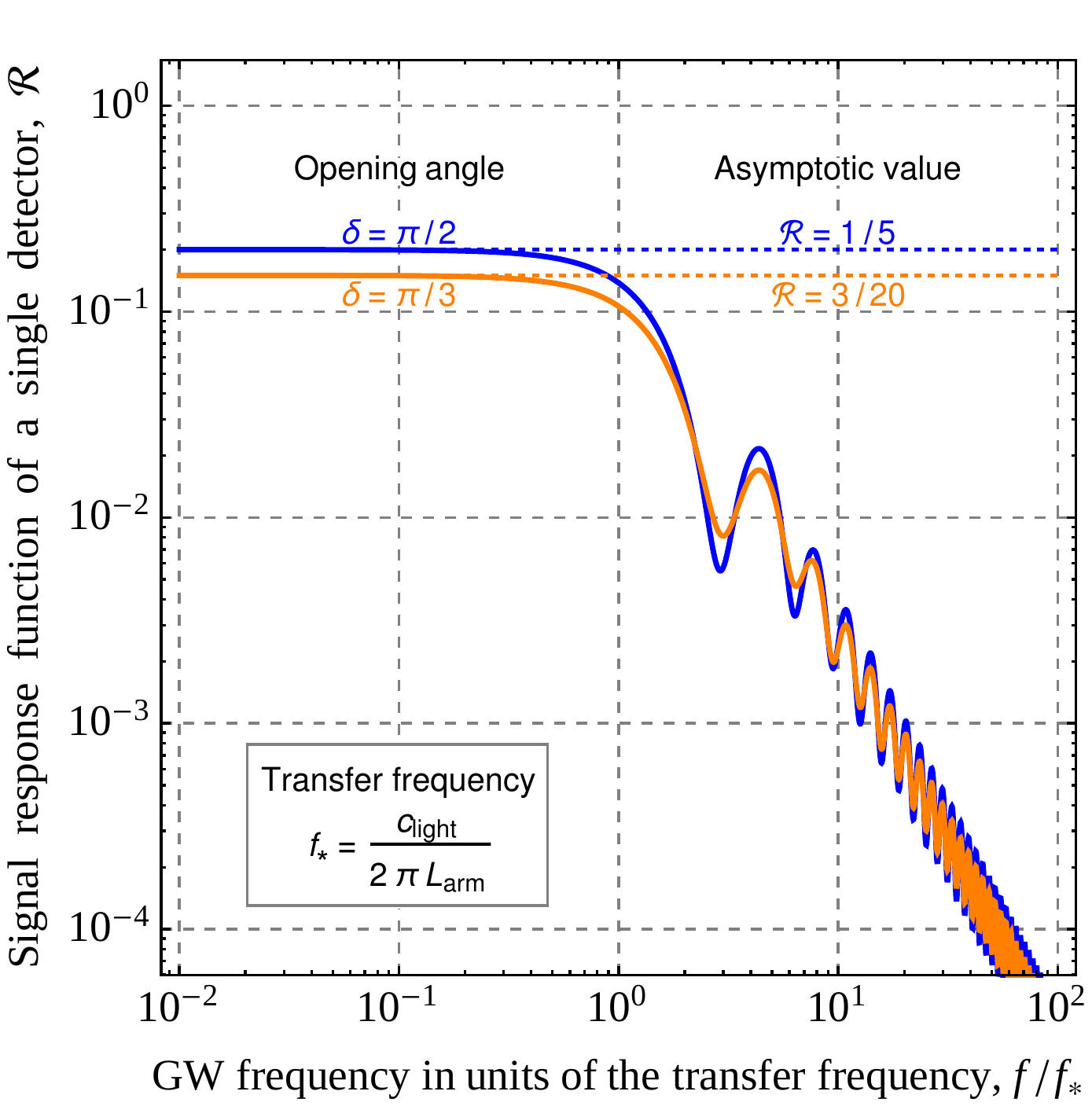}

\caption{Signal response function for an equal-arm Michelson interferometer.}
\label{fig:response}
\end{center}
\end{figure}


Let us now turn to the transfer functions (\textit{i.e.}, signal response and overlap reduction functions) of specific experiments.%
\footnote{For recent work on transfer functions for GW experiments, see Refs.~\cite{Blaut:2012zz,Liang:2019pry,Zhang:2019oet}.
These papers also discuss the response to vector and scalar GW polarization states, which only occur in models of modified gravity.}
We begin by computing the signal response function $\mathcal{R}_I$ for a single equal-arm Michelson interferometer.
A closed analytic expression for $\mathcal{R}_I$ does unfortunately not exist; however, an explicit integral representation can be found in Ref.~\cite{Larson:1999we},
\begin{align}
\label{eq:R}
\mathcal{R}_I\left(f\right) & = \frac{1}{4\,u^2}\left[\left(1 + \cos^2 u\right)\left(\frac{1}{3} - \frac{2}{u^2}\right) + \sin^2 u + \frac{4}{u^3} \sin u \cos u - \frac{1}{4\pi}\,\mathcal{I}\left(u,\delta\right)\right] \,, \\\nonumber
\mathcal{I}\left(u,\delta\right) & = \int_0^{2\pi}d\epsilon \int_0^{\pi}d\theta\: \sin\theta\left(1 - \frac{2\,\sin^2\delta \sin^2\epsilon}{1-\cos^2\theta'}\right)\left[\left(\cos u -\cos u_\theta\right)\left(\cos u -\cos u_{\theta'}\right) \right.
\\ \nonumber
& \left.\times\cos\theta\cos\theta' + \left(\sin u - \cos\theta \sin u_\theta\right)\left(\sin u - \cos\theta' \sin u_{\theta'}\right)\right] \,,
\end{align}
where $\delta$ denotes again the interferometer's opening angle and $u_\theta$, $u_{\theta'}$, and $\theta'$ are defined as
\begin{align}
u_\theta = u\,\cos\theta \,,\quad u_{\theta'} = u\,\cos\theta' \,,\quad \cos\theta' = \cos\delta\cos\theta + \cos\epsilon\sin\delta\sin\theta \,.
\end{align}
$u = \pi\,f/f_{\rm fsr} = f/f_*$ in Eq.~\eqref{eq:R} measures the GW frequency $f$ in units of the interferometer's \textit{free spectral range} (FSR), $f_{\rm fsr} = c_{\rm light}/\left(2 L_{\rm arm}\right)$, where $c_{\rm light}$ denotes the speed of light and $L_{\rm arm}$ is the length of the interferometer arms.
Instead of the FSR frequency $f_{\rm fsr}$, one sometimes also encounters $f_* = f_{\rm fsr}/\pi$, which is referred to as the transfer frequency~\cite{Cornish:2018dyw}.
Below, we will be interested in CE's and LISA's signal response functions.
Given their current design concepts, we find the following transfer frequencies for these two interferometers, 
\begin{align}
\label{eq:CELISA}
& \textrm{CE:}   & \delta & = \frac{\pi}{2} \,, & L_{\rm arm}^{\rm CE}   & = 4.0 \times 10^4\,\textrm{m} \,, & f_*^{\rm CE}   & \simeq 1193 \,\textrm{Hz} \,, \\\nonumber
& \textrm{LISA:} & \delta & = \frac{\pi}{3} \,, & L_{\rm arm}^{\rm LISA} & = 2.5 \times 10^9\,\textrm{m} \,, & f_*^{\rm LISA} & \simeq 19.09 \,\textrm{Hz} \,.
\end{align}


The integral function $\mathcal{I}\left(u,\delta\right)$ in Eq.~\eqref{eq:R} does not admit a closed analytic form and needs to be evaluated numerically.
In Fig.~\ref{fig:response}, we present our numerical results for both $\delta = \pi/2$ and $\delta = \pi/3$.
As expected, the signal response function approaches a constant value in the small-frequency limit for both opening angles, $\mathcal{R}_I \rightarrow 1/5\,\sin^2\delta$ [see the discussion below Eq.~\eqref{eq:gammaIJ}].
At larger frequencies, $\mathcal{R}_I$ is subject to sinusoidal modulations with period $f_{\rm fsr} = \pi f_*$, while its overall amplitude drops off like $1/f^2$.
Ignoring the oscillations at high frequencies, $\mathcal{R}_I$ can be well approximated by rational fit functions for both CE and LISA,
\begin{align}
\mathcal{R}_{\rm CE}\left(f\right) \simeq \frac{1/5}{1 + 0.67\left(f/f_*^{\rm CE}\right)^2} \,,\quad \mathcal{R}_{\rm LISA}\left(f\right) \simeq \frac{2\times3/20}{1 + 0.54\left(f/f_*^{\rm LISA}\right)^2} \,.
\end{align}
Here, we multiplied $\mathcal{R}_{\rm LISA}$ by an extra factor of $2$ to account for the fact that LISA's six active laser links will allow one to construct two independent data streams at low frequencies that can be used for an auto-correlation measurement~\cite{Cornish:2018dyw}.
This factor is sometimes missed in the literature.
CE, on the other hand, envisions an L-shaped interferometer similar to aLIGO, in which case only one data channel will be available for an auto-correlation measurement.


\subsubsection*{Overlap reduction functions for detector pairs in a detector network}


\begin{table}
\begin{center}
\renewcommand{\arraystretch}{1.22}
\caption{Geometrical data describing the angular separation and relative orientation of the six detector pairs that can be formed within the HLVK detector network (see Refs.~\cite{Nishizawa:2009bf,Himemoto:2017gnw} for details).}
\label{tab:angles}
\bigskip
\begin{tabular}{|r||rrrrrr|}
\hline
 & HL & HV & LV & HK & LK & VK \\
\hline\hline
$\beta$  [deg] & 27.2 & 79.6 & 76.8 & 72.4 & 99.2 & 86.6 \\
$\Delta$ [deg] & 62.2 & 55.1 & 83.1 & 25.6 & 68.1 &  5.6 \\
$\delta$ [deg] & 45.3 & 61.1 & 26.7 & 89.1 & 42.4 & 28.9 \\
\hline
\end{tabular}
\end{center}
\end{table}

\begin{figure}
\begin{center}

\includegraphics[width=0.45\textwidth]{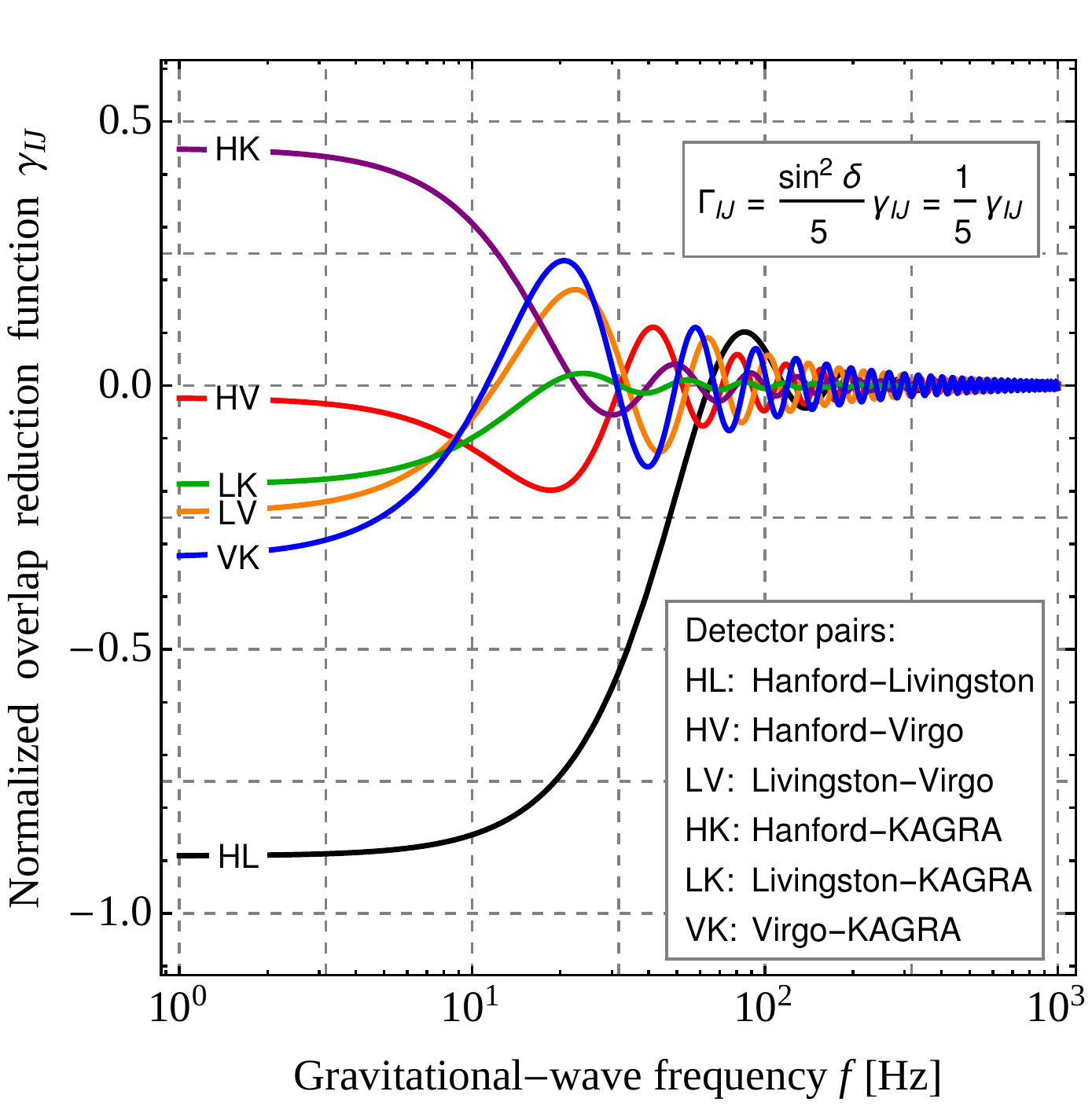}\hfill
\includegraphics[width=0.45\textwidth]{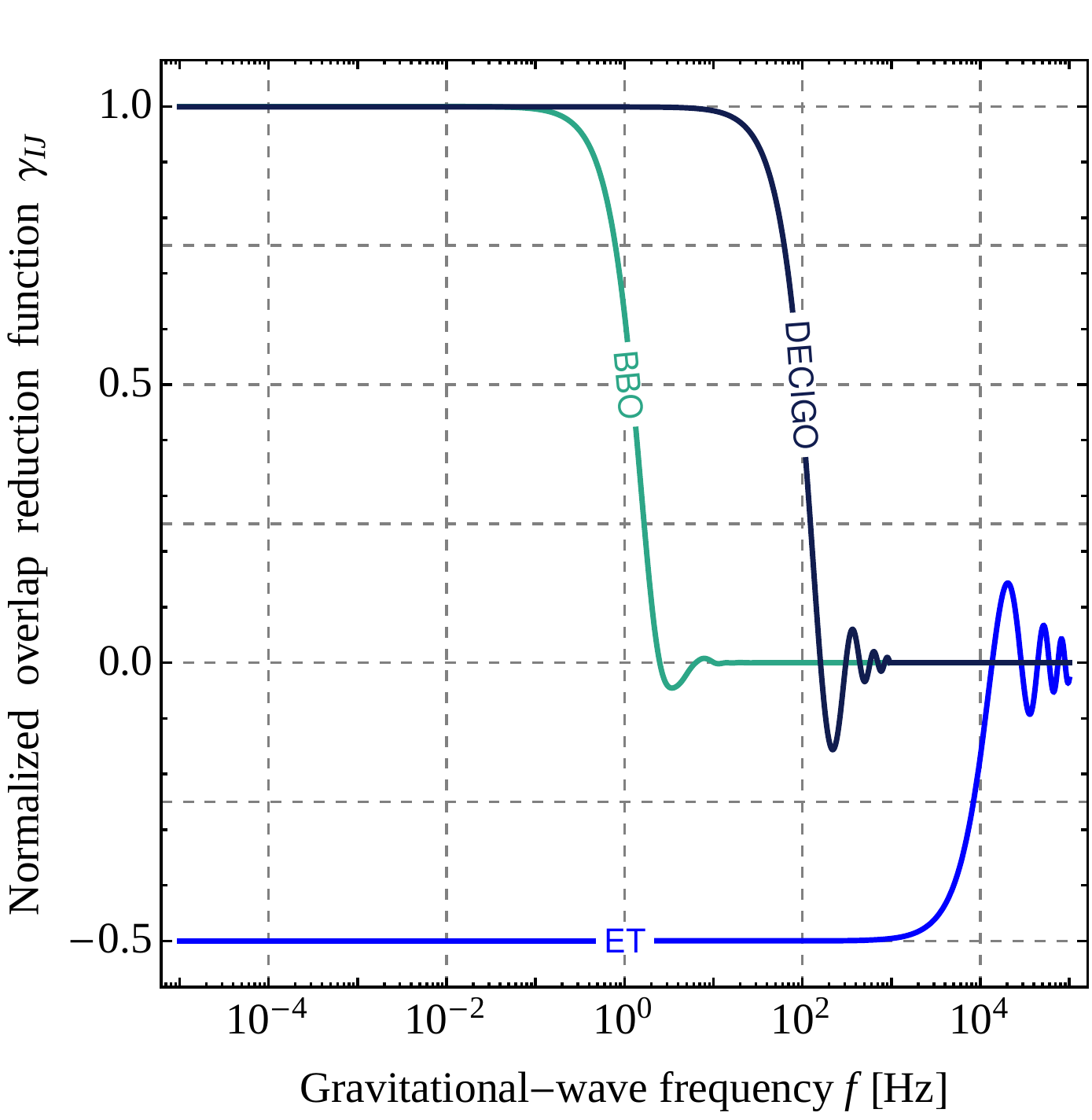}

\caption{Overlap reduction functions for pairs of L-shaped (left) and triangular (right) detectors.}
\label{fig:norf}
\end{center}
\end{figure}

A cross-correlation measurement using a detector network requires knowledge of the overlap reduction functions $\Gamma_{IJ}$ for all detector pairs that can be formed within the network.
For L-shaped ground-based interferometers, it is possible to derive an analytic expression for $\Gamma_{IJ}$, or alternatively, the normalized overlap reduction function $\gamma_{IJ}$ [see Eq.~\eqref{eq:gammaIJ}]~\cite{Nishizawa:2009bf,Himemoto:2017gnw},
\begin{align}
\label{eq:gammaL}
\gamma_{IJ}\left(f\right) =
& - \frac{1}{8} \left[3\,j_0\left(\bar{u}\right) - \frac{45}{7}\,j_2\left(\bar{u}\right) + \frac{169}{112}\,j_4\left(\bar{u}\right)\right] \:\,\,\cos\left(4\Delta\right) \\ \nonumber
& + \frac{1}{8} \left[4\,j_0\left(\bar{u}\right) - \frac{40}{7}\,j_2\left(\bar{u}\right) - \frac{108}{112}\,j_4\left(\bar{u}\right)\right] \:\,\,\cos\left(4\Delta\right) \cos \beta \\ \nonumber
& - \frac{1}{8} \left[\phantom{1}\,j_0\left(\bar{u}\right) + \,\,\frac{5}{7}\,\,j_2\left(\bar{u}\right) + \frac{3}{112}\,j_4\left(\bar{u}\right)\right] \left[\cos\left(4\Delta\right) \cos\left(2\beta\right) - 8\,\cos\left(4\delta\right) \cos^4\left(\frac{\beta}{2}\right)\right] \,,
\end{align}
where $j_n$ ($n=0,2,4$) represents the spherical Bessel function of the first kind of order $n$.
The frequency dependence in Eq.~\eqref{eq:gammaL} is encoded in $\bar{u} = f/\bar{f}_*$, where $\bar{f}_*$ is now defined as
\begin{align}
\bar{f}_* = \frac{c_{\rm light}}{2\pi\,d_{\oplus}\,\sin\left(\beta/2\right)} \,.
\end{align}
In this expression, $d_\oplus \simeq 12\,742\,\textrm{km}$ denotes the mean diameter of the earth, and $\beta$ is the angle between the detectors $I$ and $J$ in geocentric coordinates.
Similarly, $\Delta$ and $\delta$ quantify the orientation of the two detectors relative to the great circle on which they both lie.
For more details on the definitions of $\beta$, $\Delta$, and $\delta$, see Ref.~\cite{Nishizawa:2009bf}, in particular, Fig.~5 in this work.
Eq.~\eqref{eq:gammaL} can be used to compute the overlap reduction functions for all six detector pairs that can be formed in the four-detector network consisting of aLHO, aLLO, aVirgo, and KAGRA: \textit{aLHO-aLLO} (HL), \textit{aLHO-aVirgo} (HV), \textit{aLLO-aVirgo} (LV), \textit{aLHO-KAGRA} (HK), \textit{aLLO-KAGRA} (LK), and \textit{aVirgo-KAGRA} (VK).
The values of the geometrical angles $\beta$, $\Delta$, and $\delta$ for these six detector pairs are listed in Refs.~\cite{Nishizawa:2009bf,Himemoto:2017gnw} (see Tab.~\ref{tab:angles}).
The resulting normalized overlap reduction functions are shown in the left panel of Fig.~\ref{fig:norf}.


For all other detector pairs that we are interested in (\textit{i.e.}, for experiments that are not based on the cross-correlation of L-shaped ground-based interferometers), we compile  explicit numerical results for the respective overlap reduction functions from the literature:

\smallskip\noindent\textbf{ET} will consist of three V-shaped ground-based Michelson interferometers with opening angle $\delta = \pi/3$ in a triangular configuration, \textit{i.e.}, rotated with respect to each other by an angle $\omega = 2\pi/3$.
Consequently, ET will allow one to perform cross-correlation measurements in a network of three identical and co-located (but not co-aligned) detectors.
The overlap reduction function for a pair of ET detectors can be found in Ref.~\cite{Regimbau:2012ir}.
We extract the function graph from Fig.~8 of Ref.~\cite{Regimbau:2012ir} and rescale it by a factor $\sin^2\beta = 3/4$, in order to adjust its overall normalization, $\gamma_{IJ} \rightarrow \cos\left(2\omega\right) = -1/2$ in the small-frequency limit.%
\footnote{In Ref.~\cite{Regimbau:2012ir}, the normalized overlap reduction function is proportional to $\sin^2\beta$, whereas in our convention, this factor is explicitly factored out, such that it only appears in $\Gamma_{IJ} = 1/5\,\sin^2\beta\,\gamma_{IJ}$ [see Eq.~\eqref{eq:gammaIJ}].}
The normalized overlap reduction function thus obtained is shown in the right panel of Fig.~\ref{fig:norf}.


\smallskip\noindent\textbf{DECIGO} is envisioned to consist of two satellite-borne triangular \textit{Fabry--P\'erot} (FP) interferometers with opening angle $\delta = \pi/3$ in a hexagonal configuration (\textit{i.e.}, $\omega = \pi$).
The overlap reduction function for this pair of detectors has been computed in Ref.~\cite{Kuroyanagi:2014qza} (see also Appendix~B of Ref.~\cite{Kudoh:2005as}).
In our analysis, we shall use the numerical results obtained in Ref.~\cite{Kuroyanagi:2014qza}, which were kindly provided to us by Sachiko Kuroyanagi (see the right panel of Fig.~\ref{fig:norf}).
As expected, the normalized overlap reduction approaches unity in the small-frequency limit.
Similarly as in the case of LISA, we assume that DECIGO will allow one to construct two independent data streams.
At small frequencies, DECIGO's overlap reduction function therefore approaches the same value as LISA's signal response function, $\Gamma_{IJ} \rightarrow 2/5\sin^2\beta = 3/10$.
This value is a good approximation up to frequencies of $\mathcal{O}\left(100\right)\,\textrm{Hz}$, which is due to the fact that DECIGO's relatively short arm length, $L_{\rm arm}^{\rm DECIGO} = 1000\,\textrm{km}$, results in a large characteristic frequency, $f_*^{\rm DECIGO} = c_{\rm light}/\big(2 \pi L_{\rm arm}^{\rm DECIGO}\big) \simeq 47.71\,\textrm{Hz}$.
The laser travel distance in DECIGO's FP cavity is by contrast much larger than $L_{\rm arm}^{\rm DECIGO}$, which renders DECIGO most sensitive in the deci-Hertz range; hence DECIGO's name.
In many cases, it thus suffices to simply assume a constant overlap reduction function, $\Gamma_{IJ} \simeq 3/10$.


\smallskip\noindent\textbf{BBO} is planned to have a similar geometry as DECIGO (\textit{i.e.}, two satellite-borne triangular interferometers with $\delta = \pi/3$ and $\omega = \pi$).
It is, however, supposed to have a larger arm length, $L_{\rm arm}^{\rm BBO} = 50\,000\,\textrm{km}$, and utilize Michelson instead of FP interferometers.
The overlap reduction function of the two BBO units has been calculated in Ref.~\cite{Thrane:2013oya}, and the corresponding numerical results are available from Ref.~\cite{Romano:2013aaa}.
We plot the normalized overlap reduction function in the right panel of Fig.~\ref{fig:norf}.
Following Ref.~\cite{Thrane:2013oya}, we assume a single data channel, such that $\gamma_{IJ} \rightarrow 1$ and $\Gamma_{IJ}\rightarrow 1/5\sin^2\beta = 3/20$ in the small-frequency limit.


\subsubsection*{Overlap reduction functions for pulsar pairs in a pulsar timing array}


The overlap reduction function for two pulsars $I$ and $J$ in a PTA is of the form~\cite{Moore:2014eua,Hazboun:2019vhv},
\begin{align}
\Gamma_{IJ}\left(f\right) = \mathcal{R}_I\left(f\right)\zeta_{IJ}\left(\psi\right) \,,
\end{align}
with $\mathcal{R}_I$ being the response function for a single pulsar (\textit{i.e.}, $J=I$) and $\zeta_{IJ}$ denoting the Hellings--Downs factor~\cite{Hellings:1983fr} for two pulsars that are separated by an angle $\psi$ in the sky,
\begin{align}
\mathcal{R}_I\left(f\right) = \frac{1}{12\pi^2f^2} \,,\quad \zeta_{IJ}\left(\psi\right) = \frac{1}{2}\left[\delta_{IJ} + 1 + c_\psi\left(3\,\ln c_\psi-\frac{1}{2}\right)\right] \,.
\end{align}
Here, we introduced $c_\psi = \left(1-\cos\psi\right)/2$ and made sure that $\zeta_{IJ}$ is properly normalized, \textit{i.e.}, $\zeta_{II} = 1$ for a single pulsar.
Note that $\Gamma_{IJ}$ and $\mathcal{R}_I$ are now no longer dimensionless as in the case of the interferometer experiments.
Both quantities are expressed in units of $\textrm{Hz}^{-2}$, which is line with the fact that the timing noise of a pulsar has dimension $\textrm{Hz}^{-3}$ (see Sec.~\ref{subsec:detector}).


For future experiments, the pulsar distribution in the sky is still unknown.
In order to estimate their sensitivity, it is therefore reasonable to replace $\zeta_{IJ}$ by an expectation value that reflects the expected distribution of pulsars in the PTA.
A common choice, \textit{e.g.}, is to replace $\zeta_{IJ}$ by its \textit{root-mean-square} (RMS) expectation value assuming a flat prior on $\cos\psi$,
\begin{align}
\bar{\zeta}_{\rm rms} = \left[\frac{1}{2}\int_{-1}^1 d\cos\psi\: \zeta_{IJ}^2\left(\psi\right) \right]^{1/2} = \frac{1}{4\sqrt{3}} \simeq 0.144 \,,
\end{align}
whereas a flat prior on the distribution of pulsars in three-dimensional position space yields
\begin{align}
\label{eq:zetarms}
\zeta_{\rm rms} \simeq 0.147 \,.
\end{align}
This is the value that we shall employ in our analysis.
In order to estimate the sensitivities of IPTA and SKA, we will thus work with the following overlap reduction function,
\begin{align}
\label{eq:GPTA}
\Gamma_{IJ}\left(f\right) = \frac{\zeta_{\rm rms}}{12\pi^2f^2} \,.
\end{align}


\subsection{Detector noise power spectra}
\label{subsec:detector}


In addition to the transfer functions discussed in the previous section, we also require the intrinsic noise spectra $D_{\rm noise}^I$ for each detector.
The noise spectra of aLIGO, aVirgo, KAGRA, and CE can be downloaded from LIGO's \textit{Document Control Center} (DCC) [\href{https://dcc.ligo.org}{dcc.ligo.org}].
In Tab.~\ref{tab:noise}, we list the DCC documents that give access to the respective data files.
Note that, for aLIGO and aVirgo, we also consider sensitivities representative for O2 in addition to their envisioned design sensitivities.
The noise spectrum of ET can be downloaded from Ref.~\cite{ET:2018aaa} (see also DCC document P1600143~\cite{CE:2016aaa}).
In our analysis, we shall employ the ET-D-sum noise curve.
Moreover, as regards the noise spectra of the future space-based interferometers and PTA experiments, we will make use of the following analytical estimates:


\begin{table}
\begin{center}
\renewcommand{\arraystretch}{1.22}
\caption{Documents on \href{https://dcc.ligo.org}{dcc.ligo.org} that allow one to download the numerical data for the detector noise spectra of aLIGO, aVirgo, KAGRA, and CE according to their design and O2 sensitivities.}
\label{tab:noise}
\bigskip
\begin{tabular}{|l||l|l||}
\hline
       & Design sensitivity            & Representative sensitivity for observing run 2                         \\
\hline\hline
aLIGO  & T1800044~\cite{Ligo:2018aaa}  & G1801950~\cite{LHO:2017aaa} (aLHO), G1801952~\cite{LLO:2017aaa} (aLLO) \\
aVirgo & P1200087~\cite{Virgo:2018aaa} & P1800374~\cite{Virgo:2017aaa}                                          \\
KAGRA  & P1200087~\cite{Virgo:2018aaa} & ------                                                                 \\
CE     & P1600143~\cite{CE:2016aaa}    & ------                                                                 \\
\hline
\end{tabular}
\end{center}
\end{table}


\smallskip\noindent\textbf{LISA:} Following Ref.~\cite{Cornish:2018dyw}, we write LISA's noise spectrum as a sum of two contributions,
\begin{align}
\label{eq:DLISA}
D_{\rm noise}^{\rm LISA}\left(f\right) = \frac{1}{\left(L_{\rm arm}^{\rm LISA}\right)^2} \left[D_{\rm oms}^{\rm LISA}\left(f\right) + \frac{2}{\left(2\pi f\right)^4}\left(1 + \cos^2\left(\frac{f}{f_*^{\rm LISA}}\right)\right)D_{\rm acc}^{\rm LISA}\left(f\right)\right] \,,
\end{align}
where LISA's arm length $L_{\rm arm}^{\rm LISA}$ and characteristic frequency $f_*^{\rm LISA}$ are given in Eq.~\eqref{eq:CELISA} and where $D_{\rm oms}^{\rm LISA}$ and $D_{\rm acc}^{\rm LISA}$ account for the noise in the \textit{optical metrology system} (OMS) (\textit{i.e.}, position noise) and the acceleration noise of a single test mass, respectively,
\begin{align}
D_{\rm oms}^{\rm LISA}\left(f\right) & \simeq \left(1.5 \times 10^{-11}\,\textrm{m}\right)^2 \left[1 + \left(\frac{2\,\textrm{mHz}}{f}\right)^4\right] \textrm{Hz}^{-1} \,, \\\nonumber
D_{\rm acc}^{\rm LISA}\left(f\right) & \simeq \left(3\times10^{-15}\,\textrm{m}\,\textrm{s}^{-2}\right)^2 \left[1 + \left(\frac{0.4\,\textrm{mHz}}{f}\right)^2\right] \left[1 + \left(\frac{f}{8\,\textrm{mHz}}\right)^4\right] \textrm{Hz}^{-1} \,.
\end{align}


\smallskip\noindent\textbf{DECIGO:} According to Ref.~\cite{Kuroyanagi:2014qza}, DECIGO's noise spectrum receives three contributions,
\begin{align}
D_{\rm noise}^{\rm DECIGO}\left(f\right) = D_{\rm shot}^{\rm DECIGO}\left(f\right) +  D_{\rm rad}^{\rm DECIGO}\left(f\right) +  D_{\rm acc}^{\rm DECIGO}\left(f\right) \,,
\end{align}
which respectively quantify shot noise, radiation pressure noise, and acceleration noise,
\begin{align}
D_{\rm shot}^{\rm DECIGO}\left(f\right) & = \frac{\hbar c_{\rm light} \pi \lambda}{P_{\rm eff}} \left(\frac{1}{4FL_{\rm arm}^{\rm DECIGO}}\right)^2 \left[1 + \left(\frac{f}{f_*^{\rm DECIGO}}\right)^2\right] \,, \\\nonumber
D_{\rm rad}^{\rm DECIGO}\left(f\right) & = \frac{\hbar P}{c_{\rm light} \pi \lambda}\left(\frac{16F}{M L_{\rm arm}^{\rm DECIGO}}\right)^2\left(\frac{1}{2\pi f}\right)^4\left[1 + \left(\frac{f}{f_*^{\rm DECIGO}}\right)^2\right]^{-1} \,, \\\nonumber
D_{\rm acc}^{\rm DECIGO}\left(f\right) & = \frac{\hbar P}{c_{\rm light} \pi \lambda}\left(\frac{16F}{3M L_{\rm arm}^{\rm DECIGO}}\right)^2\left(\frac{1}{2\pi f}\right)^4 \,.
\end{align}
In our analysis, we shall assume the following values for DECIGO's characteristic experimental variables~\cite{Yagi:2011wg}:
arm length $L_{\rm arm}^{\rm DECIGO} = 1000\,\textrm{km}$, laser output power $P = 10\,\textrm{W}$, laser wavelength $\lambda = 532\,\textrm{nm}$, mirror mass $M = 100\,\textrm{kg}$, and mirror radius $R = 0.5\,\textrm{m}$.
These values determine the finesse $F$ of the FP cavity as well as the effective laser output power $P_{\rm eff}$,
\begin{align}
F = \frac{\pi\left(r_E\,r_F\right)^{1/2}}{1 - r_E\,r_F} \simeq 10.18 \,,\quad P_{\rm eff} = \left(\frac{r_E\,t_F^2}{1 - r_E\,r_F}\right)^2 P \simeq 6.68\,\textrm{W} \,,
\end{align}
where the quantities $r_E$, $r_F$, and $t_F$ are given as follows (see Ref.~\cite{Kuroyanagi:2014qza} for details),
\begin{align}
r_E & = r_{Em} r_G \,,\quad r_F = r_{Fm} r_G \,,\quad t_F = \left(r_G^2 - r_{Fm}^2\right)^{1/2} \,, \\\nonumber
r_G & = 1 - \exp\left(-\frac{2\pi R^2}{\lambda L_{\rm arm}^{\rm DECIGO}}\right) \,,\quad r_{Em}^2 = 0.9999 \,,\quad r_{Fm}^2 = 0.67 \,.
\end{align}


\begin{figure}
\begin{center}

\includegraphics[width=0.95\textwidth]{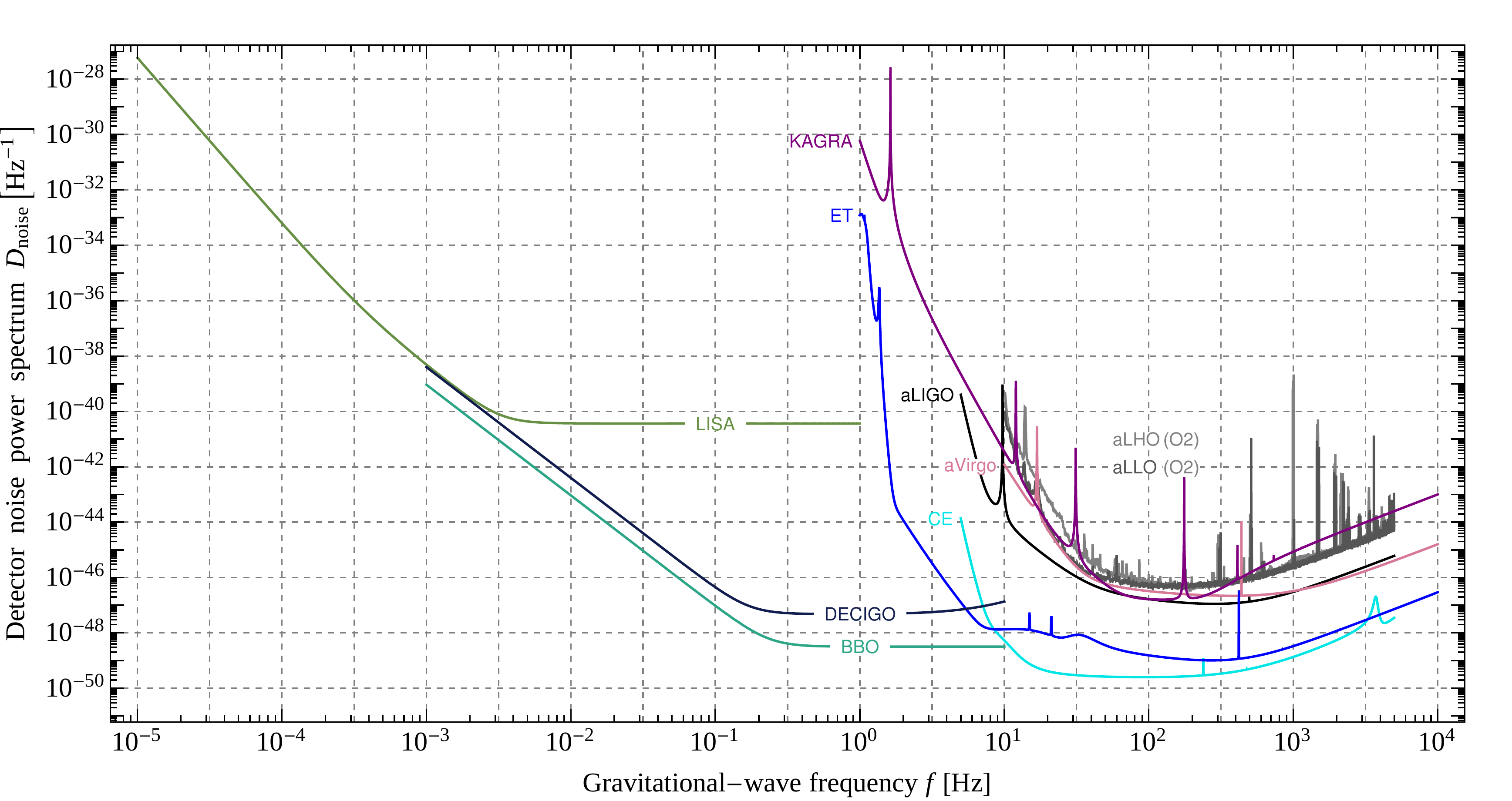}
\caption{Detector noise power spectra of present and upcoming interferometer experiments.}
\label{fig:noise}
\end{center}
\end{figure}


\smallskip\noindent\textbf{BBO:} Following Ref.~\cite{Thrane:2013oya}, we assume a BBO noise spectrum of the form
\begin{align}
D_{\rm noise}^{\rm BBO}\left(f\right)  = \frac{4}{\left(L_{\rm arm}^{\rm BBO}\right)^2}\left[D_{\rm oms}^{\rm BBO}\left(f\right) + \frac{1}{\left(2\pi f\right)^4}\,D_{\rm acc}^{\rm BBO}\left(f\right)\right] \,,
\end{align}
where $L_{\rm arm}^{\rm BBO} = 50\,000\,\textrm{km}$ and which is similar to LISA's noise spectrum in the sense that it also receives contributions from both position and acceleration noise, $D_{\rm oms}^{\rm BBO}$ and $D_{\rm acc}^{\rm BBO}$,
\begin{align}
D_{\rm oms}^{\rm BBO}\left(f\right) \simeq \left(1.4 \times 10^{-17}\,\textrm{m}\right)^2\textrm{Hz}^{-1} \,,\quad D_{\rm acc}^{\rm BBO}\left(f\right) \simeq \left(3 \times 10^{-17}\,\textrm{m}\,\textrm{s}^{-2}\right)^2\textrm{Hz}^{-1} \,.
\end{align}


\smallskip\noindent\textbf{PTA:} For the future PTA experiments IPTA and SKA, we respectively assume a network of $N$ pulsars in which each pulsar is monitored with a white timing noise of the form~\cite{Thrane:2013oya}
\begin{align}
\label{eq:DPTA}
D_{\rm noise}^{\rm PTA}\left(f\right) = 2\,T\sigma_t^2 \,.
\end{align}
Here, $1/T$ is the cadence of the timing observations and $\sigma_t$ denotes the RMS error of the timing residuals.
We shall assume a future IPTA data set based on $N = 20$, $T = 2\,\textrm{weeks}$, and $\sigma_t = 100\,\textrm{ns}$, which goes beyond the timing precision achieved in the first IPTA data release~\cite{Verbiest:2016vem}.
As for SKA, we are even more optimistic, assuming an ambitious PTA with $N = 50$, $T = 1\,\textrm{week}$, and $\sigma_t = 30\,\textrm{ns}$.
These values  are inspired by the assumptions made in Refs.~\cite{Janssen:2014dka,Moore:2014lga}.
We thus obtain the following timing noise spectra for IPTA and SKA,
\begin{align}
D_{\rm noise}^{\rm IPTA}\left(f\right) \simeq 2.4 \times 10^{-8}\,\textrm{Hz}^{-3} \,,\quad D_{\rm noise}^{\rm SKA}\left(f\right) \simeq 1.1 \times 10^{-9}\,\textrm{Hz}^{-3} \,.
\end{align}


\smallskip\noindent In Fig.~\ref{fig:noise}, we plot the noise spectra of all present and future interferometer experiments that we consider this paper.
The timing noise of PTA experiments, which has dimension $\textrm{Hz}^{-3}$ instead of $\textrm{Hz}^{-1}$ and which is frequency-independent in the idealized case, is not shown.


\subsection{Strain noise power spectra}
\label{subsec:strain}


Based on the transfer functions and detector noise spectra introduced in Secs.~\ref{subsec:transfer} and \ref{subsec:detector}, we are now able to compute the (effective) strain noise power spectra of all auto- and cross-correlation searches for a SGWB signal that we are interested in [see Eqs.~\eqref{eq:DRS} and \eqref{eq:Seff}].


\smallskip\noindent\textbf{HL network:} The strain noise spectrum of a cross-correlation measurement using the aLIGO detectors in Hanford and Livingston at design sensitivity, $D_{\rm noise}^{\rm aLHO} = D_{\rm noise}^{\rm aLLO} = D_{\rm noise}^{\rm aLIGO}$, reads
\begin{align}
S_{\rm noise}^{\rm eff, HL}\left(f\right) & = \left(\frac{\Gamma_{\rm HL}^2\left(f\right)}{D_{\rm noise}^{\rm aLHO}\left(f\right) D_{\rm noise}^{\rm aLLO}\left(f\right)}\right)^{-1/2} = \left|\frac{D_{\rm noise}^{\rm aLIGO}\left(f\right)}{\Gamma_{\rm HL}\left(f\right)}\right| \,.
\end{align}


\smallskip\noindent\textbf{HLV network:} Adding the aVirgo detector to this network results in
\begin{small}
\begin{align}
S_{\rm noise}^{\rm eff, HLV}\left(f\right) = \left(\frac{\Gamma_{\rm HL}^2\left(f\right)}{D_{\rm noise}^{\rm aLHO}\left(f\right) D_{\rm noise}^{\rm aLLO}\left(f\right)} + \frac{\Gamma_{\rm HV}^2\left(f\right)}{D_{\rm noise}^{\rm aLHO}\left(f\right) D_{\rm noise}^{\rm aVirgo}\left(f\right)} + \frac{\Gamma_{\rm LV}^2\left(f\right)}{D_{\rm noise}^{\rm aLLO}\left(f\right) D_{\rm noise}^{\rm aVirgo}\left(f\right)}\right)^{-1/2} \hspace{-0.45cm}.
\end{align}
\end{small}%
Here, when considering the effective strain noise of this network at design sensitivity, we can again set $D_{\rm noise}^{\rm aLHO} = D_{\rm noise}^{\rm aLLO} = D_{\rm noise}^{\rm aLIGO}$.
However, when evaluating $S_{\rm noise}^{\rm eff}$ for the representative O2 noise spectra (see Tab.~\ref{tab:noise}), we have to explicitly distinguish between $D_{\rm noise}^{\rm aLHO}$ and $D_{\rm noise}^{\rm aLLO}$.


\smallskip\noindent\textbf{HLVK network:} Finally, adding KAGRA to the network yields an effective strain noise
\begin{small}
\begin{align}
& S_{\rm noise}^{\rm eff, HLVK}\left(f\right) = \left(\frac{\Gamma_{\rm HL}^2\left(f\right)}{D_{\rm noise}^{\rm aLHO}\left(f\right) D_{\rm noise}^{\rm aLLO}\left(f\right)} + \frac{\Gamma_{\rm HV}^2\left(f\right)}{D_{\rm noise}^{\rm aLHO}\left(f\right) D_{\rm noise}^{\rm aVirgo}\left(f\right)} + \frac{\Gamma_{\rm LV}^2\left(f\right)}{D_{\rm noise}^{\rm aLLO}\left(f\right) D_{\rm noise}^{\rm aVirgo}\left(f\right)}\right. \\\nonumber
& + \left. \frac{\Gamma_{\rm HK}^2\left(f\right)}{D_{\rm noise}^{\rm aLHO}\left(f\right) D_{\rm noise}^{\rm KAGRA}\left(f\right)} + \frac{\Gamma_{\rm LK}^2\left(f\right)}{D_{\rm noise}^{\rm aLLO}\left(f\right) D_{\rm noise}^{\rm KAGRA}\left(f\right)} + \frac{\Gamma_{\rm VK}^2\left(f\right)}{D_{\rm noise}^{\rm aVirgo}\left(f\right) D_{\rm noise}^{\rm KAGRA}\left(f\right)} \right)^{-1/2} \,,
\end{align}
\end{small}%
which we will only evaluate at design sensitivity, such that $D_{\rm noise}^{\rm aLHO} = D_{\rm noise}^{\rm aLLO} = D_{\rm noise}^{\rm aLIGO}$.
For more information on the HLV and HLVK networks, see also Refs.~\cite{Aasi:2013wya,LIGOScientific:2019hgc}.


\smallskip\noindent\textbf{CE:} The strain noise for an auto-correlation measurement solely using the CE detector reads
\begin{align}
S_{\rm noise}^{\rm CE}\left(f\right) = \frac{D_{\rm noise}^{\rm CE}\left(f\right)}{\mathcal{R}_{\rm CE}\left(f\right)} \,.
\end{align}


\smallskip\noindent\textbf{ET:} By contrast, a cross-correlation measurement using the three-detector ET network yields
\begin{align}
S_{\rm noise}^{\rm eff, ET}\left(f\right) = \frac{1}{\sqrt{3}}\left|\frac{D_{\rm noise}^{\rm ET}\left(f\right)}{\Gamma_{\rm ET}\left(f\right)}\right| \,,
\end{align}
assuming the same instrumental noise for all three detectors.
ET's three-detector configuration thus results in an enhancement of $\Gamma_{\rm ET}$ by a factor of $\sqrt{3}$ (see also Refs.~\cite{Regimbau:2012ir,Chan:2018csa}).


\smallskip\noindent\textbf{LISA:} Following Refs.~\cite{Caprini:2015zlo,Caprini:2019egz,Thrane:2013oya}, we consider an idealized auto-correlation search based on
\begin{align}
S_{\rm noise}^{\rm LISA}\left(f\right) = \frac{D_{\rm noise}^{\rm LISA}\left(f\right)}{\mathcal{R}_{\rm LISA}\left(f\right)} \,.
\end{align}
That is, we assume a perfect subtraction of instrumental noise thanks to real-time noise monitoring, and we neglect astrophysical foregrounds such as confusion noise from unresolved galactic and extragalactic white-dwarf binaries.
This foreground signal can be distinguished from the cosmological SGWB based on its spectral shape.
Moreover, it becomes increasingly less important over time, as more and more individual white-dwarf binaries are explicitly resolved in the course of the mission (see Refs.~\cite{Cornish:2018dyw,Cornish:2017vip} for details).
Nonetheless, it would certainly be interesting to refine our analysis in this paper by including the confusion noise from white-dwarf binaries.
We leave this task for future work.
For further studies on LISA's ability to measure a SGWB signal and identify its spectral shape, see also Refs.~\cite{Karnesis:2019mph,Caprini:2019pxz,Smith:2019wny}.


\smallskip\noindent\textbf{DECIGO, BBO:} The strain noise of a DECIGO\,/\,BBO cross-correlation measurement reads
\begin{align}
S_{\rm noise}^{\rm eff, I}\left(f\right) = \left|\frac{D_{\rm noise}^{\rm I}\left(f\right)}{\Gamma_{\rm I}\left(f\right)}\right| \,,\quad I = \textrm{DECIGO, BBO} \,.
\end{align}


\smallskip\noindent\textbf{NANOGrav, PPTA, EPTA:} The effective noise spectra of NANOGrav, PPTA, and EPTA can be found in Fig.\ 3 of Ref.~\cite{Arzoumanian:2018saf}, Fig.\ 2 of Ref.~\cite{Shannon:2015ect}, and Fig.\ 13 of Ref.~\cite{Lentati:2015qwp}, respectively.
These figures indicate the current sensitivity reach of the three experiments in terms of the characteristic strain amplitude $h_{\rm c}$, which is related to the strain power $S_h$ via $h_{\rm c} = \left(fS_h\right)^{1/2}$.
The main idea behind the quantity $h_{\rm c}$ is that it represents the typical amplitude of GWs on a logarithmic frequency scale.
To see this, consider a GW signal described by a strain power spectrum $S_{\rm signal}$ and make the replacement $S_{\rm signal}\rightarrow h_{\rm c}^2/f$ in the strain variance in Eq.~\eqref{eq:sigmah},
\begin{align}
\sigma_h^2 = \big<h_{ij}^{\vphantom{*}}h_{ij}^*\big> = \int_0^\infty d\ln f\:\frac{d\big<h_{ij}^{\vphantom{*}}h_{ij}^*\big>}{d\ln f} = 2\,\int_0^\infty d\ln f\:h_{\rm c}^2\left(f\right) \,.
\end{align}
This relation between GW strain power and characteristic amplitude allows us to convert the characteristic strain noise amplitudes in Refs.~\cite{Arzoumanian:2018saf,Shannon:2015ect,Lentati:2015qwp} to effective strain noise spectra,
\begin{align}
S_{\rm noise}^{\rm eff, I} = \frac{\left(h_{\rm c}^{\rm I}\left(f\right)\right)^2}{f} \,,\quad I = \textrm{NANOGrav, PPTA, EPTA} \,.
\end{align}
In Sec.~\ref{subsec:example}, we use these effective strain noise spectra to draw PLISCs for NANOGrav, PPTA, and EPTA (see Fig.~\ref{fig:plis}), assuming the following total observing times~\cite{Arzoumanian:2018saf,Shannon:2015ect,Lentati:2015qwp},
\begin{align}
t_{\rm obs}^{\rm NANOGrav} = 11.42\,\textrm{yr} \,,\quad t_{\rm obs}^{\rm PPTA} = 10.82\,\textrm{yr} \,,\quad t_{\rm obs}^{\rm EPTA} = 17.66\,\textrm{yr} \,. 
\end{align}
Refs.~\cite{Arzoumanian:2018saf,Shannon:2015ect,Lentati:2015qwp} also directly present upper limits on the strength of a power-law SGWB signal as a function of the spectral index $p$.
We explicitly check that these limits are consistent with the amplitudes $\Omega_p$ in Eq.~\eqref{eq:OPLISp}, which we require for the construction of the PLISCs.


\smallskip\noindent\textbf{IPTA, SKA:} We estimate the effective strain noise spectra of the future PTA experiments IPTA and SKA based on the transfer function in Eq.~\eqref{eq:GPTA} and the timing noise in Eq.~\eqref{eq:DPTA},
\begin{align}
\label{eq:SeffPTA}
S_{\rm noise}^{\rm eff, PTA} = \frac{D_{\rm noise}^{\rm PTA}\left(f\right)}{\mathcal{R}_{\rm PTA}\left(f\right)}\left(\sum_{J>I}\zeta_{IJ}^2\right)^{-1/2} = \left[\frac{2}{N\left(N-1\right)}\right]^{1/2}\frac{D_{\rm noise}^{\rm PTA}\left(f\right)}{\zeta_{\rm rms}\mathcal{R}_{\rm PTA}\left(f\right)} \,,
\end{align}
where $D_{\rm noise}^{\rm PTA}\left(f\right) = 2\,T\sigma_t^2$ and $\mathcal{R}_{\rm PTA}\left(f\right) = 1/\left(12\pi^2f^2\right)$ and where we assumed the same Hellings--Downs factor for all pulsar pairs, $\zeta_{IJ} = \zeta_{\rm rms}$ [see Eq.~\eqref{eq:zetarms}].
In Sec.~\ref{subsec:example}, we use the estimate in Eq.~\eqref{eq:SeffPTA} to draw the PLISCs for IPTA and SKA, assuming an effective observing time of $t_{\rm obs}^{\rm eff} = 20\,\textrm{yr}$.
Here, $t_{\rm obs}^{\rm eff}$ accounts for the fact that the scaling behavior of the SNR with $t_{\rm obs}$ changes for large values of $t_{\rm obs}$~\cite{Siemens:2013zla}.
We define $t_{\rm obs}^{\rm eff}$ such that the SNR in the weak-signal regime after an effective observing time $t_{\rm obs}^{\rm eff}$ [see Eq.~\eqref{eq:rho}] equals the true SNR in the intermediate- or strong-signal regime after an actual observing time $t_{\rm obs}$.
A more detailed discussion of the respective $t_{\rm obs}$ values for IPTA and SKA is left for future work.


\bibliographystyle{JHEP}
\bibliography{arxiv_3.bib}


\end{document}